\documentclass[12pt]{article}
\pdfoutput=1
\usepackage{amssymb}
\usepackage{epsfig}

\usepackage{graphicx}

\def\L{\mathcal L}
\def\e{\varepsilon}

\newcommand{\wt}{\widetilde}

\begin{document}

\def\a{\alpha}
\def\b{\beta}
\def\c{\chi}
\def\d{\delta}
\def\e{\epsilon}
\def\f{\phi}
\def\g{\gamma}
\def\h{\eta}
\def\i{\iota}
\def\j{\psi}
\def\k{\kappa}
\def\l{\lambda}
\def\m{\mu}
\def\n{\nu}
\def\o{\omega}
\def\p{\pi}
\def\q{\theta}
\def\r{\rho}
\def\s{\sigma}
\def\t{\tau}
\def\u{\upsilon}
\def\x{\xi}
\def\z{\zeta}
\def\D{\Delta}
\def\F{\Phi}
\def\G{\Gamma}
\def\J{\Psi}
\def\L{\Lambda}
\def\O{\Omega}
\def\P{\Pi}
\def\Q{\Theta}
\def\S{\Sigma}
\def\U{\Upsilon}
\def\X{\Xi}

%Varletters
\def\ve{\varepsilon}
\def\vf{\varphi}
\def\vr{\varrho}
\def\vs{\varsigma}
\def\vq{\vartheta}

\def\dg{\dagger}                                     % hermitian conjugate
\def\ddg{\ddagger}                                   % double dagger
\def\wt#1{\widetilde{#1}}                    % big tilde
\def\mt{\widetilde{m}_1}
\def\mti{\widetilde{m}_i}
\def\rt{\widetilde{r}_1}
\def\mtt{\widetilde{m}_2}
\def\mttt{\widetilde{m}_3}
\def\rtt{\widetilde{r}_2}
\def\mb{\overline{m}}
\def\VEV#1{\left\langle #1\right\rangle}        % < >
\def\be{\begin{equation}}
\def\ee{\end{equation}}
\def\ds{\displaystyle}
\def\ra{\rightarrow}

\def\bea{\begin{eqnarray}}
\def\eea{\end{eqnarray}}
\def\NO{\nonumber}
\def\Bar#1{\overline{#1}}

% Journal abbreviations (preprints)

\def\pl#1#2#3{Phys.~Lett.~{\bf B {#1}} ({#2}) #3}
\def\np#1#2#3{Nucl.~Phys.~{\bf B {#1}} ({#2}) #3}
\def\prl#1#2#3{Phys.~Rev.~Lett.~{\bf #1} ({#2}) #3}
\def\pr#1#2#3{Phys.~Rev.~{\bf D {#1}} ({#2}) #3}
\def\zp#1#2#3{Z.~Phys.~{\bf C {#1}} ({#2}) #3}
\def\cqg#1#2#3{Class.~and Quantum Grav.~{\bf {#1}} ({#2}) #3}
\def\cmp#1#2#3{Commun.~Math.~Phys.~{\bf {#1}} ({#2}) #3}
\def\jmp#1#2#3{J.~Math.~Phys.~{\bf {#1}} ({#2}) #3}
\def\ap#1#2#3{Ann.~of Phys.~{\bf {#1}} ({#2}) #3}
\def\prep#1#2#3{Phys.~Rep.~{\bf {#1}C} ({#2}) #3}
\def\ptp#1#2#3{Progr.~Theor.~Phys.~{\bf {#1}} ({#2}) #3}
\def\ijmp#1#2#3{Int.~J.~Mod.~Phys.~{\bf A {#1}} ({#2}) #3}
\def\mpl#1#2#3{Mod.~Phys.~Lett.~{\bf A {#1}} ({#2}) #3}
\def\nc#1#2#3{Nuovo Cim.~{\bf {#1}} ({#2}) #3}
\def\ibid#1#2#3{{\it ibid.}~{\bf {#1}} ({#2}) #3}

\title{
%{\normalsize \mbox{ }\hfill
%\begin{minipage}{3cm}
%MPP-2005-118
%\end{minipage}}\\
\vspace*{1mm}
\bf Supersymmetric  $SO(10)$-inspired leptogenesis and a new $N_2$-dominated scenario}
\author{
{\Large Pasquale Di Bari and Michele Re Fiorentin}
\\
{\it Physics and Astronomy}, 
{\it University of Southampton,} \\
{\it  Southampton, SO17 1BJ, U.K.}
}

\maketitle \thispagestyle{empty}

\vspace{-6mm}
%\centerline{\date{\today}}

\begin{abstract}
We study the supersymmetric extension of $SO(10)$-inspired thermal leptogenesis
showing the constraints on neutrino parameters and on the reheat temperature $T_{\rm RH}$
that derive from the condition of successful leptogenesis from next-to-lightest 
right handed (RH) neutrinos ($N_2$) decays and the more stringent ones 
when independence of the initial conditions (strong thermal leptogenesis) is superimposed.  
In the latter case, the increase of the lightest right-handed 
neutrino ($N_1$) decay parameters helps the  wash-out of a pre-existing asymmetry 
and constraints relax compared to the non-supersymmetric case. 
We find significant changes especially in the case of large $\tan\b$ values $(\gtrsim 15)$.
In particular, for normal ordering, the atmospheric mixing angle 
can now be also maximal. The lightest left-handed neutrino mass is still constrained within the range 
$10 \lesssim m_1/{\rm meV} \lesssim 30$ (corresponding to $75\lesssim \sum_i m_i/{\rm meV} \lesssim 120$). 
Inverted ordering is still disfavoured, but an allowed region satisfying strong thermal leptogenesis 
opens up at large $\tan\b$ values.  We  also study in detail the lower bound on $T_{\rm RH}$ finding 
$T_{\rm RH}\gtrsim 1 \times 10^{10}\,{\rm GeV}$   
independently of the initial $N_2$ abundance.  
%and $T^{\rm min}_{\rm RH}\simeq 4 \times 10^{9}\,{\rm GeV}$
%admitting some fine tuning in the seesaw formula.
Finally, we propose a new $N_2$-dominated scenario where the  $N_1$ mass is lower than the sphaleron freeze-out temperature. In this case there is no $N_1$ 
wash-out and we find  $T_{\rm RH} \gtrsim 1\times 10^{9}\,{\rm GeV}$.
These results indicate that $SO(10)$-inspired thermal leptogenesis can be made compatible
with the upper bound from the gravitino problem, an important result in light of 
the role often played by supersymmetry in the quest of a realistic model of fermion masses.
\end{abstract}

\newpage
\section{Introduction}

There is no evidence so far of new physics at the electroweak scale or below, in particular not of the kind that would be required in order to address  the problem of the  matter-antimatter asymmetry of the Universe within the Standard Model.
\footnote{The recent diphoton excess reported by the ATLAS and CMS collaborations \cite{diphoton},
if confirmed, might or might not have direct relevance for baryogenesis. It might have if the excess is explained for example by a new scalar as predicted in the NMSSM that would be able to re-open electroweak baryogenesis viability \cite{pietroni}.
Or, more indirectly, the excess could be associated to a new resonance signalling the existence of
new strong dynamics that might originate within a grand-unified theory embedding leptogenesis.} 
On the other hand, 
the lightness of neutrino masses, within a minimal type I seesaw mechanism \cite{seesaw},  would point to the existence of a very high energy scale intriguingly  close to the  grand-unified scale.  This encourages the idea that the cosmological
matter-antimatter asymmetry might have been generated in the early Universe
well above the electro-weak energy  scale.  Traditional high energy leptogenesis \cite{fy} scenarios based on the minimal type I seesaw mechanism naturally realise this interpretation of the current phenomenological picture.
However, testing these scenarios is challenging, relying on the possibility to find the way to over-constrain the large seesaw parameter space  imposing successful leptogenesis within a definite  model of new physics embedding the type I seesaw mechanism. 

A traditional, and somehow paradigmatic, example of models able to embed  the type I seesaw mechanism realising leptogenesis is given by  $SO(10)$-inspired models \cite{SO10inspired,afs}. 
In these models the fermion mass matrices, including the RH neutrino Majorana mass matrix, are not independent of each other but
linked by relations that reduce the number of independent parameters establishing connections, 
for example between the quark and the lepton sector. 
In particular, the Dirac neutrino masses are typically not too different from the up quark masses.  
 Moreover the mismatch between the flavour basis, where the charged lepton mass matrix is diagonal 
 and the Yukawa basis, where the neutrino Dirac mass matrix is diagonal, 
 can be described by a unitary matrix acting on the left-handed neutrino fields 
 with mixing angles comparable to those of the CKM matrix in the quark sector. 

$SO(10)$-inspired relations  are  realised not only within traditional $SO(10)$ models \cite{SO10}
but, mentioning some recent examples, also within models combining grand-unification 
with discrete flavour symmetries  \cite{A2Z} or with extra dimensions \cite{ferruccio}.
Barring fine tuned cancellations in the seesaw formula, the resulting RH neutrino mass spectrum would be 
highly hierarchical with the RH neutrino masses proportional to the squares of the up-quark masses
with typical values $(M_1, M_2, M_3) \sim (10^{5}, 10^{11}, 10^{15})\,{\rm GeV}$. 
In this case  the final asymmetry has to be  necessarily dominantly produced by 
the $N_2$ decays,  since the contributions both 
from the $N_1$ and from the heaviest RH neutrinos ($N_3$) decays are too small to explain the observed value:
an $N_2$-dominated scenario of leptogenesis is therefore naturally realised \cite{geometry}.  
It is interesting that this scenario necessarily requires the existence of at least three RH neutrino species in order for the  $N_2$ $C\!P$ asymmetries to get a sizeable contribution from the interference
between tree level $N_2$ decays  and one loop graphs with the exchange of virtual $N_3$'s. 
Therefore, there is an intriguing convergence between  the $SO(10)$ prediction for the existence of three RH neutrino species and the requirements of $N_2$-dominated leptogenesis. 
 
 A challenging crucial aspect of this scenario is the necessity for the asymmetry produced by the $N_2$ decays
 to survive the $N_1$ wash-out. Flavour effects \cite{bcst,flavour1,flavour2} 
 greatly enhance the region in the space of parameters
 where the $N_1$ wash-out is negligible since this acts separately 
 on the three charged lepton flavours \cite{vives,bounds}.  
 In this way it has been shown that flavour effects indeed rescue $SO(10)$-inspired models
 with strong hierarchical RH neutrino spectrum \cite{riotto1}. 
 Interestingly, imposing successful $SO(10)$-inspired leptogenesis one obtains constraints on low energy neutrino 
 parameters that can be testable \cite{riotto2}. These have been also derived and explained analytically in the approximation 
 where the mismatch between the flavour and the Yukawa basis is neglected \cite{SO10decription}. 
 In particular the lightest left-handed (LH) neutrino mass is constrained within the range
 $1\,{\rm meV} \lesssim m_1 \lesssim 300\,{\rm meV}$. 
 The upper bound
 \footnote{Notice that this is more relaxed compared to the upper bound holding in the $N_1$ dominated scenario where $m_1 \lesssim 0.1\,{\rm eV}$ \cite{upperbound,pedestrians,bounds}.} has now been tested by latest cosmological results
 that place an upper bound on the sum of the neutrino masses $\sum_i m_i \lesssim 0.23 \,{\rm eV}$ \cite{planck}, 
 translating into $m_1 \lesssim 70\,{\rm meV}$.  \footnote{Future cosmological observations should be able to constraint
 $m_1\gtrsim 10\,{\rm meV}$   at $95\%$ C.L. and in this case they would test 
 most of the window allowed by $SO(10)$-inspired leptogenesis.}
 
 Another interesting constraint is placed on the atmosperic mixing angle. 
 This has to be necessarily in the second octant in the case of inverted ordered
(IO)  neutrino masses.  More stringent constraints on the low energy neutrino parameters can be obtained superimposing
additional conditions. An interesting possibility is to impose the so called {\em strong thermal condition}, 
 the requirement that the asymmetry is independent of the initial conditions. This is indeed nicely realised within 
$SO(10)$-inspired models \cite{strongSO10} and results into a `strong thermal $SO(10)$-inspired solution' characterized by
normally ordered (NO) neutrino masses, lightest neutrino mass in the range $10\,{\rm meV} \lesssim m_1 \lesssim 30\,{\rm meV}$,
atmospheric mixing angle in the first octant and Dirac phase $\d \sim -45^{\circ}$,
in very nice agreement with current best fit results from neutrino oscillation experiments global analyses \cite{global}. 

Recently it has been shown that flavour coupling \cite{bcst,spectators,flcoupling1,flavour1} can help to open new solutions \cite{flcoupling2} 
and these can be crucial to realise successful leptogenesis within specific models. 
An explicit example has been recently obtained in \cite{flcoupling3}
within a specific realistic grand unified model, the `A to Z model' \cite{A2Z},  obtaining quite definite
predictions on the atmospheric mixing angle $(\theta_{23}\sim 52^{\circ})$, Dirac phase ($\d \sim 20^{\circ}$) and on the ordering (NO). 
 Alternatively, at the expense of very highly fine tuned seesaw cancellations,  
in the vicinity of a crossing level solutions one can have a departure from
a very highly hierarchical pattern \cite{afs} in  a way that $M_1$ can be uplifted  and its $C\!P$ asymmetry strongly 
enhanced. Recently this kind of solution has been realised within  a realistic fit
of quark and neutrino parameters within $SO(10)$ models. In this case the uplift of $M_1$
is also accompanied by a simultaneous decrease of $M_3$ so that a compact spectrum is obtained 
\cite{rodejohann} and this can also lead to successful leptogenesis \cite{compact}.  
An unpleasant feature of these solutions, in addition to the very high fine tuning, 
is that, because of the uplift of $M_1$, they predict NO and  too small values 
for the neutrinoless double beta decay effective neutrino mass $m_{ee}$  
to be measured \cite{SO10decription}. 

Supersymmetric extensions of $SO(10)$-models are important since 
they offer a  traditional way to address naturalness. At the same time 
they help improving the goodness of fits of lepton and quark parameters \cite{SUSYSO10,rodejohann}.
Recently \cite{rodejohann} good fits of the fermion parameters have been obtained within $SO(10)$ models 
with hierarchical RH neutrino masses and  interestingly  IO light neutrino masses,  
leading to values of $m_{ee}$ well in the reach of next generation neutrinoless double beta decay experiments.   
However, supersymmetry is typically implemented as a local symmetry leading to supergravity
and in this case one has to worry whether successful thermal leptogenesis can be achieved with
values of  $T_{\rm RH}$ compatible with the upper bound from the solution of the gravitino problem 
\cite{gravitino}. A quite conservative model independent upper bound, $T_{\rm RH}\lesssim 10^{10}\,{\rm GeV}$,
comes from preventing Dark Matter over abundance, where the Dark Matter particle can be either the neutralino or the gravitino itself or some other
hidden sector lighter particle depending whether the gravitino is or it is not the lightest supersymmetric particle. 
\footnote{It should be noticed, however, that different ways  to circumvent even this upper bound
have been proposed. For example thanks to entropy production
diluting Dark Matter abundance \cite{yanagida} or in models with mixed axion/axino Dark Matter \cite{baer}  or yet another way to evade completely this upper bound is that the gravitino
is heavier than $\sim 10^7\,{\rm GeV}$ in a way that its life-time is so short to decay before neutralino dark matter freeze-out \cite{arkani}.}
 
In this paper we extend the study of $SO(10)$-inspired leptogenesis to the supersymmetric case,
showing how the constraints derived in the non-supersymmetric case change, with a particular focus 
on the lower bound on  $T_{\rm RH}$.  
We find that in a traditional scenario, where the lightest RH neutrino wash-out has to be taken into account, 
this can be as low as $\sim 10^{10}\,{\rm GeV}$ or even below admitting some fine tuning in the seesaw
parameters and an initial thermal $N_2$ abundance. 
These results indicate that, in those supersymmetric scenarios 
where the gravitino is heavier than $\sim 30\,{\rm TeV}$ and decays prior to the onset of BBN,
$SO(10)$-inspired thermal leptogenesis can be indeed reconciled with the gravitino problem.  
Similar analysis, though for more specific choices of the parameters, 
has been also done in \cite{marfatia}, finding a much more stringent lower bound 
$T_{\rm RH} \gtrsim 5\times 10^{11}\,{\rm GeV}$
and concluding that thermal $SO(10)$-inspired leptogenesis is incompatible with the upper bound from the gravitino problem
thus motivating a non-thermal scenario. We will comment on this difference between our results and those of \cite{marfatia}.

We also propose a new  scenario where the lightest RH neutrino mass is comparable or below the sphaleron 
freeze-out temperature $T_{\rm sph}^{\rm out}\sim 100\,{\rm GeV}$ \cite{sphalerons} in a way that the lightest RH neutrino 
wash-out occurs too late to wash-out the baryon asymmetry. In this case we show that values of $T_{\rm RH}$ as low as $\sim 10^9\,{\rm GeV}$
are possible. Therefore, our results  indicate that supersymmetric $SO(10)$-inspired thermal leptogenesis can be reconciled with the gravitino problem and is certainly not ruled out model independently. 

The paper is organised in the following way.  In Section 2 we show how the calculation of the asymmetry can be extended to a supersymmetric $N_2$-dominated scenario. In Section 3 we study $SO(10)$-inspired 
(supersymmetric) leptogenesis deriving the constraints on the low energy neutrino parameters and comparing them with those obtained in the non-supersymmetric case in \cite{riotto2,strongSO10,SO10decription}.  In Section 4 we discuss in detail
the lower bound on $T_{\rm RH}$ showing that values as low as $\simeq 1 \times 10^{10}\,{\rm GeV}$
are possible. In Section 5 we discuss a new  $N_2$-dominated scenario where the lightest RH neutrino mass is lower than the sphaleron freeze out temperature, so that the $N_1$ wash-out is absent.  We recalculate the lower bound on $T_{\rm RH}$ 
in this scenario obtaining  $T_{\rm RH}\gtrsim 1 \times 10^9\,{\rm GeV}$,
enlarging even more the region of compatibility with the gravitino problem.  
In Section 6 we draw some final remarks and conclude. 

\section{Calculation of the asymmetry within supersymmetric  $N_2$-dominated leptogenesis}

In this section we extend the calculation of the asymmetry in  the $N_2$-dominated
scenario, as rising from $SO(10)$-inspired conditions, to a supersymmetric framework.  

First of all we assume a minimal type I seesaw extension of the MSSM introducing 
three RH neutrinos $N_{i R}$ with Yukawa couplings $h$ and Majorana mass $M$.  
In the flavour basis, where  the charged lepton and the Majorana mass matrices are both diagonal, 
the masses and Yukawa couplings relevant for leptogenesis are given by the following terms in the superpotential   \cite{crv,plumacher} ($\a=e,\m,\t$)
\be
{\cal W}_{\ell+\nu+N}=  \overline{\a_L} \,\epsilon \, H_{d}\, D_{h_{\ell}}\,\a_R + 
                              H_u\,\epsilon\,\overline{\nu_{\a L}}\,h_{\nu \a i} \, N_{i R} +
                               {1\over 2} \, \overline{N^{c}_{i R}} \, D_{M} \, N_{i R}  + \mbox{\rm h.c.}\, ,
\ee 
where $ D_{h_{\ell}} \equiv {\rm diag}(h_e, h_{\mu}, h_{\tau})$,
$D_M \equiv {\rm diag}(M_1, M_2, M_3)$, with $M_1 \leq  M_2 \leq M_3$,
and $\epsilon$ is the totally anti-symmetric tensor. 

After spontaneous symmetry breaking the two neutral Higgs field vev's 
generate the Dirac masses  for the charged leptons and for the neutrinos, respectively
\be
m_{\ell} = v_{d}\,h_{\ell}   \hspace{10mm} \mbox{\rm and} \hspace{10mm}
m_{D} = v_{u}\,h_{\nu}  \,  ,
\ee
with $\tan\beta \equiv v_u/v_d$ and $v = \sqrt{v^2_u + v^2_d} \simeq 174.6 \,{\rm GeV}$, where 
$v$ is the SM Higgs vev.
The Dirac mass matrix in the flavour basis can be expressed through the  singular value decomposition
(or bi-unitary parameterisation) as 
\be
m_D = V_L^{\dagger}\,D_{m_D} \, U_R  \,  ,
\ee
where $D_{m_D} \equiv {\rm diag}(m_{D1},m_{D2},m_{D3})$ is the neutrino Dirac mass matrix 
in the Yukawa basis and $V_L$ and $U_R$ are the unitary matrices  acting respectively on the 
LH and RH neutrino fields in the transformation from the flavour basis to the Yukawa basis. 

In the seesaw limit, for $M \gg m_D$, the spectrum of neutrino mass eigenstates 
splits into a very heavy set with masses almost coinciding  with the Majorana masses $M_i$
and into a light set  $\nu_i \simeq \nu_{iL} + \nu_{i L}^c$,  with a  symmetric mass matrix $m_{\nu}$ given by the seesaw formula
\be\label{seesaw}
m_{\nu} = - m_D \, {1\over D_M} \, m_D^T  \,  .
\ee
This is diagonalised by a unitary matrix $U$,
\be\label{leptonic}
U^{\dagger} \,  m_{\nu} \, U^{\star}  =  - D_m  \,  ,
\ee
where $D_m \equiv {\rm diag}(m_1,m_2,m_3)$ with $m_1 \leq m_2 \leq m_3$,
corresponding to the PMNS leptonic mixing matrix, in a way that we can write
\be\label{diagonalseesaw}
 D_m =  U^{\dagger} \, m_D \, {1\over D_M} \, m_D^T  \, U^{\star}     \,   .
\ee
Assuming $SO(10)$-inspired conditions,  i) $I \leq V_L \leq V_{CKM}$ and 
ii) $\a_i \equiv m_{Di}/m_{\rm q_i} ={\cal O}(0.1$--$10)$, where $m_{q_i}$ are the three up quark masses, 
$m_u$, $m_c$ and $m_t$ for $i=1,2,3$ respectively, the RH neutrino masses are approximated by the
following simple analytical expressions \cite{afs,SO10decription,flcoupling3}
\be\label{Mi}
M_1 \simeq {(m_{D 1})^2 \over |(\widetilde{m}_{\nu})_{11}|} \,    , 
\hspace{5mm}
M_2 \simeq {(m_{D 2})^2 \,|(\widetilde{m}_{\nu})_{11}| \over m_1\,m_2\,m_3\, |(\widetilde{m}^{-1}_{\nu})_{33}|}\,  ,
\hspace{5mm}
M_3 \simeq (m_{D 3})^2 \, |(\widetilde{m}^{-1}_{\nu})_{33}|  \,  ,
\ee
where $\widetilde{m}_{\nu}\equiv V_L \, m_{\nu} \, V_L^T$  is the light neutrino mass matrix in the  
Yukawa basis.
These expressions show that under $SO(10)$-inspired conditions, barring fine tuned conditions on 
$(\widetilde{m}_{\nu})_{11}$ and $(\widetilde{m}^{-1}_{\nu})_{33}$
\footnote{Recently it has been noticed that such a fine tuning can be precisely quantified in terms of the orthogonal matrix \cite{flcoupling3}.}, the RH neutrino masses are highly hierarchical and in particular 
$M_1 \ll 10^9\,{\rm GeV}$ and $M_2 \gg 10^9\,{\rm GeV}$, 
in a way that  the $N_2$-dominated scenario is realised, where the asymmetry is necessarily 
produced by the $N_2$'s. 

A general calculation of the asymmetry valid for any mass regime should proceed within
a density matrix formalism \cite{bcst,flavour2,densitymatrix,densitymatrix2}. 
However, except for some transition regimes, 
the mass of the $N_2$ producing the asymmetry, $M_2$, falls within so called fully flavoured regimes where
the density matrix equation simplifies into Boltzmann equations \cite{densitymatrix2} and in this case the final asymmetry can be calculated using simple approximate analytic expressions.  

We will neglect flavour coupling effects  \cite{bcst,flavour1,flcoupling1,flcoupling2}, 
that can in some cases  produce dominant contributions to the final asymmetry \cite{flcoupling2,flcoupling3}
and have been studied in detail in the supersymmetric case in \cite{fong}, 
but we will comment in the conclusions on the impact they can have on our results. 
We will also not pursue here the case of soft leptogenesis, offering a way to lower the 
scale of leptogenesis circumventing the gravitino problem \cite{soft}. 

It is important to notice that, within a supersymmetric framework, the $N_2$-production of the
asymmetry for a fixed mass $M_2$  can occur in different fully flavoured regimes  depending
on the value of $\tan\b$ since charged lepton interaction rates involving leptons 
are  $\propto (1+\tan^2 \beta)$ \cite{flavour1}.   On the other hand since, because of our working assumptions,
one has $M_1 \ll 10^{9}\,{\rm GeV}$,  the lightest RH neutrino produced asymmetry is always negligible 
and the $N_1$ wash-out occurs always  in the three-flavoured regime independently of the value of $\tan\b$.  
We can then distinguish three fully flavoured regimes for the calculation of the asymmetry:
\begin{itemize}
\item In the unflavoured 
\footnote{Here we refer to an `unflavoured' regime rather than to a `one-flavoured' 
regime, as sometimes it is done, since we refer only to the number of charged lepton flavours.}
regime,  for $M_2 \gg 5\times 10^{11}\,{\rm GeV}\,(1+\tan^2\beta)$, the 
final $B-L$ asymmetry can be calculated using
\bea\label{unfl} 
N_{B-L}^{\rm f} & \simeq &
\left[{K_{2 e}\over K_{2}}\, \ve_{2}\,\kappa(K_{2}) +
\left(\ve_{2 e} - {K_{2 e}\over K_{2}}\, \ve_{2} \right)\, \kappa(K_{2}/2) \right]
\, e^{-{3\pi\over 8}\, K_{1 e}} \,  \\ \nonumber
& + &\left[{K_{2 \m}\over K_{2}}\,
\ve_{2}\,\kappa(K_{2}) +
\left(\ve_{2\m} - {K_{2\m}\over K_{2}}\, \ve_{2} \right)\,
\kappa(K_{2}/2) \right]
\, e^{-{3\pi\over 8}\,K_{1\m}} \,  \\ \nonumber
& + &
\left[{K_{2\t}\over K_{2}}\,\ve_{2}\,\kappa(K_{2}) 
+ \left(\ve_{2\t} - {K_{2\t}\over K_{2}}\, \ve_{2} \right)\,\kappa(K_{2}/2)\right]\,
\, e^{-{3\pi\over 8}\,K_{1 \t}} \,   .
\eea
\item In the two-(fully) flavoured regime, for  $5\times 10^{11}\,{\rm GeV}\,(1+\tan^2\beta) \gg M_2 \gg 5\times 10^{8}\,{\rm GeV}\,(1+\tan^2\beta)$,  the  final $B-L$ asymmetry can  be calculated using
\bea\label{twofl} \nonumber
N_{B-L}^{\rm f} & \simeq &
\left[{K_{2e}\over K_{2\tau_2^{\bot}}}\,\ve_{2 \tau_2^{\bot}}\kappa(K_{2 \tau_2^{\bot}}) 
+ \left(\ve_{2e} - {K_{2e}\over K_{2\tau_2^{\bot}}}\, \ve_{2 \tau_2^{\bot}} \right)\,\kappa(K_{2 \tau_2^{\bot}}/2)\right]\,
\, e^{-{3\pi\over 8}\,K_{1 e}}+ \\ \nonumber
& + &\left[{K_{2\mu}\over K_{2 \tau_2^{\bot}}}\,
\ve_{2 \tau_2^{\bot}}\,\kappa(K_{2 \tau_2^{\bot}}) +
\left(\ve_{2\mu} - {K_{2\mu}\over K_{2\tau_2^{\bot}}}\, \ve_{2 \tau_2^{\bot}} \right)\,
\kappa(K_{2 \tau_2^{\bot}}/2) \right]
\, e^{-{3\pi\over 8}\,K_{1 \mu}}+ \\
& + &\ve_{2 \tau}\,\kappa(K_{2 \tau})\,e^{-{3\pi\over 8}\,K_{1 \tau}} \,  ,
\eea
where we indicated with $\tau_2^{\bot}$ the electron plus muon component of the 
quantum flavour states produced by the $N_2$-decays defining 
$K_{2\tau_2^{\bot}}\equiv K_{2e}+K_{2\mu}$, $\ve_{2\tau_2^{\bot}}\equiv \ve_{2e}+\ve_{2\mu}$.
\item Finally, in the three-flavoured regime, for $M_2 \ll 5\times 10^{8}\,{\rm GeV}\,(1+\tan^2\beta)$,
the  final $B-L$ asymmetry can be be calculated using
\bea\label{threefl} 
N_{B-L}^{\rm f}  \simeq 
         \ve_{2 e}\,\kappa(K_{2 e}) \, e^{-{3\pi\over 8}\,K_{1 e}} +  \ve_{2 \mu}\,\kappa(K_{2 \mu}) \, e^{-{3\pi\over 8}\,K_{1 \mu}} 
 + \ve_{2 \tau}\,\kappa(K_{2 \tau})\,e^{-{3\pi\over 8}\,K_{1 \tau}} \,  .
\eea
\end{itemize}
As we discussed, in the transition regimes, about $M_2\sim 5\times 10^{11}\,{\rm GeV}\,(1+\tan^2\beta)$
and $M_2 \sim 5\times 10^{8}\,{\rm GeV}\,(1+\tan^2\beta)$, the asymmetry  should be calculated
using density matrix equations. We will describe these transition regimes switching abruptly from 
one fully flavoured regime to another at the two given values of $M_2$. 
We will also comment on the impact of this `step approximation'. 

The total and flavoured decay parameters, $K_i$ and $K_{i\a}$ respectively, can be still written as
\be
K_{i\a} \equiv {\G(T=0) \over H(T= M_i)}  = {|m_{D \a i}|^2 \over m_{\star}^{MSSM}\, M_i}  
\hspace{5mm}\mbox{\rm and} \hspace{5mm}
K_i = \sum_{\a}\,K_{i\a}= {(m^{\dagger}_D \, m_D)_{ii} \over m_{\star}^{MSSM}\, M_i} \,  ,
\ee
but the  equilibrium neutrino mass is now given by \cite{proceedings,marfatia}
\be\label{mstar}
m_{\star}^{MSSM} \equiv {8\, \pi^{5/2}\,\sqrt{g_{\star}^{MSSM}} \over 3\,\sqrt{5}}\,
{v^2_u \over M_{Pl}} = {1\over 2}\,\sqrt{g_{\star}^{MSSM}\over g_{\star}^{SM}}\, m_{\star}^{SM} 
\,\sin^2\b \simeq 0.78\times 10^{-3}\,{\rm eV} \, \sin^2\b \,  ,
\ee
having taken into account that: i)  RH neutrinos and sneutrinos have a doubled number of decay channels 
compared to the SM case that simply doubles the rates and ii) the  number of ultra-relativistic
degrees of freedom $g_{\star}^{MSSM} = 915/4$.  This implies that the (total and flavoured) decay 
parameters are $\sim \sqrt{2}$ larger than in the SM. 
\footnote{We will assume that the number of ultra-relativistic degrees of freedom stays constant between
$N_2$ production at $T\sim M_2$ and $N_1$ wash-out occurring at $T\sim M_1$. However, in general one can think of (model dipendent) supersymmetric models where $M_1$ is low enough that some supersymmetric degrees of freedom associate to heavier particles get suppressed in between. In Section 5 we will consider
a scenario with $M_1 \lesssim 100\,{\rm GeV}$ but in that case we will point out that the 
$N_1$ wash-out at all can be neglected so in any case this point has no relevance.}
This will clearly tend to enhance the wash-out both at the production,  depending on the $K_{2\a}$'s, 
and from the lightest RH neutrinos, depending on the $K_{1\a}$'s.  The wash-out at the production 
is described by the efficiency factor $\k(K_{2\a})$ that for an initial thermal 
$N_2$ abundance can be calculated as \cite{pedestrians,predictions} 
\be\label{kappa}
\k(K_{2\a}) = 
{2\over z_B(K_{2\a})\,K_{2\a}}
\left(1-e^{-{K_{2\a}\,z_B(K_{2\a})\over 2}}\right) \,  , \;\; z_B(K_{2\a}) \simeq 
2+4\,K_{2\a}^{0.13}\,e^{-{2.5\over K_{2\a}}} \,   .
\ee
For an initial vanishing $N_2$ abundance this is the sum of a negative and a positive contribution \cite{pedestrians},
\be
\k(K_{2\a},K_2)   =\k_{-}^{\rm f}(K_2,K_{2\a})+ \k_{+}^{\rm f}(K_2,K_{2\a}) \, ,
\ee
that are approximated  by the following expressions \cite{predictions}
\be\label{k-}
\k_{-}^{\rm f}(K_2,K_{2\a})\simeq
-{2\over p_{2\a}^{0}}\ e^{-{3\,\pi \over 8}\,K_{2\a} }
\left(e^{{p_{2\a}^{0}\over 2}\,\overline{N}(K_2)} - 1 \right) 
\ee
and
\be\label{k+}
\k_{+}^{\rm f}(K_2,K_{2\a})\simeq
{2\over z_B(K_{2\a})\,K_{2\a}}
\left(1-e^{-{K_{2\a}\,z_B(K_{2\a})\,\overline{N}(K_2)\over 2}}\right) \, ,
\ee
where
\begin{equation}\label{nka}
\overline{N}(K_2)\equiv {N(K_2)\over\left(1 + \sqrt{N(K_2)}\right)^2} \, ,
\end{equation}
and $p^0_{2\a} = K_{2\a}/K_2$ is the tree level probability that the 
lepton  quantum state  produced by a $N_2$-decay  
is measured as an $\a$ flavour eigenstate.
If the asymmetry is produced in the strong wash-out regime, the two expressions
converge to the same asymptotic limit and there is no dependence on the initial $N_2$ abundance. 

The other important modification to be taken into account, compared to the non-supersymmetric case,
is that now there are also more interference terms  contributing to the $C\!P$ asymmetries 
and one obtains \cite{crv}
\bea \label{eps2al}
\varepsilon_{2 \alpha}
        &\,=\,& {3\over 8\,\pi}\,\frac{M_2\,m_{\rm atm}}{v^2}\,\, \,
 \sum\limits_{j \neq 2}  \, \left(
{\cal I}_{2j}^\alpha \, {\xi(M_j^2/M_2^2)}
+ {2\over 3}\,{\cal J}_{2j}^\alpha \, \frac{M_j/M_2}{M_j^2/M_2^2 -1} \right) \, ,
\eea
where we defined \cite{2RHN},
\bea \label{eq:calIJ}
{\cal I}_{2j}^\alpha \equiv
{{\rm Im} \Big[ \big(m_D^\dagger \big)_{i \alpha} \, \big(m_D \big)_{\alpha j}
\big(m_D^\dagger m_D \big)_{ij} \Big]\over M_2\,M_j\,\mtt \, m_{\rm atm}}~,~~
{\cal J}_{2j}^\alpha \equiv {{\rm Im}
\Big[\big (m_D^\dagger \big)_{i \alpha} \, \big(m_D \big)_{\alpha j} \big(m_D^\dagger m_D \big)_{ji} \Big]
\over  M_2\,M_j\,\mtt \, m_{\rm atm}} \, ,
\eea
with $\mtt \equiv (m_D^{\dag}\, m_D)_{2 2}/M_2$, and
\be
\xi(x)=\frac{x}{3}\,\left[\ln\left(\frac{1+x}{x}\right)-\frac{2}{1-x}\right] \,  .
\ee
In the hierarchical RH neutrino mass limit  one has $\xi(x) \ra 1$ and moreover  
terms $\propto {\cal I}_{21}^\alpha\,\xi(M^2_1/M^2_2), {\cal J}_{21}^\alpha,  {\cal J}_{23}^\alpha$
are strongly suppressed in the $N_2$-dominated scenario so that the
$\ve_{2\a}$'s  can be approximated simply by 
\be\label{eps2al}
\ve_{2\a} \simeq  {3\over 8\,\pi}\frac{\,M_2\,m_{\rm atm}}{v^2}\, {\cal I}_{23}^\alpha  \,  .
\ee
Compared to the SM case, for a given set of values of the seesaw parameters, the $C\!P$ asymmetries are double. Finally the baryon-to-photon number ratio can be calculated from the final $B-L$ asymmetry produced 
by the RH neutrinos (or sneutrinos),  as $\eta_B \simeq d^{MSSM}\, N_{B-L}^{\rm f}$, where \cite{proceedings}
\be
d^{MSSM} =    
2\,\left({a_{\rm sph}\over N_{\g}^{\rm rec}}\right)^{MSSM} 
\simeq 0.89 \times 10^{-2} \simeq 0.92\, d^{SM} \,   ,
\ee
having taken into account a factor $2$ from the sum of the asymmetry generated by RH neutrinos and sneutrinos, the 
sphaleron conversion coefficient $a^{MSSM}_{\rm sph} = 8/23$ \cite{ks88} and that the number of photons
at recombination is given by $(N_{\g}^{\rm rec})^{MSSM}=4\,g_{\star}^{MSSM}/(3\,g_{\star}^{\rm rec})\simeq 78$
since $g_{\star}^{MSSM}=915/4$,  $g_{\star}^{\rm rec}\simeq 3.91$ and one has to consider that
in the portion of co-moving volume containing one RH neutrino in ultra-relativistic equilibrium there are $4/3$ photons.

In the non supersymmetric case, and in the approximation $V_L \simeq I$, 
the solutions are tauon dominated \cite{riotto1} and, as shown in \cite{SO10decription}, the asymmetry is well described by a full analytical expression. We can extend this analytical expression, for the tauon 
contribution to the final asymmetry, to  the supersymmetric case with the simple modifications we discussed
\footnote{In addition we are correcting a typo that we found in \cite{SO10decription} where instead of the
term ${|(m^{-1}_{\n})_{\m\t}|^2 / |(m^{-1}_{\n})_{\t\t}|^2}$ there is, incorrectly, its inverse.}, 
obtaining
\bea\label{NBmLf}
\left.N_{B-L}^{\rm f} \right|_{V_L=I} & \simeq &  
{3\over 8\,\pi}\, {\a_2^2\,m_c^2 \over v^2}\, {|m_{\nu ee}|\,
(|(m^{-1}_{\nu})_{\t \t}|^2 + |(m^{-1}_{\nu})_{\m \t}|^2)^{-1} \over m_1\,m_2\,m_3}\,
{|(m^{-1}_{\n})_{\m\t}|^2\over |(m^{-1}_{\n})_{\t\t}|^2}\,\sin\a_L    \\  \nonumber
& \times & \kappa\left({m_1\,m_2\,m_3 \over m_{\star}}\, 
{|(m_{\nu}^{-1})_{\m \t}|^2 \over |m_{\nu ee}|\, |(m_{\nu}^{-1})_{\t \t}|} \right)  \\  \nonumber
& \times & 
e^{-{3\pi\over 8}\,{|m_{\nu e\t}|^2 \over m_{\star}\,|m_{\nu ee}|}  }  \,   ,
\eea
with
\be
\a_L =  {\rm Arg}\left[m_{\nu ee}\right]  - 2\,{\rm Arg}[(m^{-1}_{\nu})_{\m\t}] + \pi -2\,(\rho+\s)  \,  .
\ee 
We will have of course to check whether the tau dominance, holding for $V_L=I$ in the non-supersymmetric case, still holds in the supersymmetric case.  

Finally, we also want to give the expression for the relic value of a 
pre-existing asymmetry and the condition for its wash-out (strong thermal leptogenesis condition)
that we will superimpose to the successful leptogenesis condition, extending 
the results found in the non-supersymmetric case \cite{strongSO10}.

If the production occurs in the unflavoured regime, for $M_2 \gg 5 \times 10^{11}\,{\rm GeV}\,(1+\tan^2\b)$,
then it is impossible to realise successful strong thermal leptogenesis since the $N_2$
wash-out cannot suppress completely the pre-existing asymmetry in any of the three (charged lepton) flavours. 
The pre-existing asymmetry can be only washed-out by the lightest RH neutrinos in all three flavours 
\cite{engelhard} but in this way
it also suppresses the $N_2$ produced asymmetry and one cannot attain successful leptogenesis.
On the other hand if the $N_2$ production occurs in the two fully-flavoured regime, 
for $5 \times 10^{11}\,{\rm GeV}\,(1+\tan^2\b) \gg M_2 \gg 5 \times 10^{8}\,{\rm GeV}\,(1+\tan^2\b)$, then
the relic value of the pre-existing $B-L$ asymmetry is given by
\be
N^{\rm p,f}_{B-L}= N_{\D_\t}^{\rm p,f} + N_{\D_\m}^{\rm p,f} + N_{\D_e}^{\rm p,f} \,  ,
\ee
where
\bea\label{finalpas}
N_{\D_\t}^{\rm p,f} & = & 
p^0_{{\rm p}\t}\,  e^{-{3\pi\over 8}\,(K_{1\t}+K_{2\t})} \, N_{B-L}^{\rm p,i} \,  , \\  \nonumber
N_{\D_\m}^{\rm p,f} & = & 
(1-p^0_{{\rm p}\t})\,\,e^{-{3\pi\over 8}\,K_{1\m}} \,\left[ 
p^0_{\mu\t_2^{\bot}}\, p^0_{p\t^\bot_2}\,
e^{-{3\pi\over 8}\,(K_{2e}+K_{2\m})} + (1-p^0_{\m \t_2^{\bot}})\,(1-p^0_{p\t^\bot_2}) \, \right] 
\, N_{B-L}^{\rm p,i}
 ,  \\  \nonumber
N_{\D_e}^{\rm p,f}& = & 
(1-p^0_{{\rm p}\t})\,\,e^{-{3\pi\over 8}\,K_{1e}} \,\left[ p^0_{e\t_2^{\bot}}\,
 p^0_{p\t^\bot_2}\,e^{-{3\pi\over 8}\,(K_{2e}+K_{2\m})} + 
 (1-p^0_{e \t_2^{\bot}})\,(1-p^0_{p\t^\bot_2}) \, \right] 
\, N_{B-L}^{\rm p,i} \,   .
\eea
In this case imposing $K_{2\tau}, K_{1\mu}, K_{1e} \gg 1 $ and $K_{1\t} \lesssim 1$ 
one can wash-out the pre-existing asymmetry but not the  tauonic component of the $N_2$ produced 
asymmetry \cite{problem}. 
This is the case holding in the non-supersymmetric case or in the supersymmetric case for
small $\tan\b$ values. On the other hand for sufficiently large $\tan\b$ values, 
such that $M_2 \ll 5 \times 10^{8}\,{\rm GeV}\,(1+\tan^2\b)$,
the production occurs in the thee-flavoured regime and in this case
the relic value of the final flavoured asymmetries are simply modified 
by the replacement $K_{1\mu}\ra K_{1\m}+K_{2\m}$ and $K_{1e}\ra K_{1e}+K_{2e}$
in the exponentials.  In this way the conditions for the wash-out of the pre-existing 
asymmetry are now less stringent since one has to impose $K_{2\t}, K_{1\m}+K_{2\m}, K_{1e}+K_{2e} \gg 1$,
so that one can also have $K_{2\m} \gg 1$ and $K_{1\m} \lesssim 1$, 
washing-out the pre-existing asymmetry and having a final muon (instead of tauon) dominated asymmetry,
a new situation compared to the non-supersymmetric case.

\section{Constraints on the low energy neutrino parameters}

In the $SO(10)$-inspired scenario of leptogenesis that we described, the asymmetry formally depends on the nine parameters in the low energy neutrino mass matrix, on  the six parameters in the matrix $V_L$ 
and on the three $\a_i$.  
As we discussed the 3 RH neutrino masses $M_i$ and the RH neutrino mixing matrix $U_R$ can be expressed
in terms of these parameters.  However, since the finally asymmetry is dominated by the $N_2$ contribution,
the dependence on $\a_1$ and $\a_3$ cancels out (this can be seen analytically in 
the eq.~(\ref{NBmLf}) for $V_L=I$ but the result remains true for a generic $V_L$) and this is crucial to understand why one gets constraints on the low energy neutrino parameters.

We have numerically calculated the final asymmetry and imposed the condition of successful leptogenesis in the $SO(10)$-inspired case producing scatter plots 
in the space of parameters for $(\a_1,\a_2,\a_3)=(1,5,1)$. 
As in \cite{riotto1,riotto2,SO10decription}, for the up quark masses at the leptogenesis scale we
adopted the values $m_u =1\,{\rm MeV}$, $m_c = 400\,{\rm MeV}$ and $m_t =100\,{\rm GeV}$ 
\cite{fusaoka}.
We verified that indeed constraints do not depend on $\a_1$ and $\a_3$ but only on $\a_2$ as in the non-supersymmetric case \cite{riotto2}. 
\footnote{This statement is true under the implicit assumption that $\a_1$ is not that large  
that $M_1$ becomes larger than $10^9\,{\rm GeV}$ or or $\a_3$ that small to make 
$M_3/M_2 \lesssim 2$.}
The value $\a_2=5$ can be considered a close-to-maximum value in a way that the constraints obtained for this value
have to be regarded close to the most conservative ones.  
Moreover this value has been used as a benchmark value both in the non-supersymmetric case \cite{riotto1,riotto2} and 
also in \cite{marfatia}, allowing us a useful comparison among the results.

We have to distinguish `small $\tan\beta$ values'
for which the production, as in the non-supersymmetric case, occurs in the two-flavoured regime, from
`large $\tan\beta$ values', for which the production occurs in the three-flavoured regime.  Since for 
successful $SO(10)$-inspired leptogenesis one typically has  $M_2 \gtrsim 10^{11}\,{\rm GeV}$ and since the transition from the
two to the three flavoured regime occurs for $M_2 \simeq 5\times 10^{8}\,{\rm GeV}(1+\tan^2\b)$, 
one can say that for $\tan\b \gtrsim 15$ the  production occurs mainly in the three flavoured regime, while 
for $\tan\b \lesssim 15$ it occurs mainly in the two-flavoured regime.
We made the calculation for two extreme values , $\tan \b=5$ and $\tan\b = 50$. In the first case the production occurs almost 
entirely in the two flavoured regime, except for very special points, 
while in the second case the production occurs mostly in the three flavoured regime.  

We also performed the scatter plots both for NO and for IO neutrino masses so that in total we have 
four cases to consider. 

In addition to successful leptogenesis, we also show the results when the condition of strong thermal leptogenesis, such that a large pre-existing asymmetry is washed-out, is  superimposed. 
As in the non-supersymmetric case \cite{strongSO10}, this singles out a sub set of the solutions 
out of those satisfying successful leptogenesis. In the scatter plots we highlight these sub sets in blue 
(light blue for $V_L=I$ and dark blue for $I \leq V_L \leq V_{CKM}$). 

For the low energy neutrino parameters we adopted the same values and ranges as in \cite{SO10decription}.
In particular for the solar neutrino mass scale $m_{\rm sol} \equiv \sqrt{m^2_2 - m^2_1} = 0.0087\,{\rm eV}$ 
and for the  atmospheric neutrino mass scale $m_{\rm atm}\equiv \sqrt{m^2_3 - m^2_1} = 0.0495\,{\rm eV}$, the best fit values found in a recent global analysis \cite{global}.  When these values are combined with
the upper bound on the sum of the neutrino masses from the {\em Planck} satellite, $\sum_i m_i < 0.23\,{\rm eV}$ ($95\%$ C.L.) \cite{planck}, one obtains an upper bound on the lightest neutrino mass
\be\label{m1ub}
m_1 \lesssim 0.07\,{\rm eV} \,  .
\ee
The mixing angles, respectively the reactor, the solar and the atmospheric ones, 
are now measured with the following best fit values and $1\s$ ($3\s$) ranges 
\cite{foglilisi2013} for NO and IO respectively,
\bea\label{expranges}
\theta_{13} & = &  8.8^{\circ}\pm 0.4^{\circ} \, \;\;  (7.6^{\circ}\mbox{--}9.9^{\circ}) 
 \;\; \mbox{\rm and} \;\;
 \theta_{13}  =    8.9^{\circ}\pm 0.4^{\circ} \, \;\; (7.7^{\circ}\mbox{--}9.9^{\circ}) \,  ,
 \\ \nonumber
\theta_{12} & = &  33.7^{\circ}\pm 1.1^{\circ} \,  \;\;  (30.6^{\circ}\mbox{--}36.8^{\circ}) 
\;\; \mbox{\rm and} \;\;
\theta_{12}  =   33.7^{\circ}\pm 1.1^{\circ} \,  \;\;  (30.6^{\circ}\mbox{--}36.8^{\circ}) \,  , 
\\ \nonumber
\theta_{23} & = &  {41.4^{\circ}}^{+1.9^{\circ}}_{-1.4^{\circ}} \,  \;\;  
(37.7^{\circ}\mbox{--}52.3^{\circ}) 
\;\; \mbox{\rm and} \;\;
\theta_{23}  =   
                 {42.4^{\circ}}^{+8.0^{\circ}}_{-1.8^{\circ}}  
                  \;\;  (38.1^{\circ}\mbox{--}52.3^{\circ})  \,  .
 \eea 
Current experimental data also start to put constraints on the
Dirac phase and the following best fit values and $1\s$ errors are found  for NO and IO respectively,
\be\label{delta}
\d/\pi = -0.61^{+0.38}_{-0.27} \,\,  
\;\;\mbox{\rm and} \;\;
\d/\pi = -0.69^{+0.29}_{-0.33} \,  ,
\ee
though all values $[-\pi,+\pi]$ are still allowed at $3\,\s$. They do not yet favour 
one of the two orderings over the other. 

\subsection{Normal ordering}

Let us first present the results for NO neutrino masses. 
As mentioned, we also discuss separately the results  for 
`low $\tan\b$' values and for `high $\tan\b$ values'.

\subsubsection{Small $\tan \b$ values ($\tan\b =5$)}

Let us first start discussing the results for $\tan \b=5$. As mentioned, this is 
a sufficiently low value for most of the allowed values of $M_2$ to
fall in the two fully flavoured regime. The results  are shown in Fig.~1. 
\begin{figure}
\begin{center}
\psfig{file=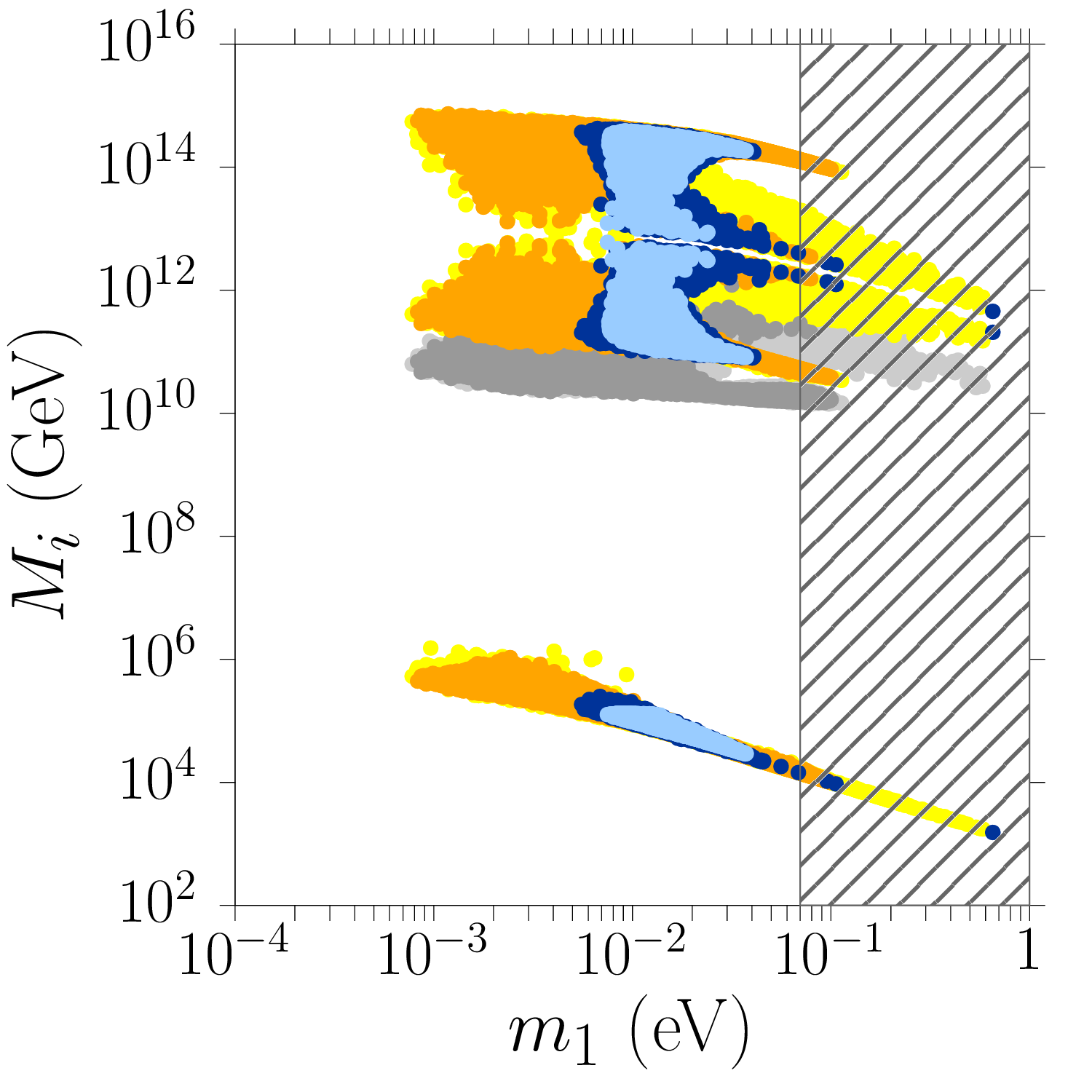,height=50mm,width=56mm}
\hspace{-7mm}
\psfig{file=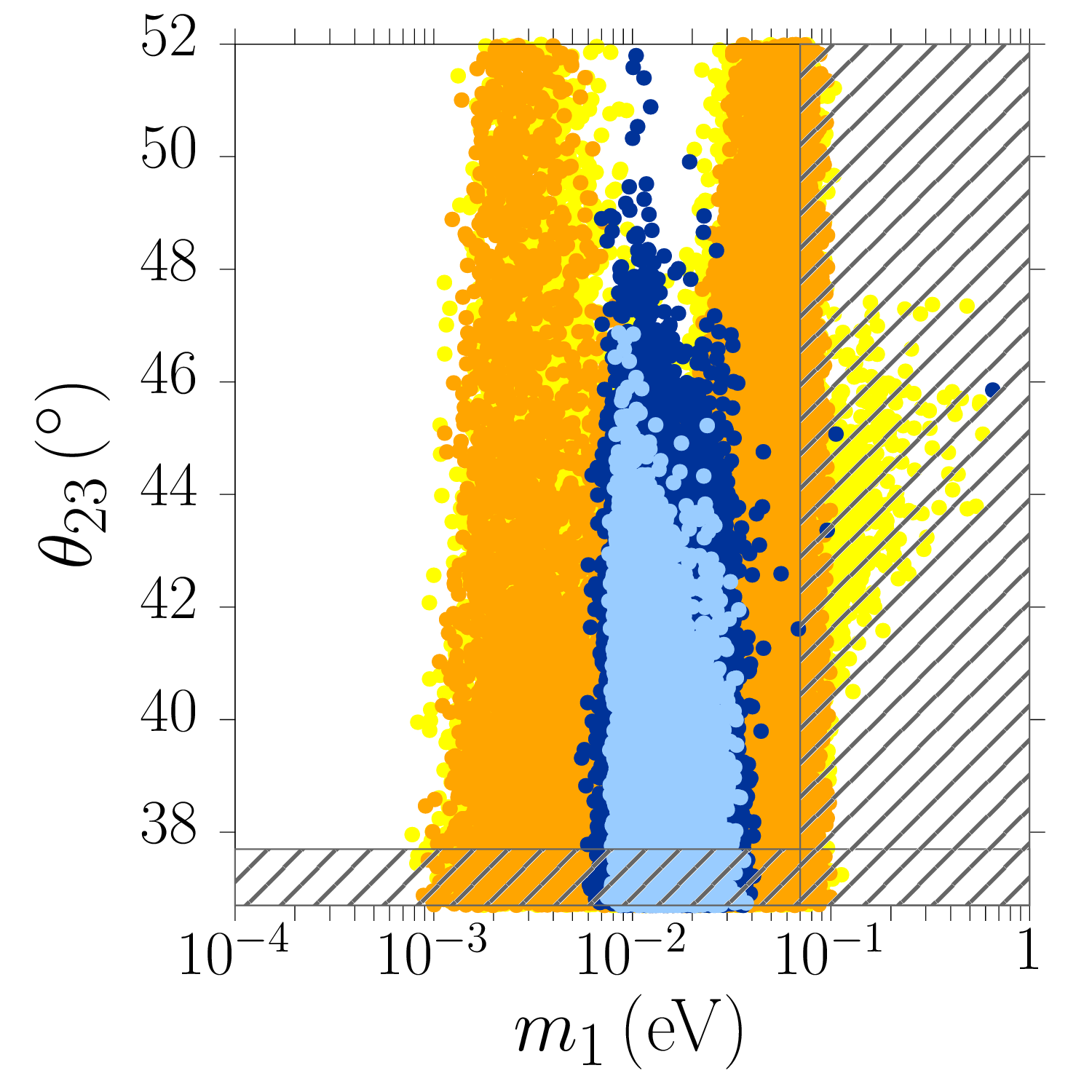,height=50mm,width=56mm}
\hspace{-7mm}
\psfig{file=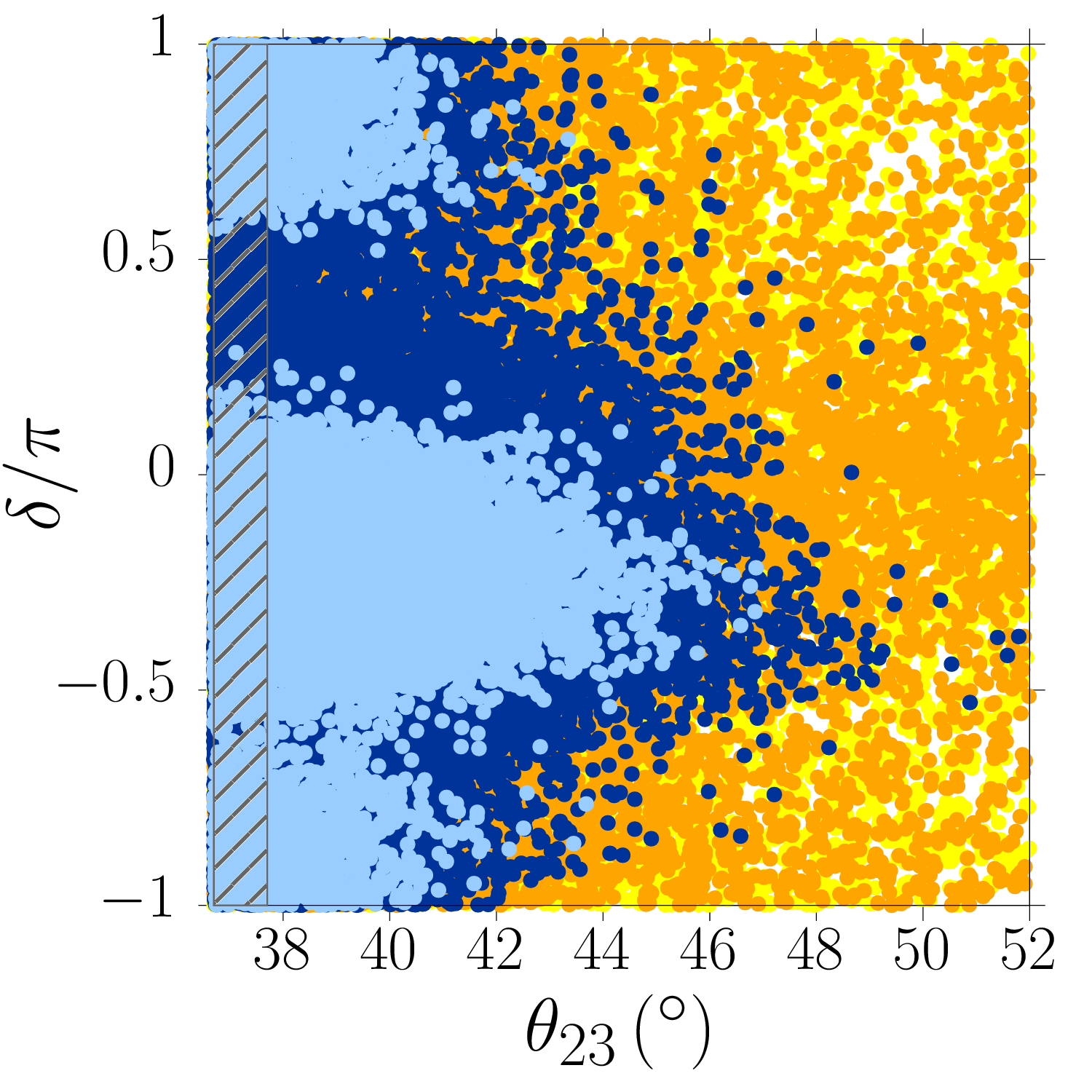,height=50mm,width=56mm} \\
\hspace{-7mm}
\psfig{file=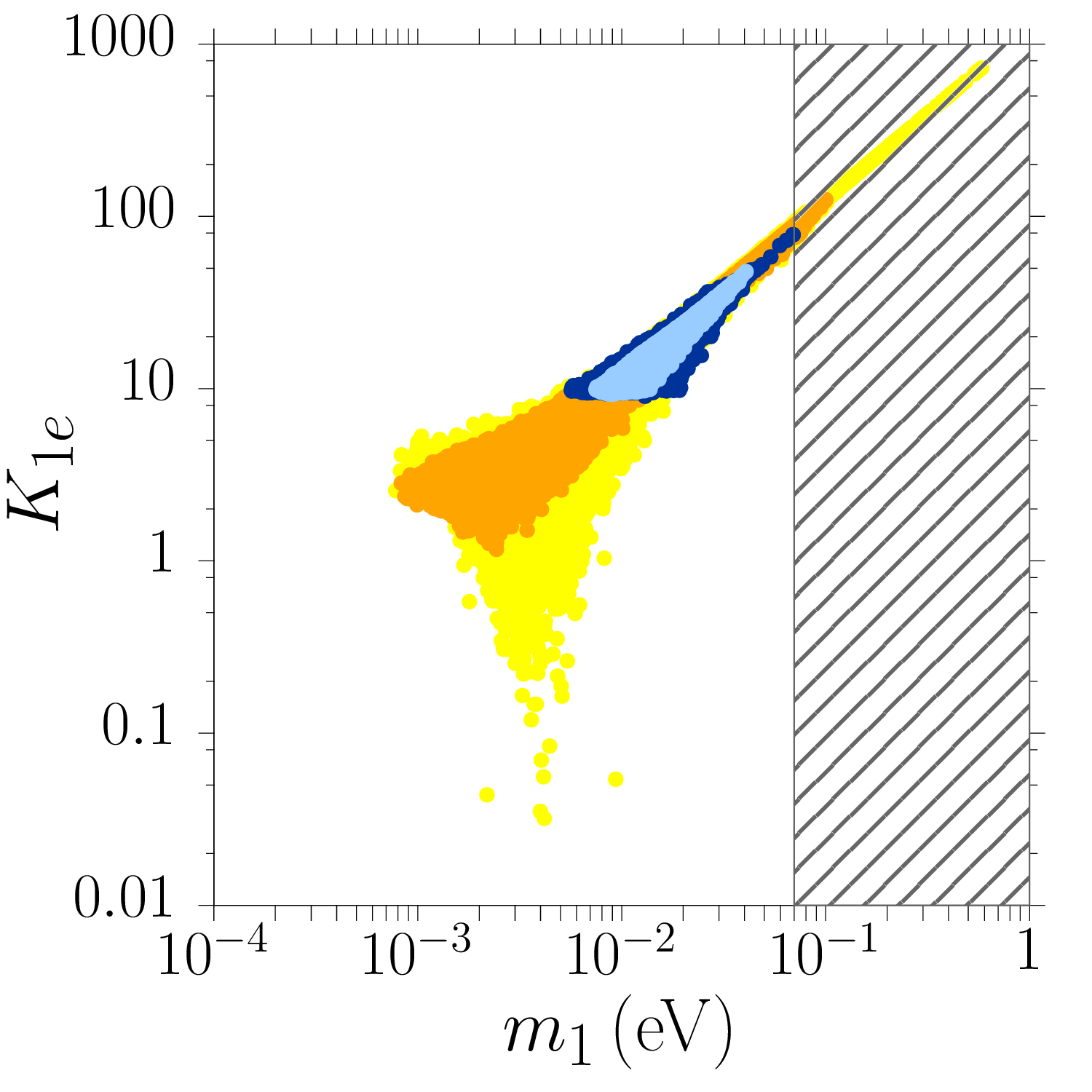,height=50mm,width=56mm}
\hspace{-7mm}
\psfig{file=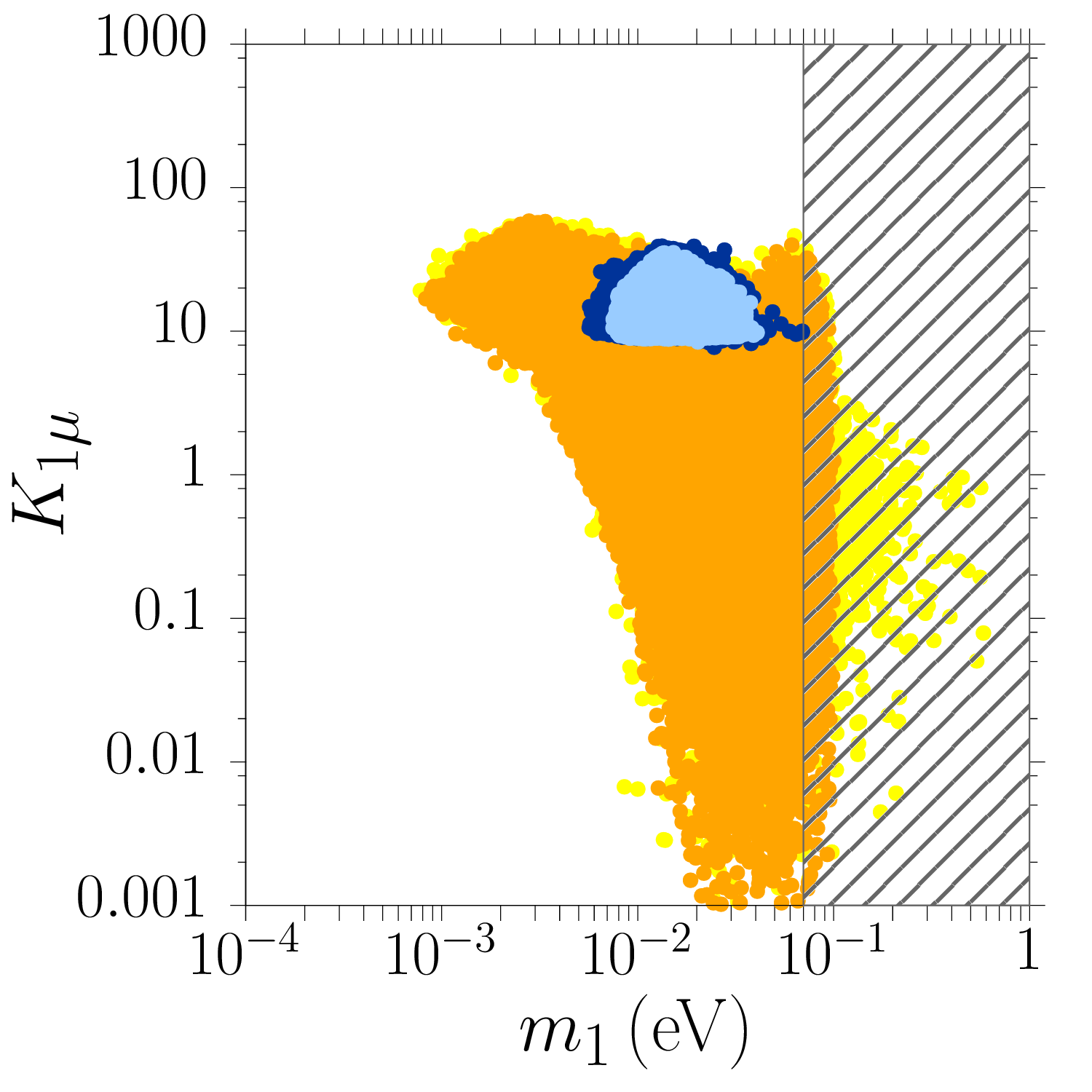,height=50mm,width=56mm}
\hspace{-7mm}
\psfig{file=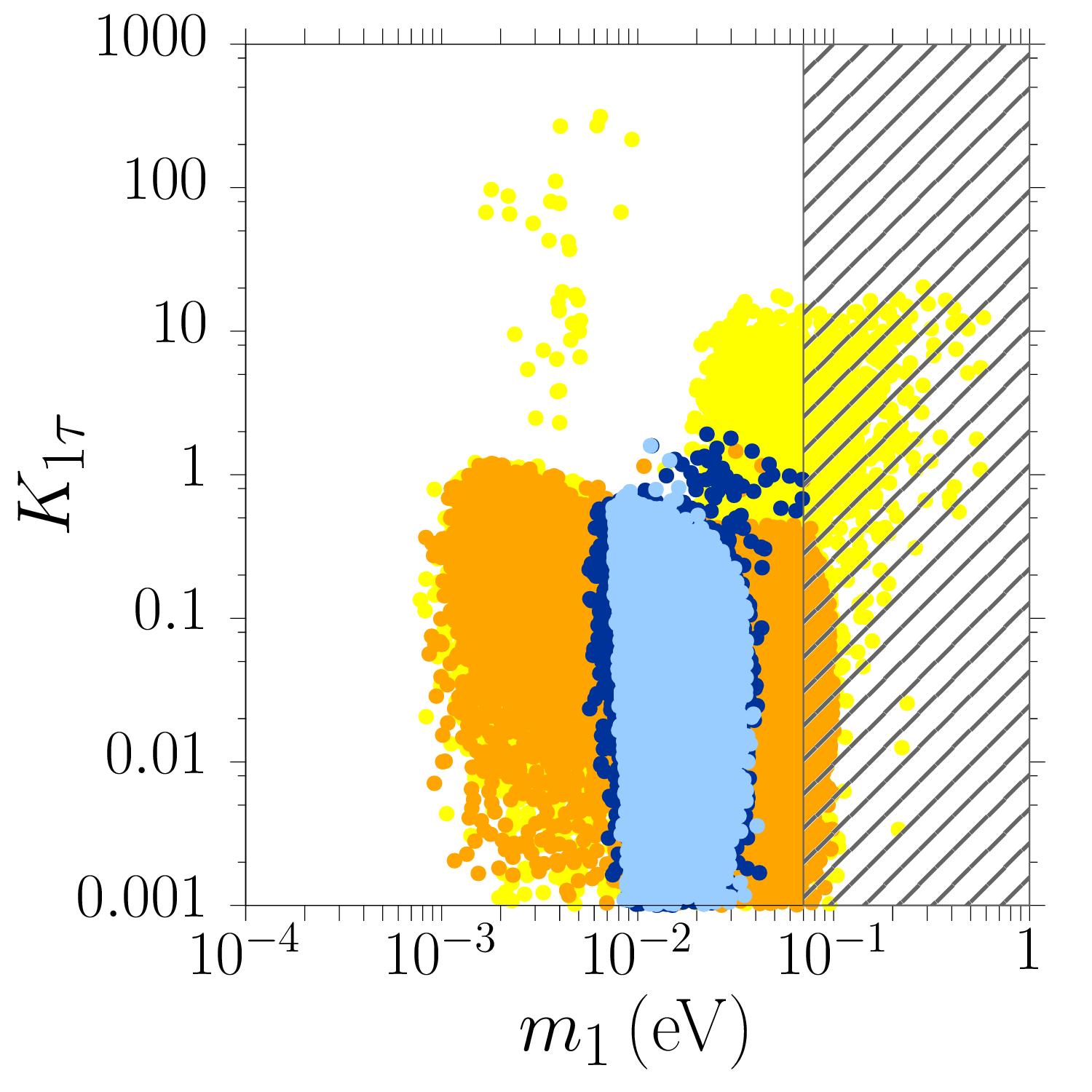,height=50mm,width=56mm}
\end{center}
\vspace{-11mm}
\caption{Scatter plots in the low energy neutrino parameter space 
projected on different selected planes for NO, $(\a_1,\a_2,\a_3) = (1,5,1)$, $M_3/M_2 > 3$, 
$\tan\b = 5$ and initial thermal $N_2$ abundance.
The yellow (orange) points respect the successful leptogenesis condition
$\eta_B^{\rm lep} > \eta_B^{\rm CMB} > 5.9 \times 10^{-10}$
for $I \leq V_L \leq V_{\rm CKM}$ ($V_L =I$) where 
$\eta_B^{\rm lep}$ is calculated from the eqs.~(\ref{unfl}), (\ref{twofl}) and (\ref{threefl})
depending on the value of $M_2$ determining the flavoured regime (mainly the two fully-flavoured regime)
using a numerical determination of  RH neutrino masses, mixing matrix and phases.  
The mixing angles vary within the $3\s$ ranges in Eqs.~(\ref{expranges}).  The dark (light) blu points are those
respecting the additional {\em strong thermal condition} for  $I \leq V_L \leq V_{\rm CKM}$ ($V_L =I$)
for an initial value of the pre-existing asymmetry $N_{B-L}^{\rm p,i}=10^{-3}$.
The dashed regions indicate either the values of $m_1$ excluded by 
the {\em Planck} upper bound $m_1 \lesssim 70\,{\rm meV}$  (cf. eq.~(\ref{m1ub})) or
the values of $\theta_{23}$ excluded by current 
data at $3\s$ (cf. eq.~(\ref{expranges})).
The grey points indicate the minimum value of
$T_{\rm RH}$.}
\label{constrNO}
\end{figure}
The yellow points are all those solutions realising successful $SO(10)$-inspired leptogenesis
for $I \leq V_L \leq V_{CKM}$ and initial thermal $N_2$ abundance. The orange points are the  subset for $V_L=I$. 
These results for $\tan \b =5$ are similar to those obtained in the non-supersymmetric case and they are 
well explained and understood \cite{riotto1,riotto2,SO10decription}.  
Even though they have been obtained for initial thermal $N_2$ abundance,
they are actually very marginally dependent of the initial $N_2$ abundance since
for the tauon-dominated solutions ($K_{1\t} \lesssim 1$) one has $K_{2\tau} \gg 1$ and for the muon-dominated
solutions ($K_{1\mu}$) one has $K_{2\t_2^{\bot}}\gg 1$ 
(except for very few points with $K_{2\t_2^{\bot}}\simeq 1$).
As we will discuss, there are also some electron dominated solutions that entirely depend
on the initial $N_2$ abundance, since $K_{2\t_2^{\bot}}\lesssim 1$, but for low $\tan\b$ values they
are marginal and do not influence the constraints on the low energy neutrino parameters. 
They correspond to the sparse point at $K_{1\t} \gg 1$ and in the range 
$1\,{\rm meV}\lesssim m_1 \lesssim 10\,{\rm meV}$. The low density indicates 
that these solutions are marginal and require some fine tuning to realise weak wash-out 
at the production, i.e. $K_{2\t_2^{\bot}} \lesssim 1$, and to enhance the $C\!P$ asymmetry $\ve_{2\a}$.

Coming back to the leading tauon dominated solutions,  
notice that in principle since the washout is stronger compared to the SM case, 
because of the smaller value of $m_{\star}^{MSSM}$ compared to $m_{\star}^{SM}$ (cf. eq.~(\ref{mstar})), 
one could think that it should be more difficult to realise the condition $K_{1\t}\lesssim 1$. However, from the analytical expression
given in \cite{SO10decription} for $V_L = I$, extended to the supersymmetric case with the 
simple replacement $m_{\star}^{SM}\ra m_{\star}^{MSSM}$,  explicitly
\be\label{K1tauexplicit}
K_{1\t} \simeq {|c_{13}\,c_{12}\,s_{12}\,s_{23}\,(m_1\,e^{2\,i\,\rho}-m_2) 
+ s_{13}\,c_{13}\,c_{23}\,(m_3\,e^{i\,(2\s-\d)} - m_2\,s^2_{12}\,e^{i\,\d} - m_1\,c^2_{12}\,e^{i\,(2\,\rho+\d)})|^2\over m_{\star}^{MSSM}\,|m_1\,c^2_{12}\,c^2_{13}\,e^{2\,i\,\rho}+ m_2\,s^2_{12}\,c^2_{13}+ m_3\,s^2_{13}\,e^{2\,i\,(\s-\d)}|} \, ,
\ee
one can see 
that the slightly lower value of $m_{\star}^{MSSM}$ plays just a  marginal
role since the condition $K_{1\t} \lesssim 1$ produces conditions on the phases marginally dependent on
$m_{\star}^{MSSM}$. Actually the increase of the asymmetry of a factor $\sim \sqrt{2}$ at the production, 
due to the doubled $C\!P$ asymmetry only partly compensated by a stronger wash-out, enlarges the 
allowed region in the plane $\theta_{23}$ vs. $m_1$ at values 
$m_1 \simeq 50\,{\rm meV}$, the so called $\tau_B$ solution and, more generally, the
allowed range of $m_1$ gets slightly wider
(for example the upper bound relaxes from $0.06\,{\rm eV}$ to  $0.1\,{\rm eV}$).

The blue points in Fig.~1 are the subset satisfying the strong thermal condition (dark blue for $I\leq V_L \leq V_{CKM}$,
light blue for $V_L =I$) for an initial pre-existing asymmetry $N^{\rm p,i}_{B-L}=10^{-3}$. 
Also in this case we can compare the results with the non-supersymmetric case.
This time there is one significant difference since in the supersymmetric case 
the strong thermal region is more extended and in particular it allows higher values of the 
atmospheric mixing angle. Indeed while in the non-supersymmetric case one has
a quite stringent upper bound on the atmospheric mixing angle $\theta_{23}\lesssim 43^{\circ}$, in the supersymmetric case this now gets relaxed to $\theta_{23} \lesssim 46^{\circ}$, a relaxation that might be relevant in view of the next expexted results  from long baseline experiments. 

This relaxation is quite well explained analyticaly for $V_L = I$ extending to the supersymmetric case
the discussion in \cite{SO10decription}.  
The upper bound on $\theta_{23}$ indeed originates from the 
requirement $K_{1e} \gg 1$ from the strong thermal condition. This condition first translates 
into a lower bound on the $0\nu\b\b$ effective neutrino mass $m_{ee}\gtrsim 8\,{\rm meV}$
and then into one on the lightest neutrino mass $m_1\gtrsim 1.3\, m_{ee} \gtrsim 10\,{\rm meV}$ 
\cite{strongSO10}. 
Since in the supersymmetric case all $K_{i\alpha}$ are $\sim \sqrt{2}$ larger, 
this requirement is now more easily satisfied and one has $m_{ee} \gtrsim 6\,{\rm meV}$ 
giving $m_1 \gtrsim 7\,{\rm meV}$, well explaining the constraints 
in the plane $m_{ee}$ vs. $m_1$ (see bottom left panel in Fig.~1), and this in turn implies indeed
$\theta_{23}\lesssim 46^{\circ}$.

In Fig.~2 we show 6 panels, for integer values of $\a_2$ from one to six, of the RH neutrino masses $M_i$
and of the minimum requested value of $T_{\rm RH}$ (we will discuss this in detail separately in Section 4). 
In these panels we have highlighted the flavour that dominates the asymmetry associating a different
colour to each flavour (blue for tauon, green for muon, red for electron).  The points
are calculated in the case $I \leq V_L \leq V_{CKM}$ and again for initial thermal $N_2$ abundance.
\begin{figure}
\begin{center}
\psfig{file=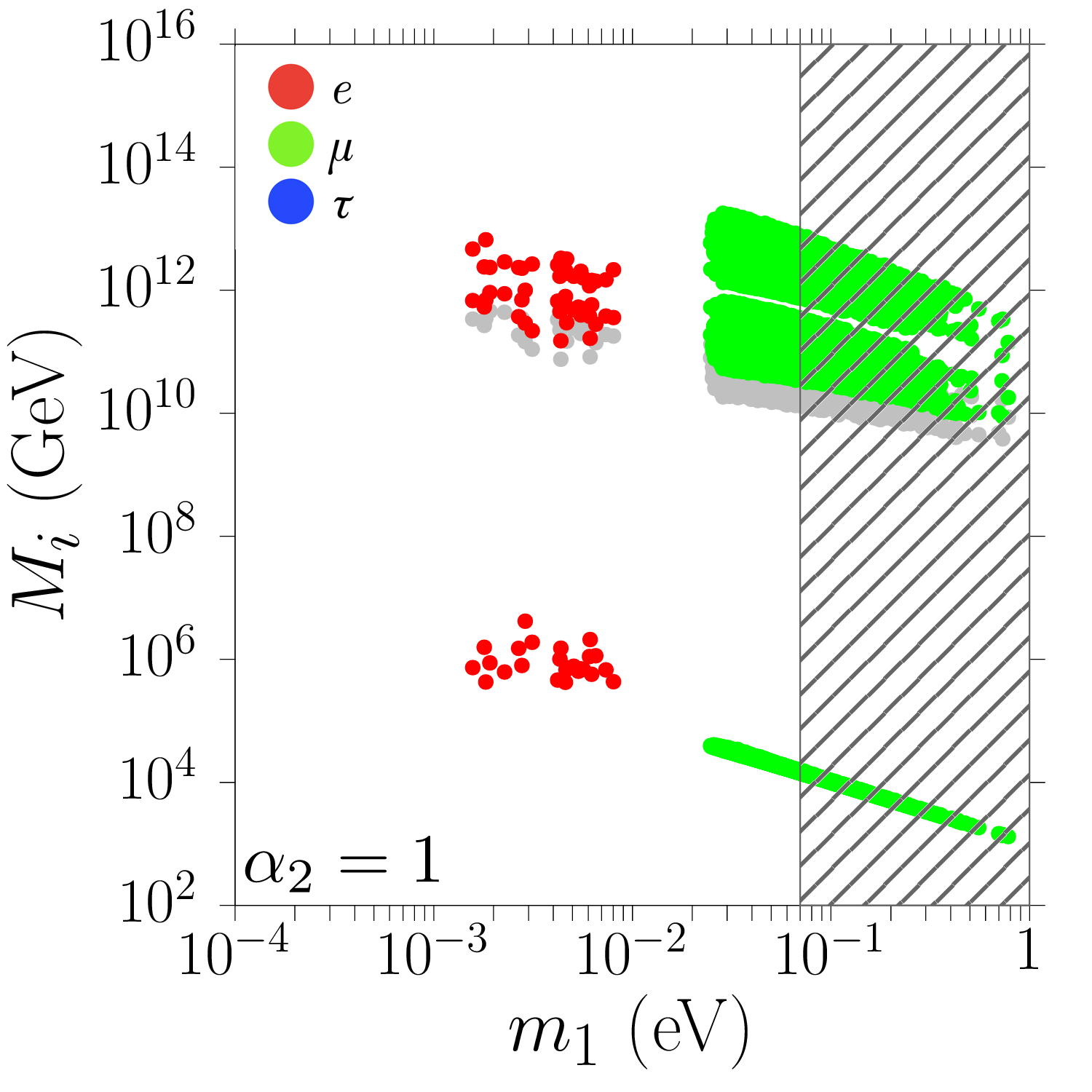,height=50mm,width=56mm}
\hspace{-7mm}
\psfig{file=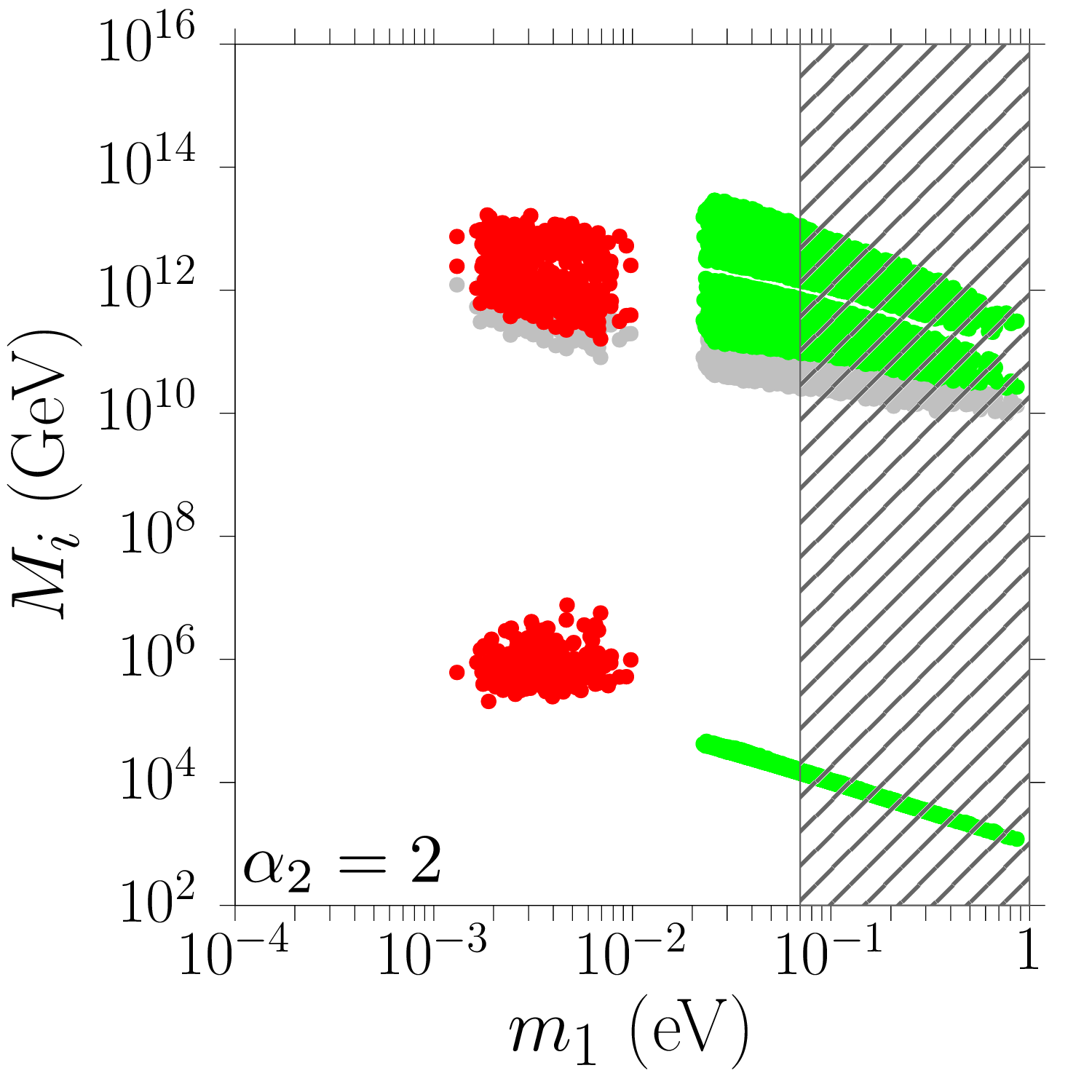,height=50mm,width=56mm}
\hspace{-7mm}
\psfig{file=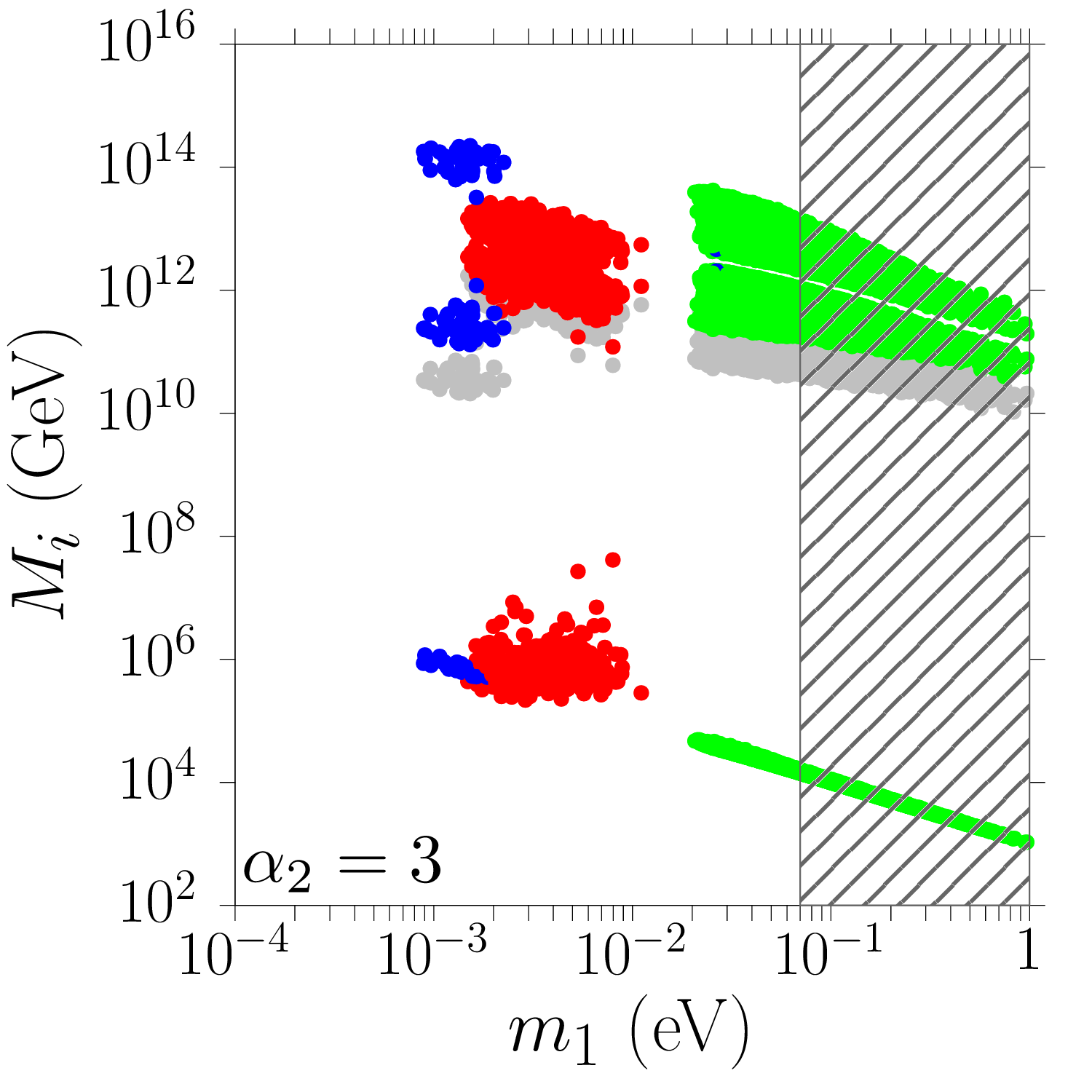,height=50mm,width=56mm}  \\
%\hspace{-7mm}
\psfig{file=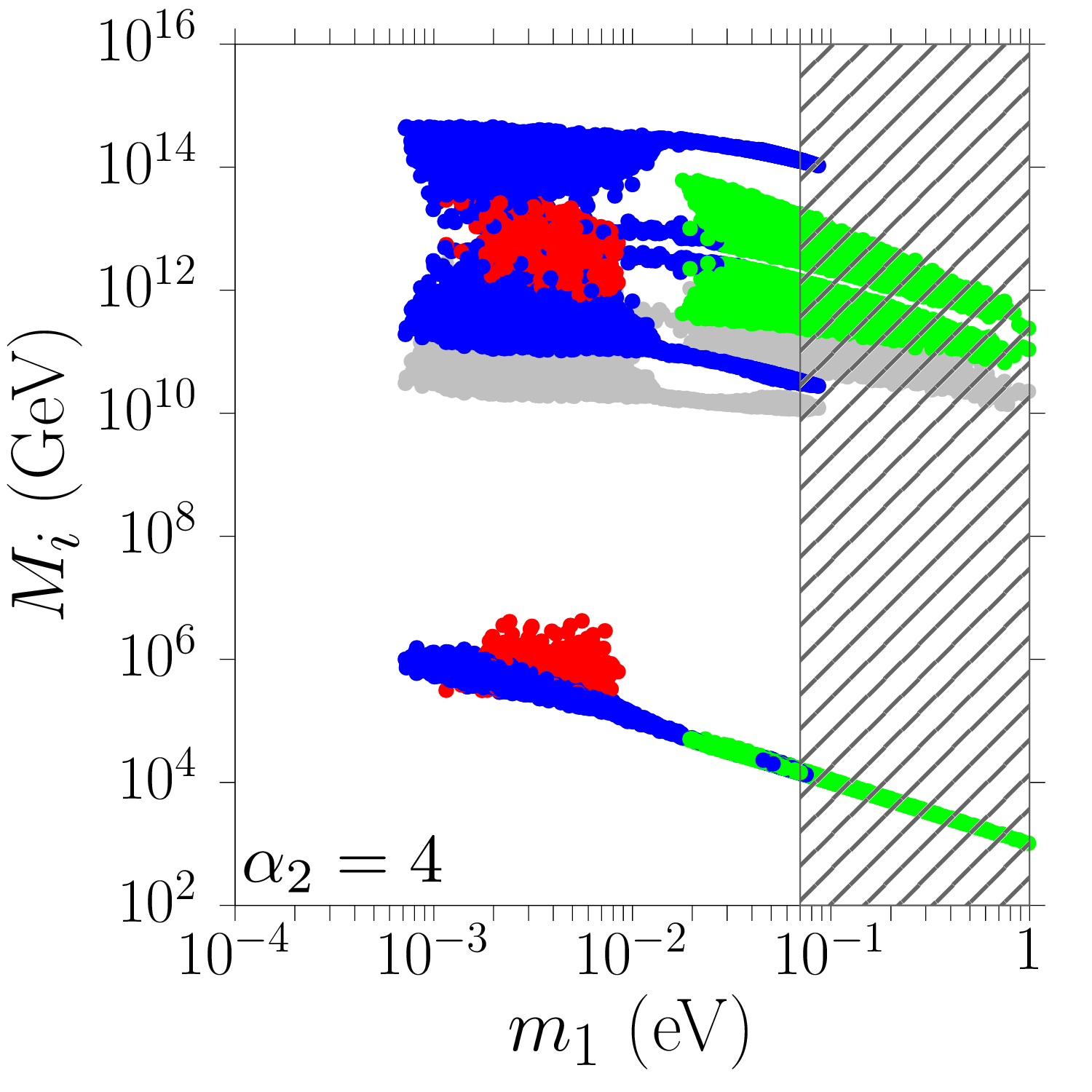,height=50mm,width=56mm}
\hspace{-7mm}
\psfig{file=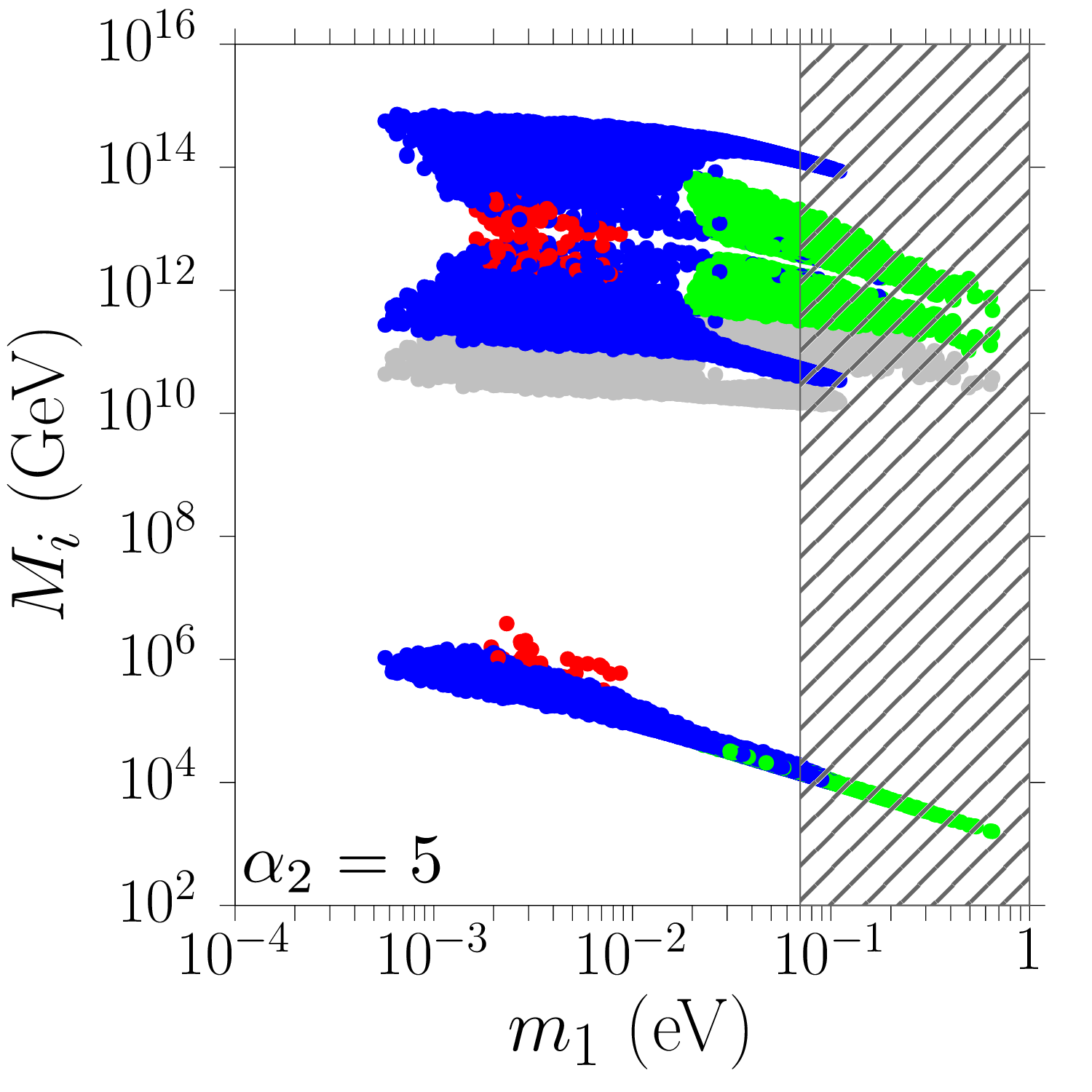,height=50mm,width=56mm}
\hspace{-7mm}
\psfig{file=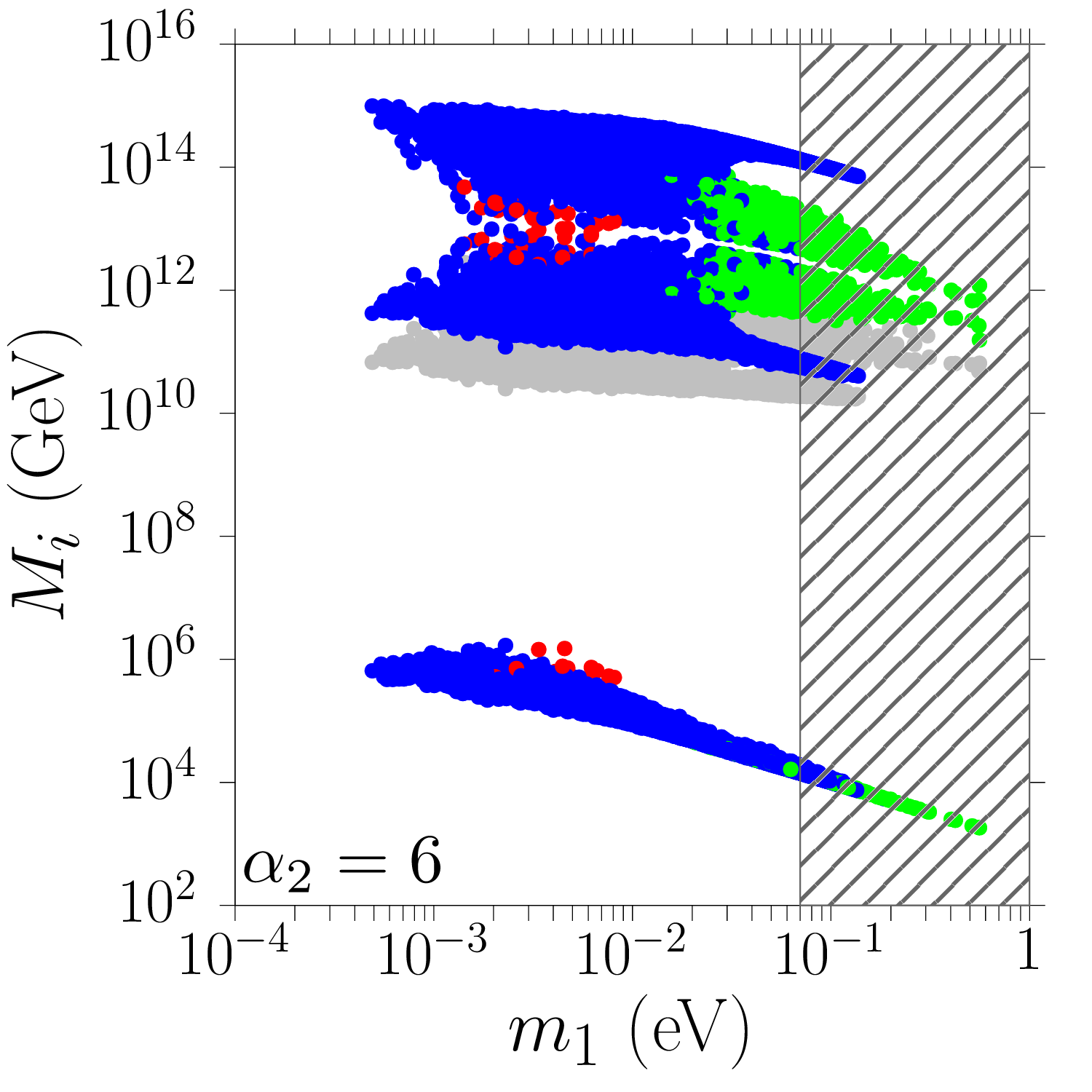,height=50mm,width=56mm}
\end{center}
\vspace{-10mm}
\caption{Scatter plots in the low energy neutrino parameter space 
projected on the plane $M_i$ vs. $m_1$ for NO, $\tan\b = 5$
and for integer $\a_2=[1,6]$ from top left to bottom right. All points respect the 
successful leptogenesis condition $\eta_B^{\rm lep} > \eta_B^{\rm CMB} > 5.9 \times 10^{-10}$
for $I \leq V_L \leq V_{\rm CKM}$.  The dashed region indicate  the value of $m_1$ excluded by 
the {\em Planck} upper bound eq.~(\ref{m1ub}).
The red, green and blue points points are those for which the final asymmetry is dominated
by the electron, muon and tauon flavour respectively. The grey points indicate the minimum value of
$T_{\rm RH}$.}
\label{fldomNO}
\end{figure}
As one can see, in addition to muon (green points)
and tauon (blue points) flavour dominated solutions,  also electron flavour dominated 
solutions are present. At low $\a_2$ values ($\a_2=1,2$) these are even the only solutions for  $m_1 \lesssim 20\,{\rm meV}$.
For $V_L = I$ the electron flavour asymmetries are many order of magnitude suppressed compared to the
muonic and even more compared to the tauonic \cite{SO10decription} but when $V_L \neq I$ this sharp flavour dominance does not hold \cite{riotto2}. In the non-supersymmetric case we have also found electron-flavour dominated solutions but in a very marginal way. This means that these solutions realise successful leptogenesis
only for very special conditions in the non-supersymmetric case and the maximum possible asymmetry is just
very slightly above the observed value. In the supersymmetric case, since the $C\!P$ asymmetries double
and the wash-out at the production is only $\sim \sqrt{2}$ stronger, the $B-L$ asymmetry 
at the production is $\sim\sqrt{2}$ higher and this helps the marginal electron-dominated solutions to be realised for a slightly wider region in parameter space in any case without really opening up new allowed regions in the low energy neutrino parameters. 
At the same time it is important to stress that since these solutions are realised 
for $K_{2\t_2^{\bot}}\lesssim 1$,
they are strongly dependent on the initial $N_2$ abundance and, in particular, they completely disappear for
initial vanishing $N_2$ abundance.  

In conclusion for low $\tan\b$ values the low energy neutrino constraints
are only slightly more relaxed than in the non-supersymmetric case and in particular, as we have seen, 
the strong thermal condition  is satisfied for slightly lower $m_1$ values. 

\subsubsection{Large $\tan \b$ values ($\tan\b =50$)}

Let us see now what happens when $\tan \b$ is large enough that the production occurs in the three-flavoured regime. There is no explicit
dependence of the asymmetry on $\tan\b$, the dependence is all encoded in the
values of $M_2$ marking the transitions between two different flavoured regimes.
The results will be the same for all $\tan\b$ values large enough to lead to a production in the three flavoured regime for all allowed values of $M_2$.  Since the condition for the three-flavoured regime is 
$M_2 \lesssim 5 \times 10^8\,{\rm GeV}\,(1+\tan^2\beta)$  
and since one expects $M_2\gtrsim 10^{10}\,{\rm GeV}$, some
solutions occurring in the three-flavoured regime are expected to appear 
for $\tan\b\gtrsim 5$. On the other hand
since there are no solutions for $M_2 \gtrsim 3\times 10^{12}\,{\rm GeV}$,
for $\tan\b \gtrsim 80$ all solutions fall in the three flavoured regime.
We choose for definiteness $\tan\b=50$. This is  sufficiently large that basically all solutions fall
in the three flavoured regime so that  constraints
on low energy neutrino data are saturated increasing $\tan\b$.
The results are shown in Fig.~3. The panels, the colour codes
and all benchmark values are the same as in Fig.~1, so that there can be a straightforward 
comparison with the results obtained for $\tan\b=5$. 
 \begin{figure}
\begin{center}
\psfig{file=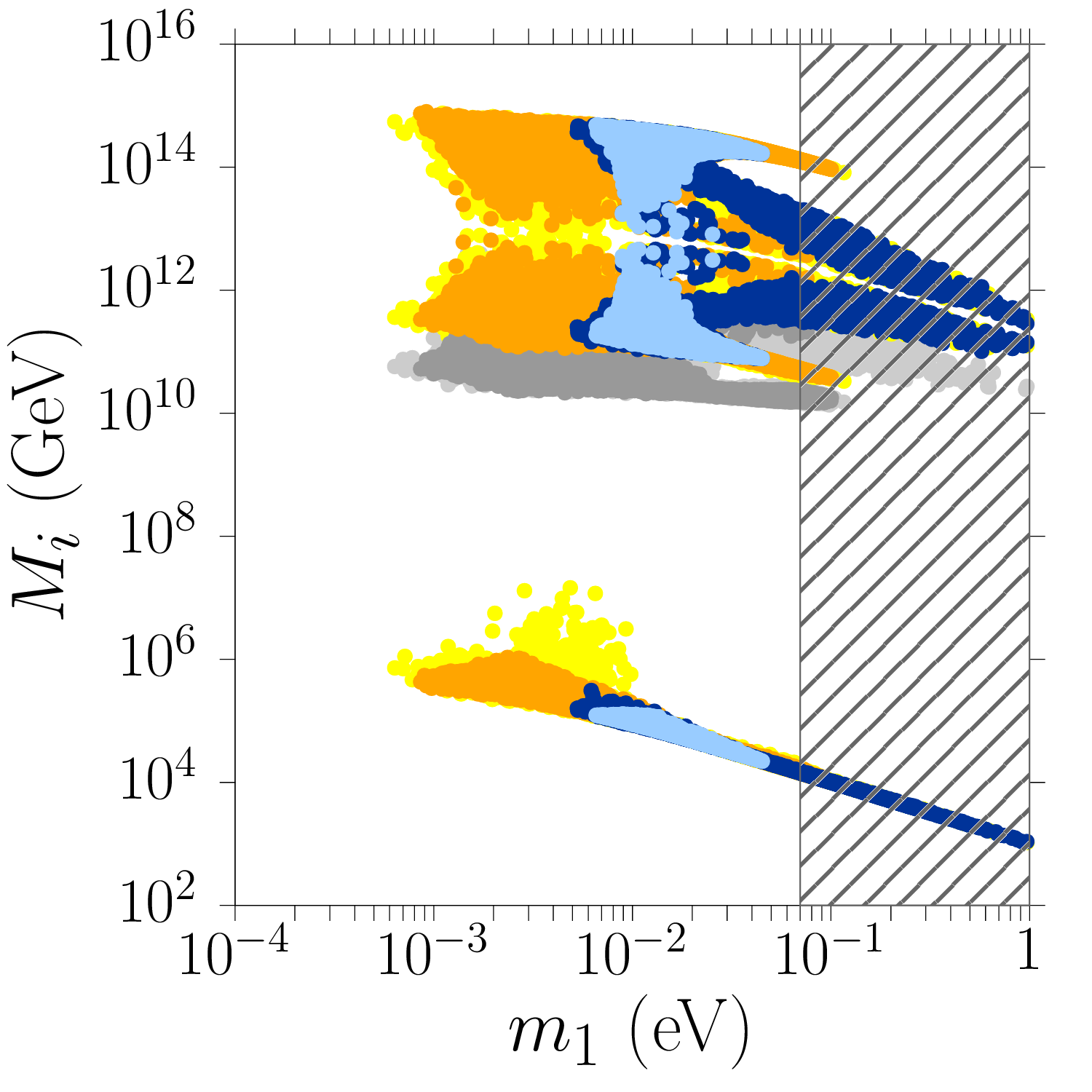,height=48mm,width=56mm}
\hspace{-7mm}
\psfig{file=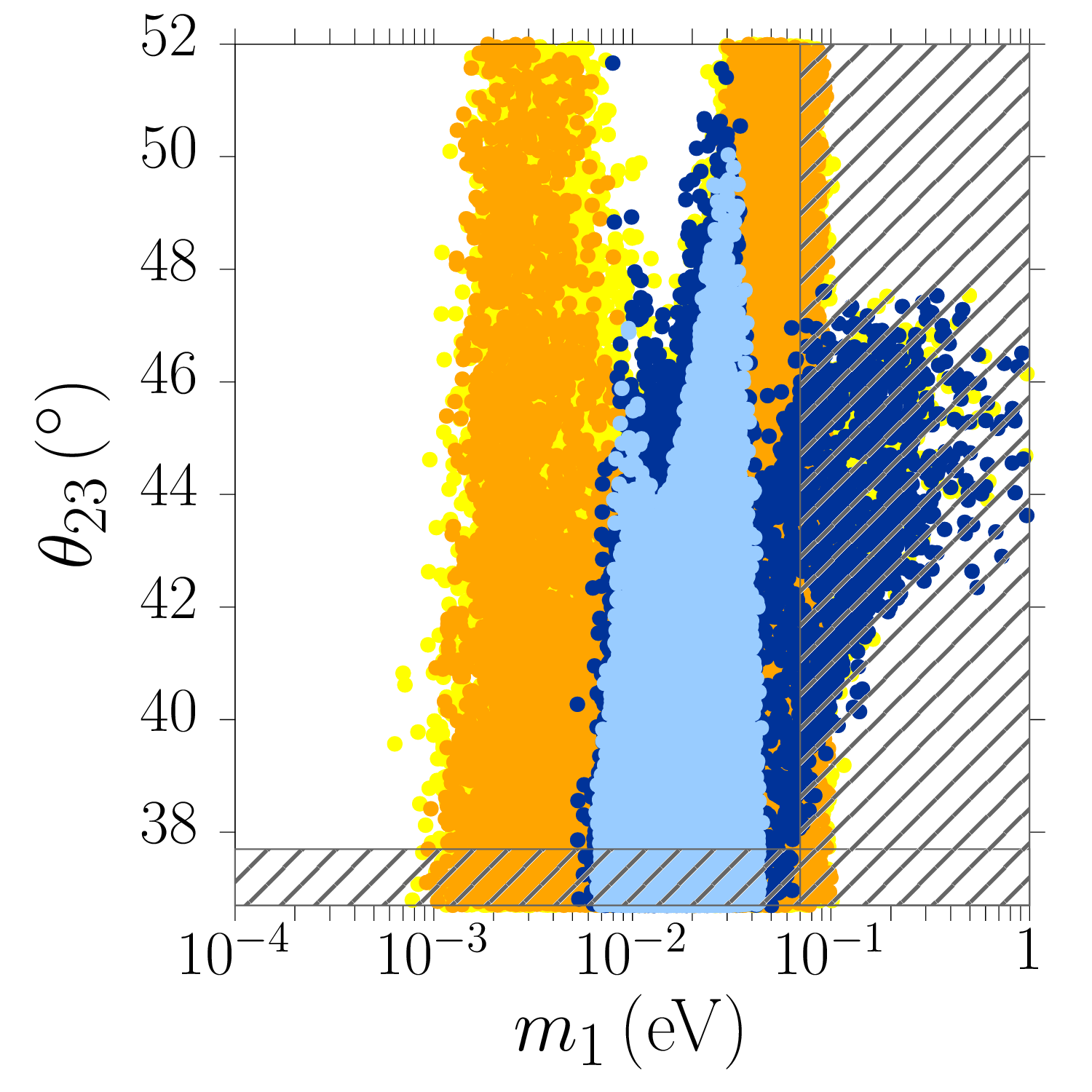,height=48mm,width=56mm}
\hspace{-7mm}
\psfig{file=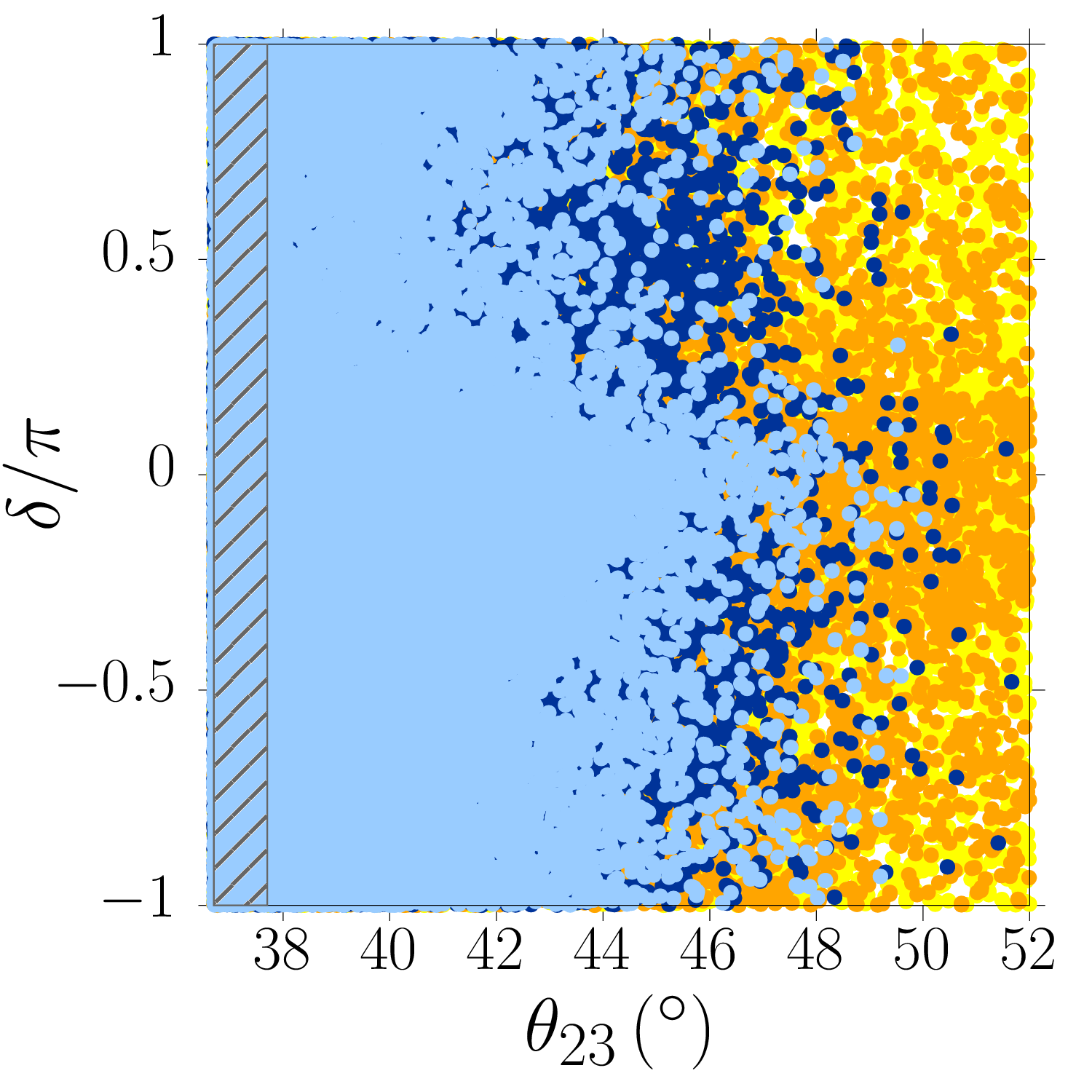,height=48mm,width=56mm}  \\
%\hspace{-7mm}
\psfig{file=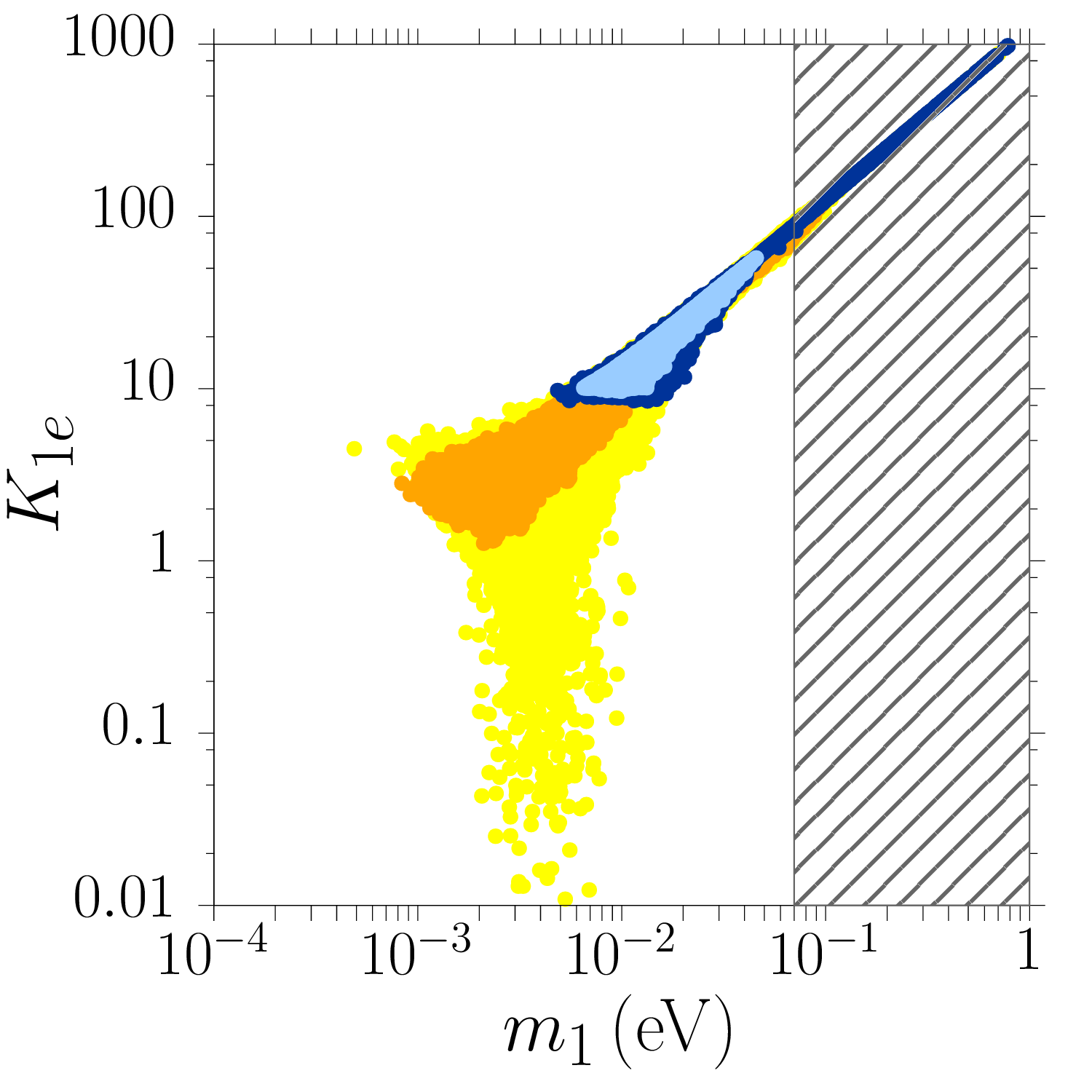,height=48mm,width=56mm}
\hspace{-7mm}
\psfig{file=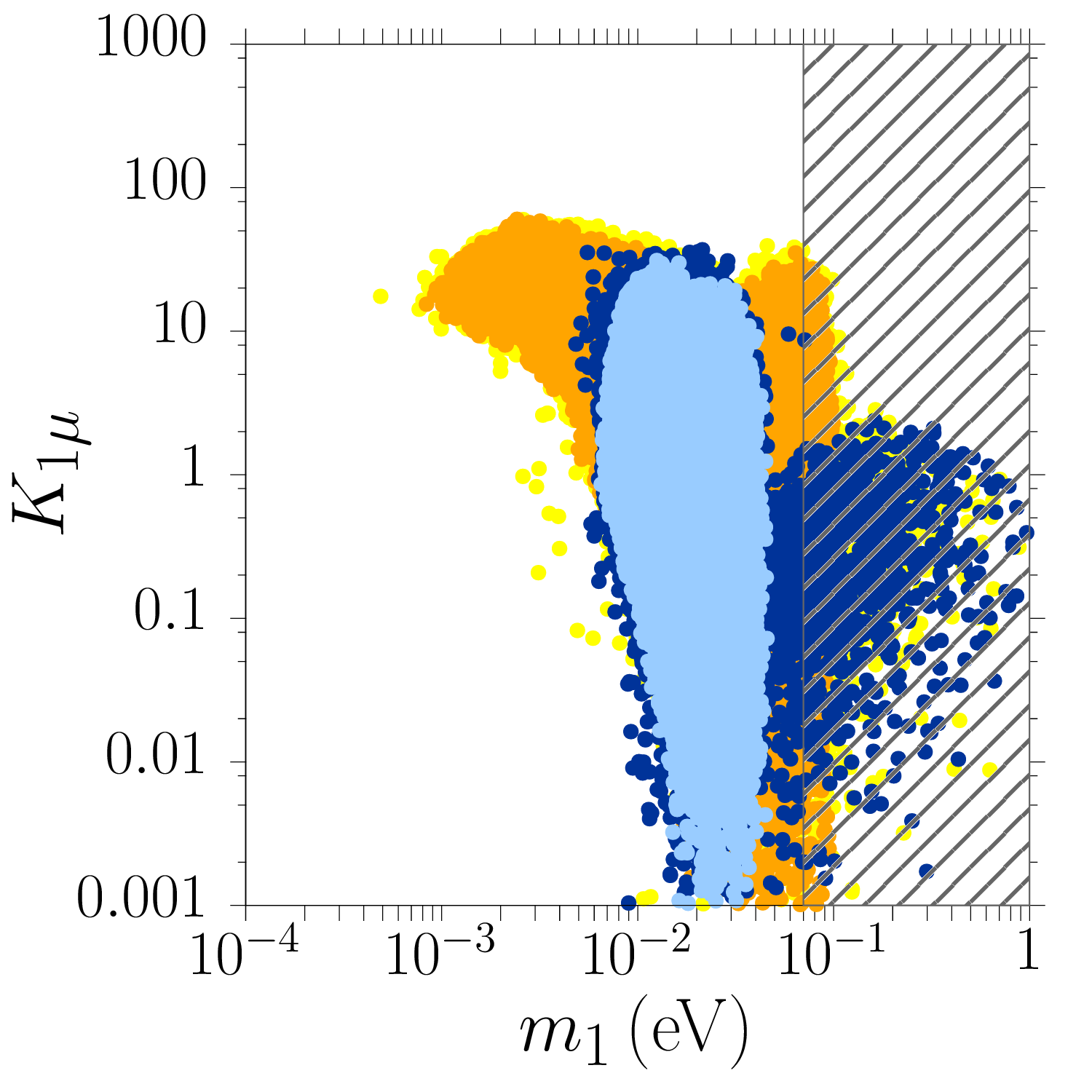,height=48mm,width=56mm}
\hspace{-7mm}
\psfig{file=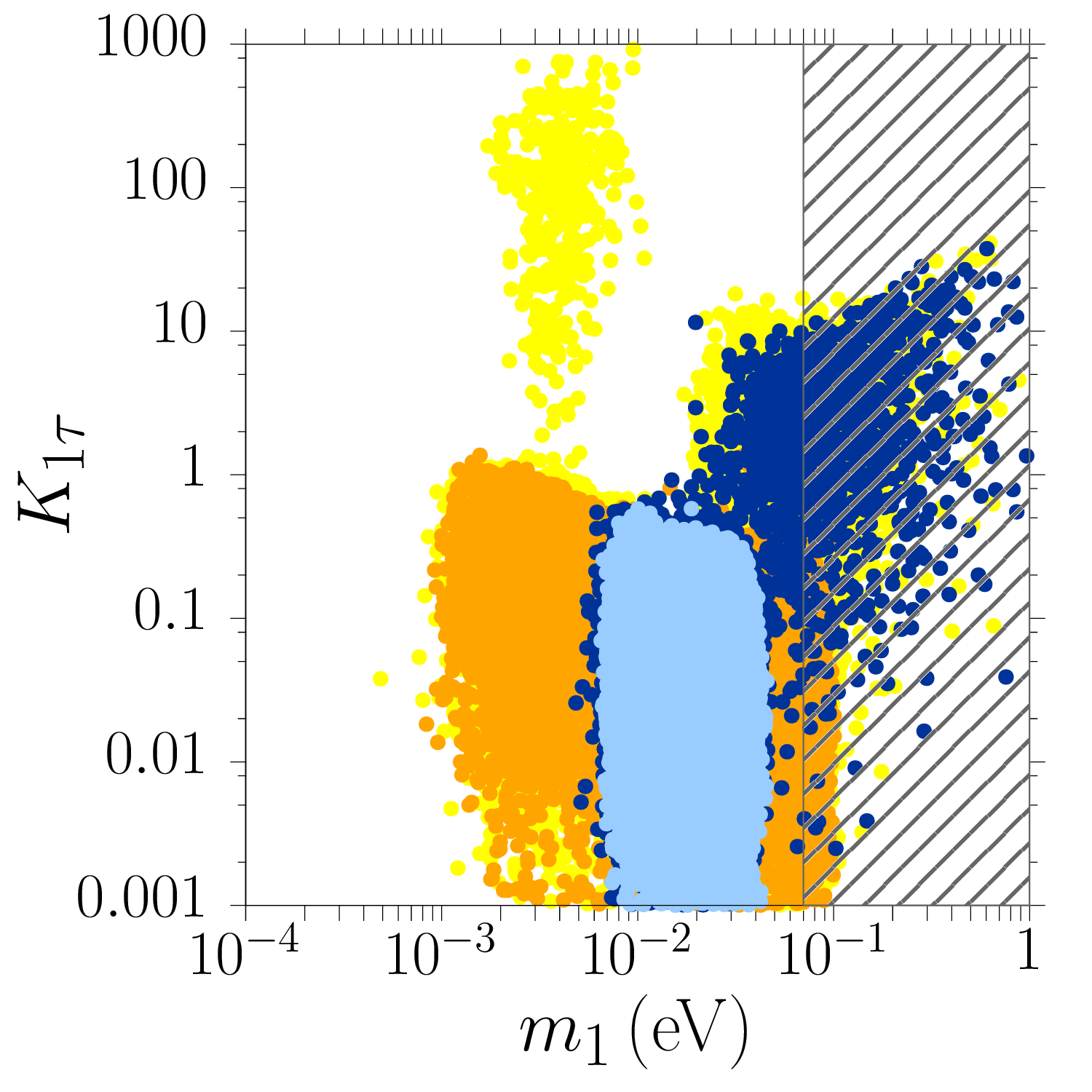,height=48mm,width=56mm}
\end{center}
\vspace{-10mm}
\caption{Same scatter plots as in Fig.~1 but for $\tan\b=50$.}
\label{constrNO}
\end{figure}
\begin{figure}
\begin{center}
\psfig{file=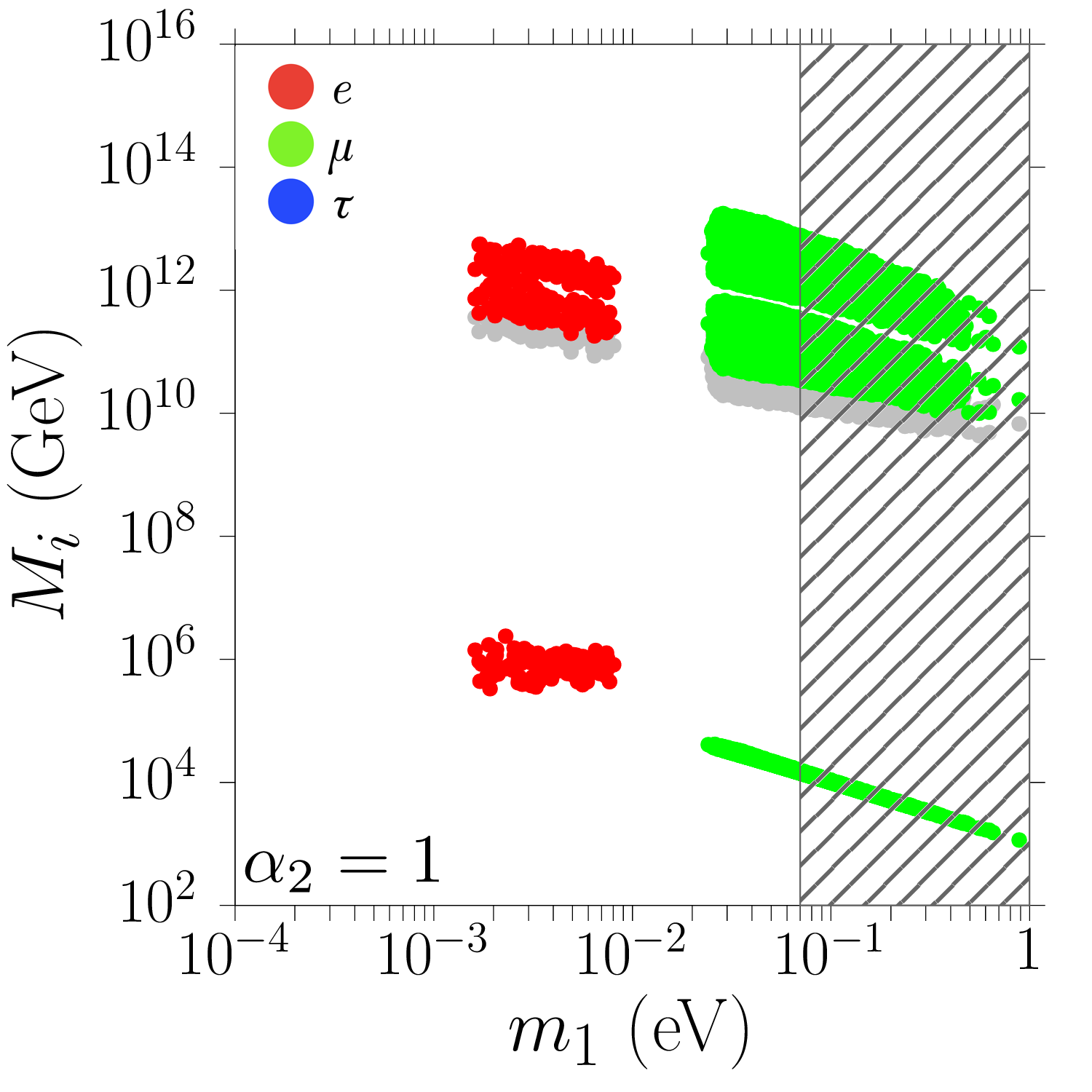,height=48mm,width=56mm}
\hspace{-7mm}
\psfig{file=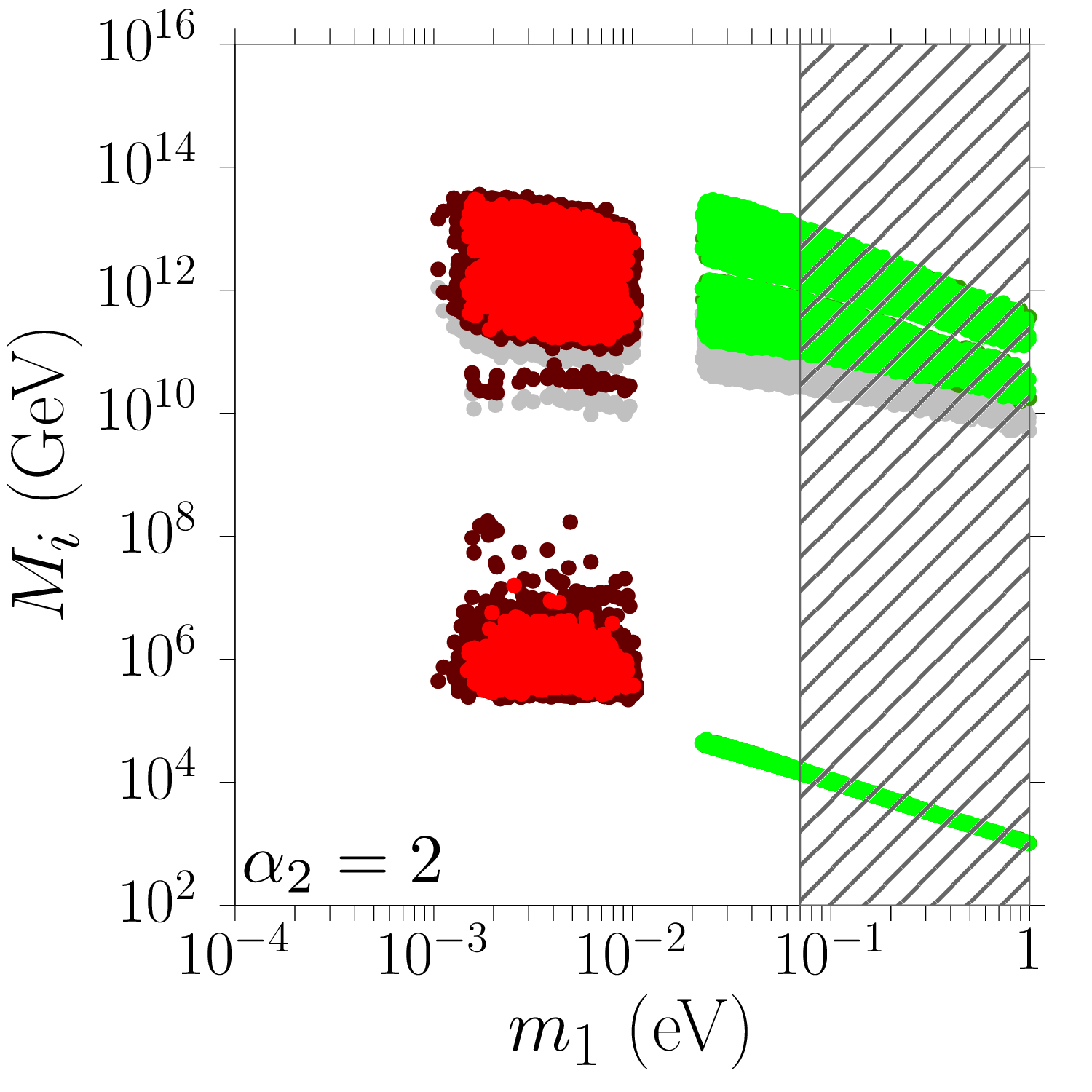,height=48mm,width=56mm}
\hspace{-7mm}
\psfig{file=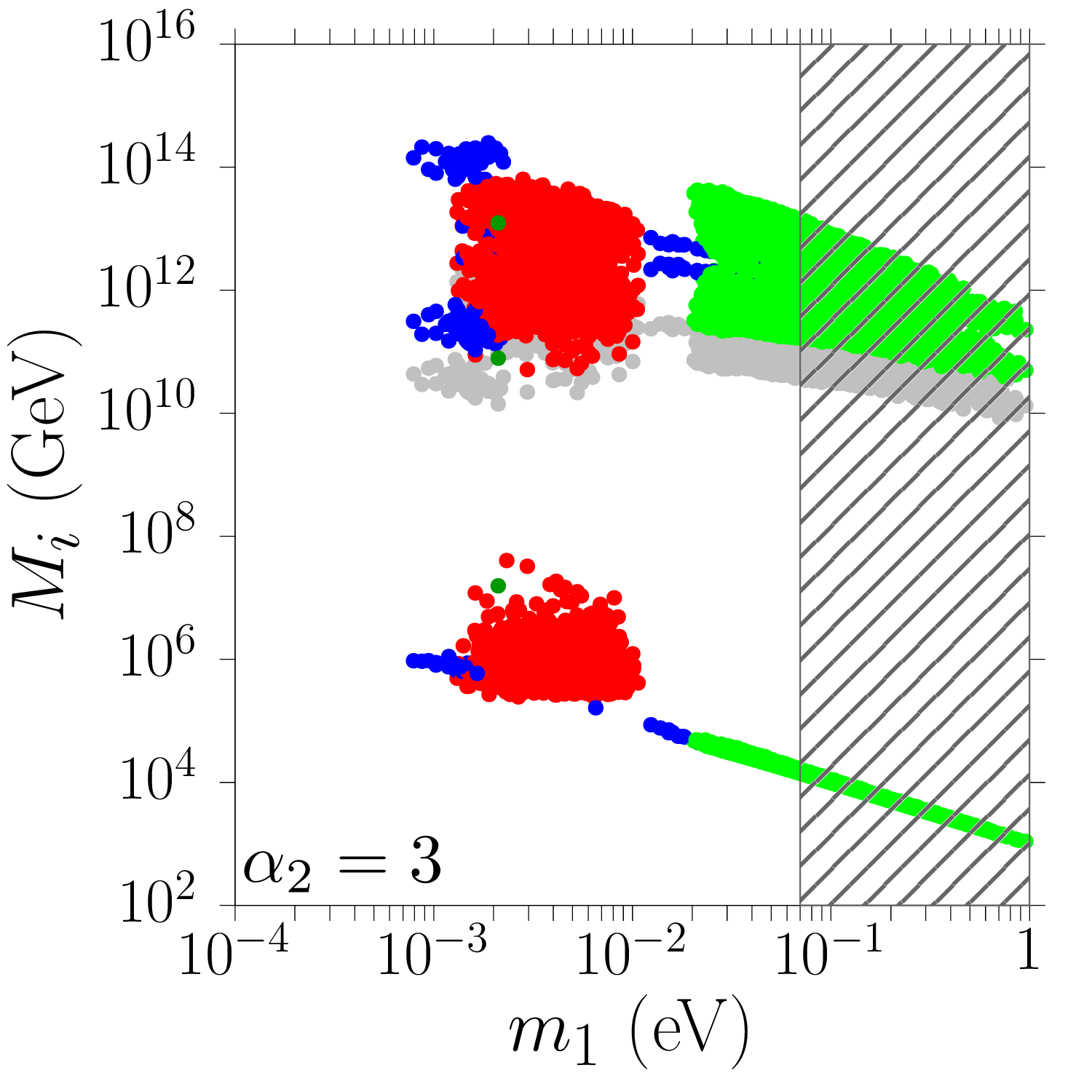,height=48mm,width=56mm}  \\
%\hspace{-7mm}
\psfig{file=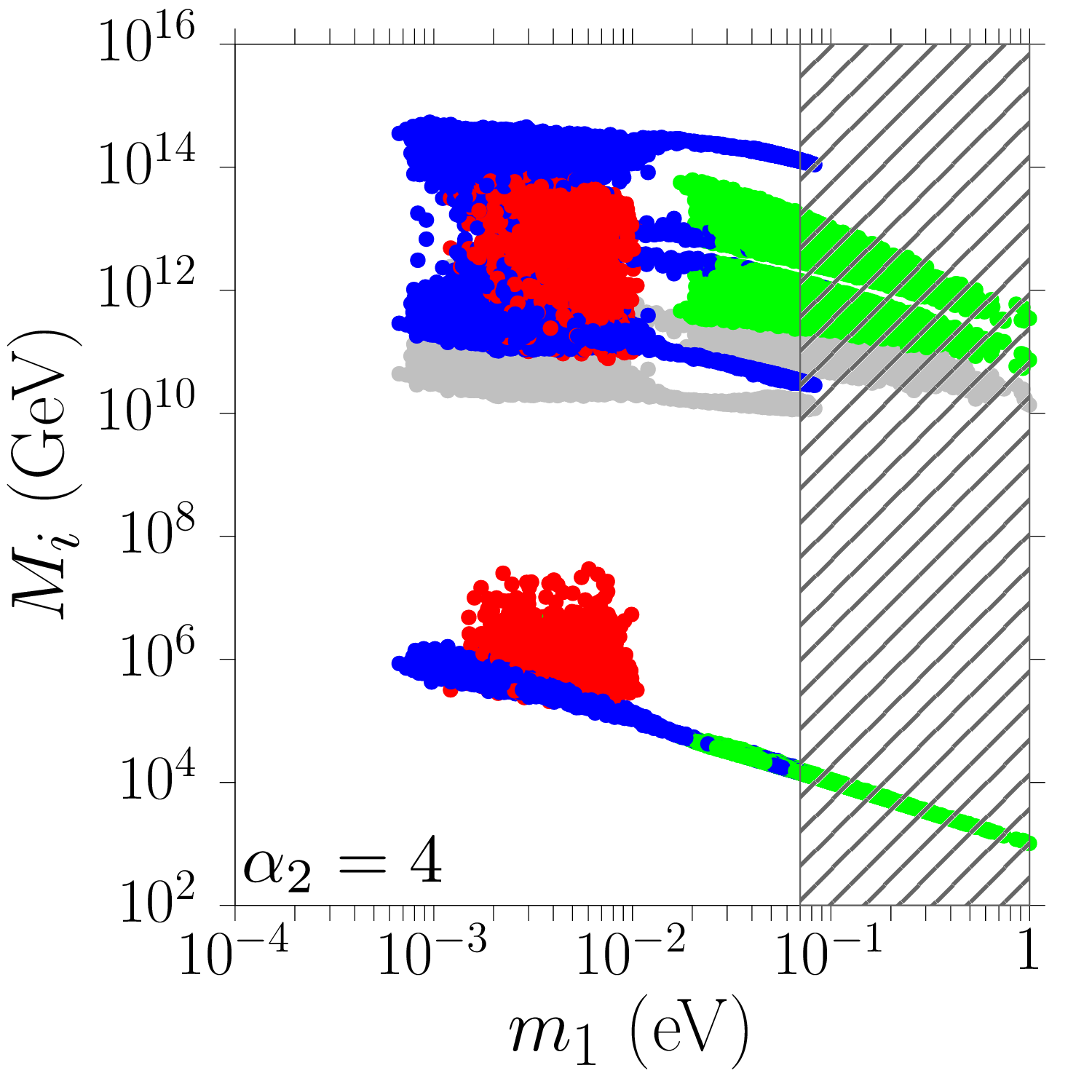,height=48mm,width=56mm}
\hspace{-7mm}
\psfig{file=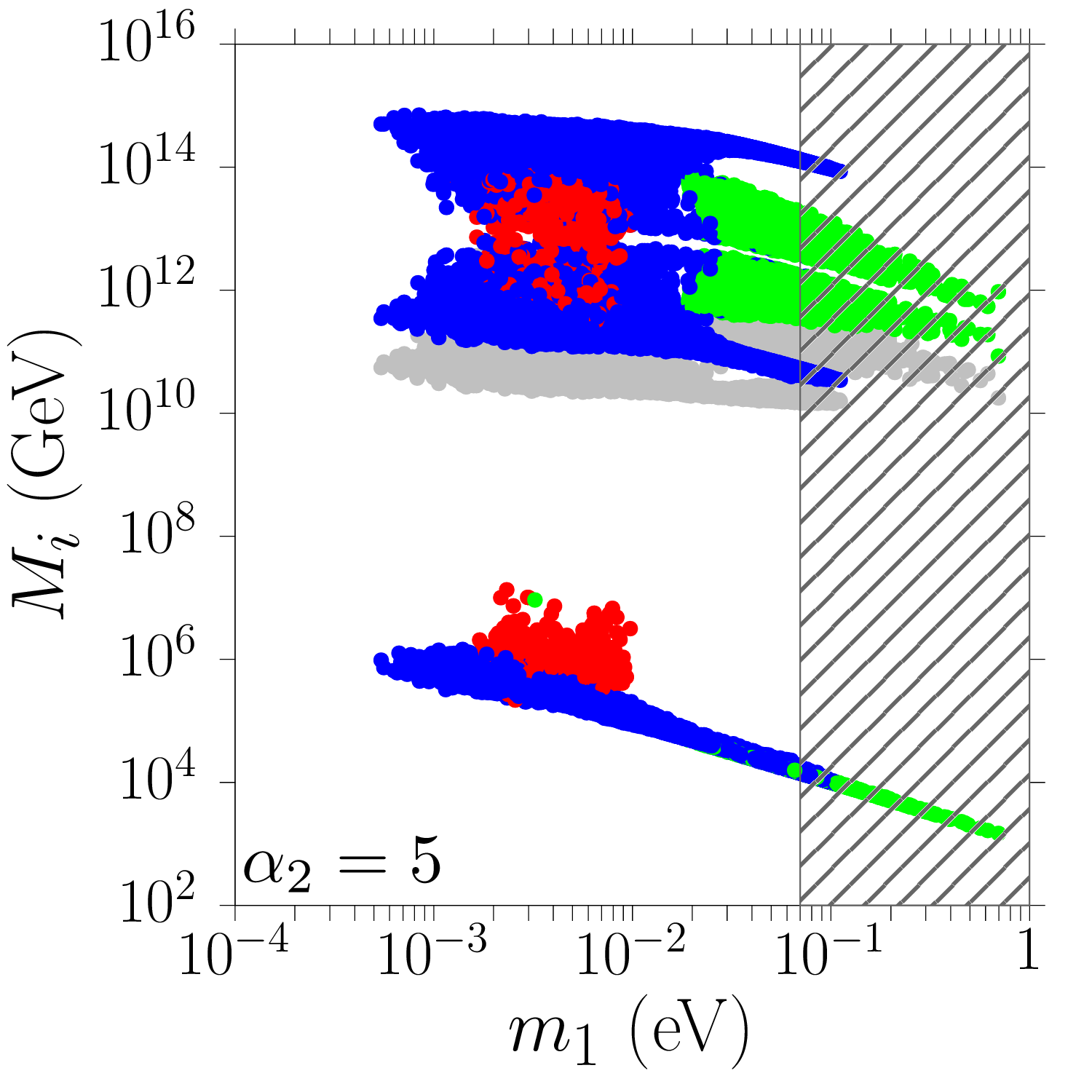,height=48mm,width=56mm}
\hspace{-7mm}
\psfig{file=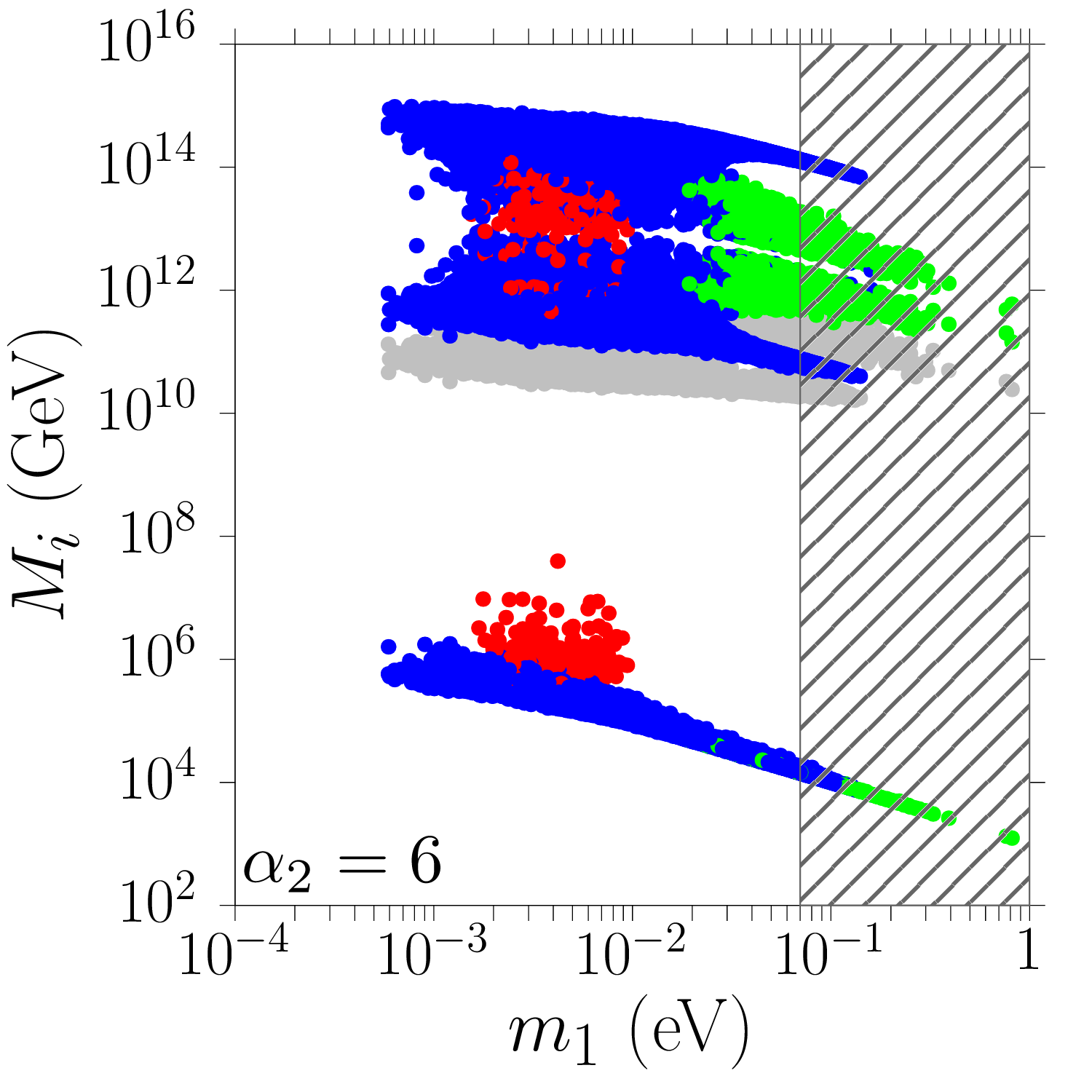,height=48mm,width=56mm}
\end{center}
\vspace{-10mm}
\caption{Scatter plots as in Fig.~2 but for  $\tan\b = 50$. In the top central panel
for $\a_2=2$, the dark (light) red points are solutions for $|\O_{ij}|^2 > 3$  ($|\O_{ij}|^2 < 3$) 
able to lower $T_{\rm RH}$ below $10^{10}\,{\rm GeV}$.}
\label{fldomNOhightan}
\end{figure}
Looking at the yellow (and orange) points, those satisfying only the successful leptogenesis condition, 
one can notice that the constraints are even more relaxed than in the previous case for $\tan\b=5$
compared to the non-supersymmetric case.
There is still a lower bound on the lightest neutrino mass $m_1 \gtrsim 1\,{\rm meV}$ that is just 
very slightly relaxed compared to non-supersymmetric case (the $C\!P$ asymmetry doubles
but the value of $K_{2\t}$ determining the wash-out at the production gets $\sim \sqrt{2}$ higher
and the two effects almost cancel out). 
This is interesting because one can conclude that the lower bound on $m_1$ is quite a stable and general feature 
of $SO(10)$-inspired models that, therefore, predict some deviation from the hierarchical limit
though this might well be below current experimental sensitivity. Indeed in the most
optimistic case cosmological observations should place a $2\s$ upper bound 
$m_1 \lesssim 10\,{\rm meV}$ \cite{wong,lowerbound}. 

From Fig.~3 it should be also noticed how the region satisfying $K_{1e} \lesssim 1$ now greatly enlarges compared
to the large $\tan\b$ case.  Indeed, if one looks at the panels in Fig.~4, 
showing again (as in Fig.~2 but now for $\tan\b=50$) what flavour dominates
the final asymmetry, one can notice how this time there are plenty of electron dominated solutions, in the range for $2 \, {\rm meV} \lesssim m_1 \lesssim 10\,{\rm meV}$, 
as anticipated. This region was very marginal, almost absent, in the
non-supersymmetric case and it was still quite marginal also for $\tan\b=5$, as discussed. 
However now, for $\tan\b=50$, it becomes quite significant  and as we will discuss in the next section, 
it allows a relaxation of the lower bound on $T_{\rm RH}$ below $10^{10}\,{\rm GeV}$ for
$1\lesssim \a_2 \lesssim 2$. 
We should however stress again that these electron dominated solutions occur in the weak wash-out regime 
at the production ($K_{2e}\lesssim 1$) and, therefore, they strongly depend on the initial $N_2$ abundance.
They exist for initial thermal $N_2$ abundance but for
 vanishing initial $N_2$ abundance they 
 completely disappear (i.e. they do not realise successful leptogenesis).
The reason why they are obtained much easier at large $\tan\b$ values compared
to low $\tan\b$ values is because now the condition of weak wash-out at the production 
is more relaxed, $K_{2e}\lesssim 1$ instead of $K_{2\t_2^{\bot}}\equiv K_{2e}+K_{2\mu} \lesssim 1$.

If we again consider the subset of points satisfying also the strong thermal condition (dark and light blue points),
we can see that, as in the low $\tan\b$ case, the region is now much more extended, even more than before. 
This happens because of the effect explained at the end of Section 2:
one can now have $K_{1\mu}\lesssim 1$ and at the same time
wash-out the pre-existing asymmetry  in all flavours imposing $K_{1e}, K_{2\m}, K_{2\t} \gg 1$. This opens up a new 
(muon dominated) region  at large values of $m_1 \gtrsim 0.05\,{\rm eV}$, though notice that this is now largely excluded by the upper bound Eq.~(\ref{m1ub}). 

\subsection{Inverted ordering}

Let us now discuss the IO case distinguishing, as  we did for NO, small $\tan\b$ values ($\lesssim 15$)
from large $\tan\b$ values ($\gtrsim 15$). 

\subsubsection{Small $\tan \b$ values}

For small $\tan\b$ values the situation is, as for NO, similar to the non-supersymmetric case though
the allowed regions are slightly more relaxed.
\begin{figure}
\begin{center}
\psfig{file=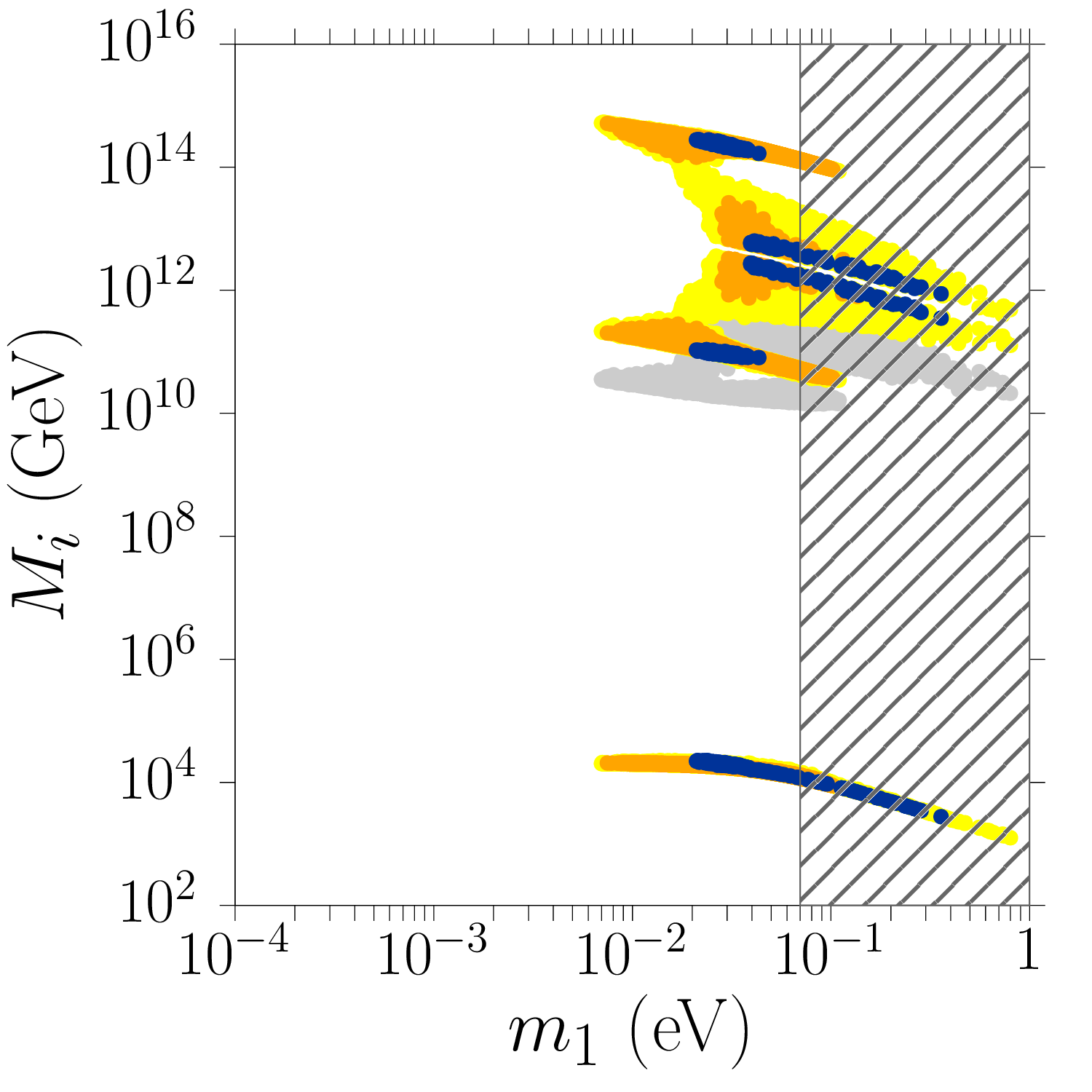,height=49mm,width=56mm}
\hspace{-7mm}
\psfig{file=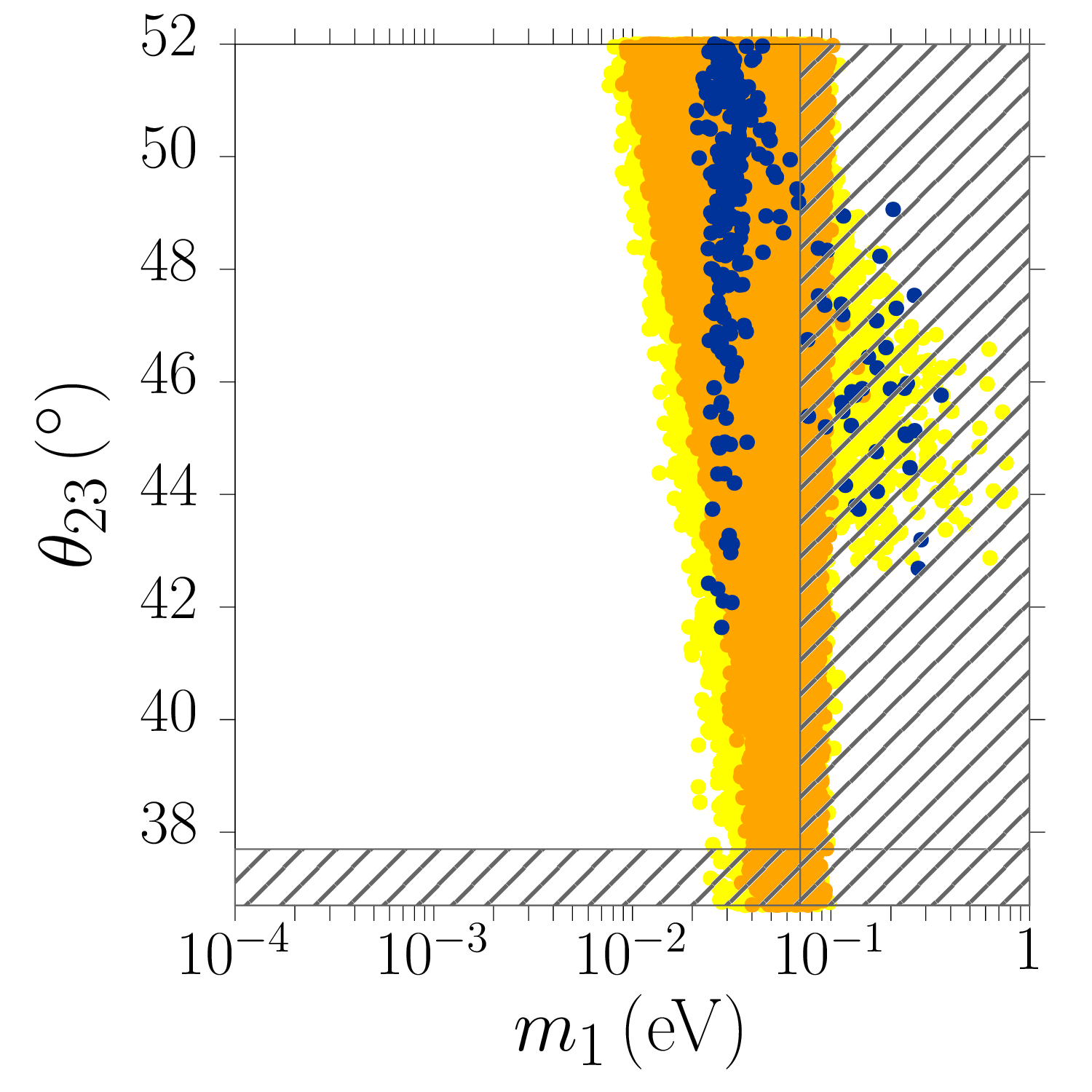,height=49mm,width=56mm}
\hspace{-7mm}
\psfig{file=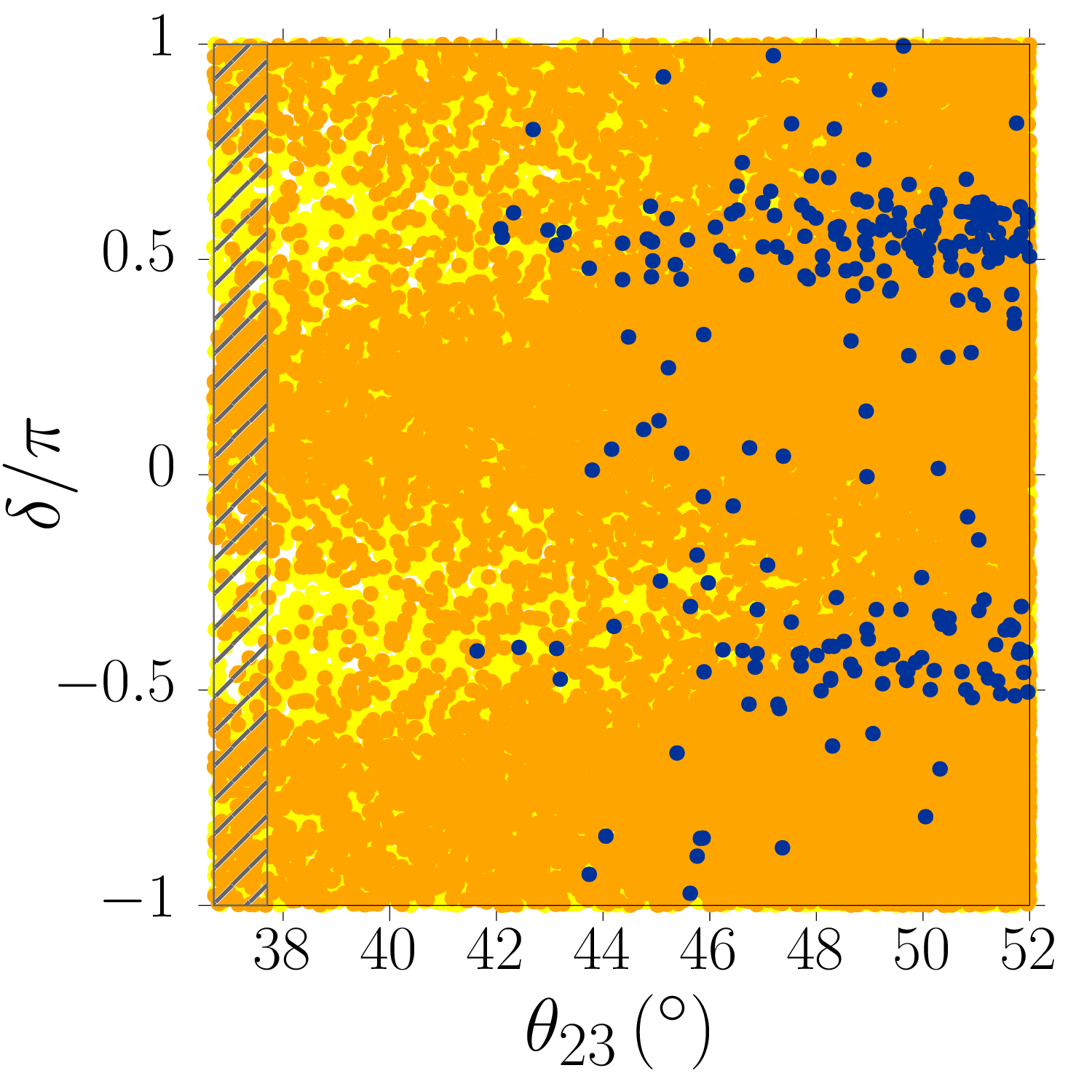,height=49mm,width=56mm} \\
\psfig{file=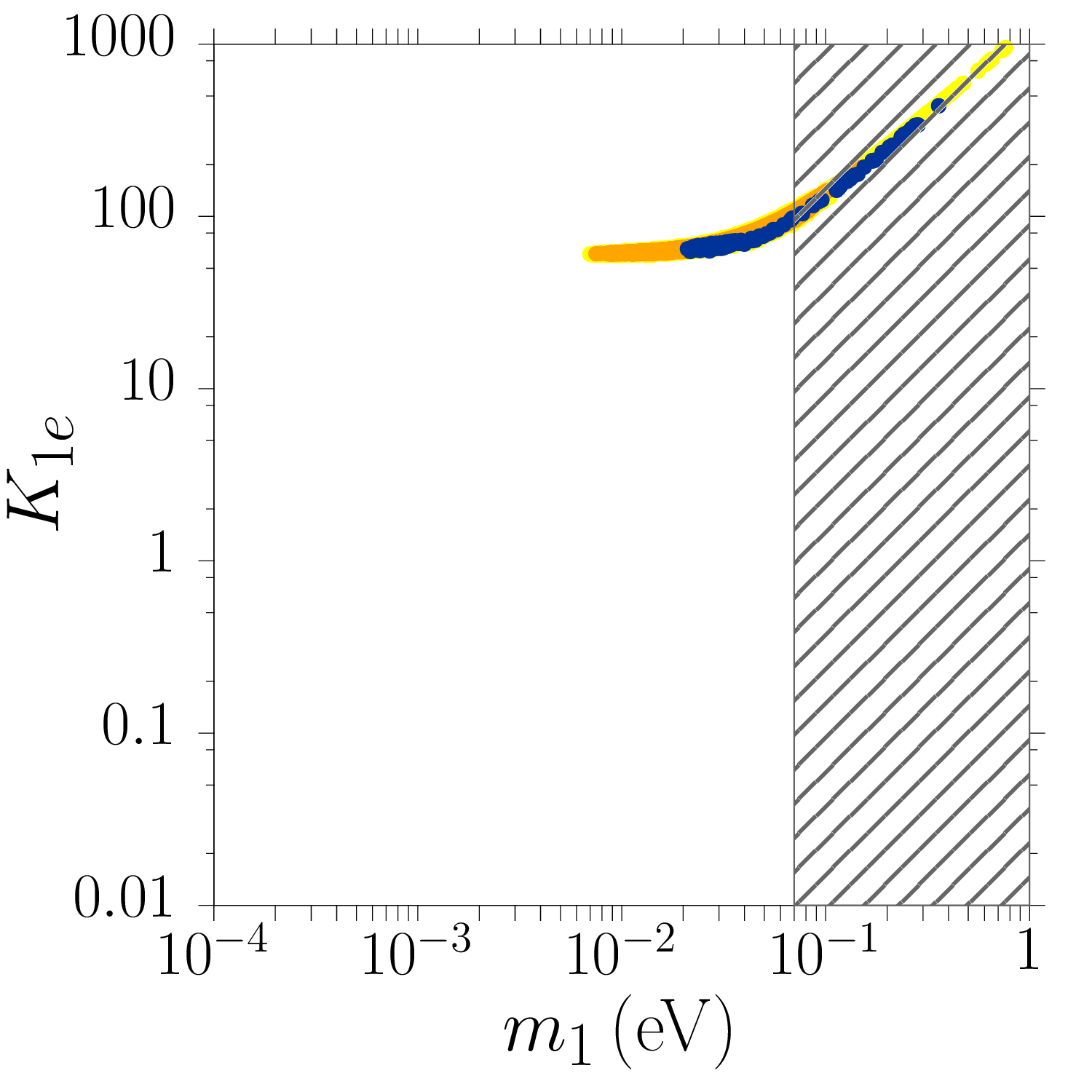,height=49mm,width=56mm}
\hspace{-7mm}
\psfig{file=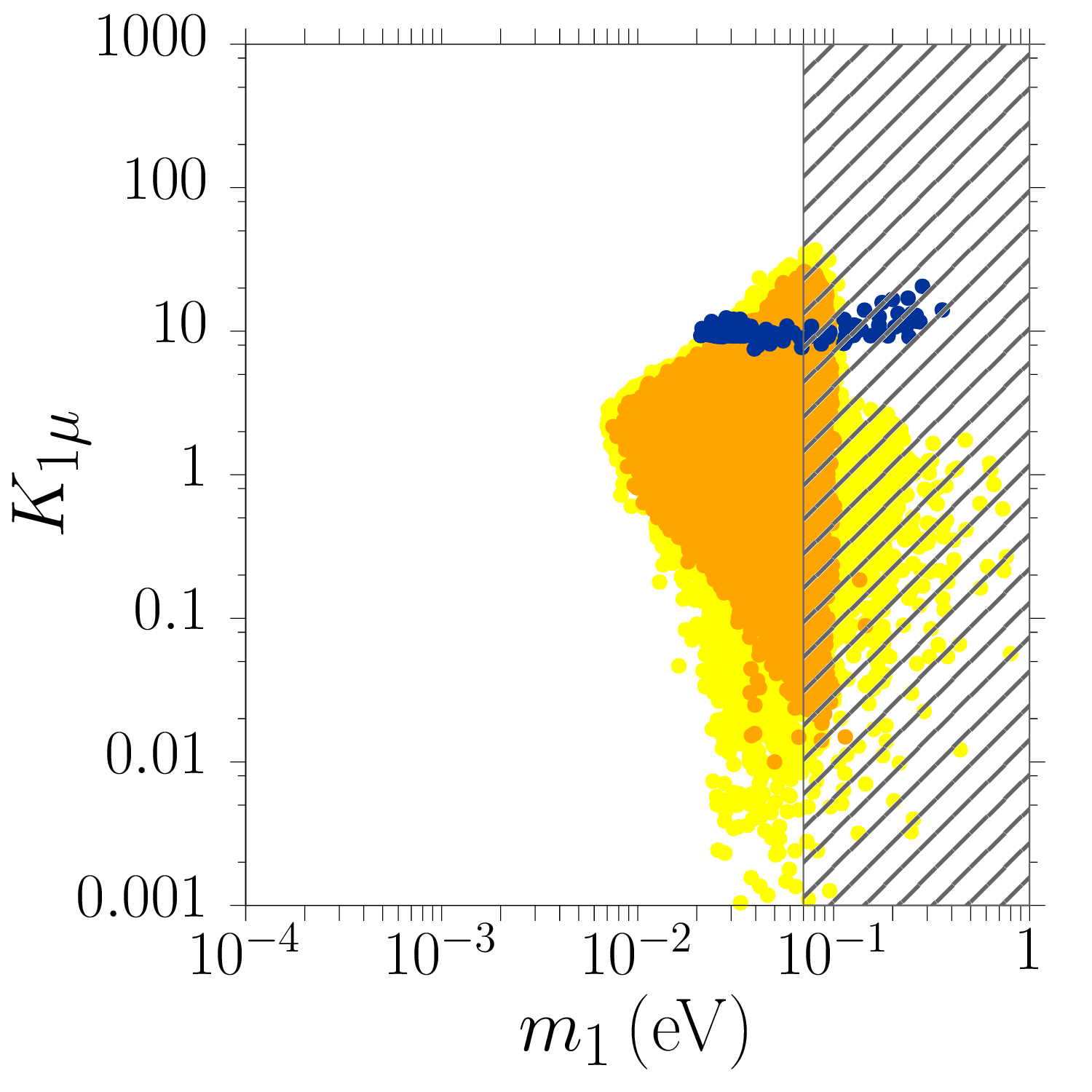,height=49mm,width=56mm}
\hspace{-7mm}
\psfig{file=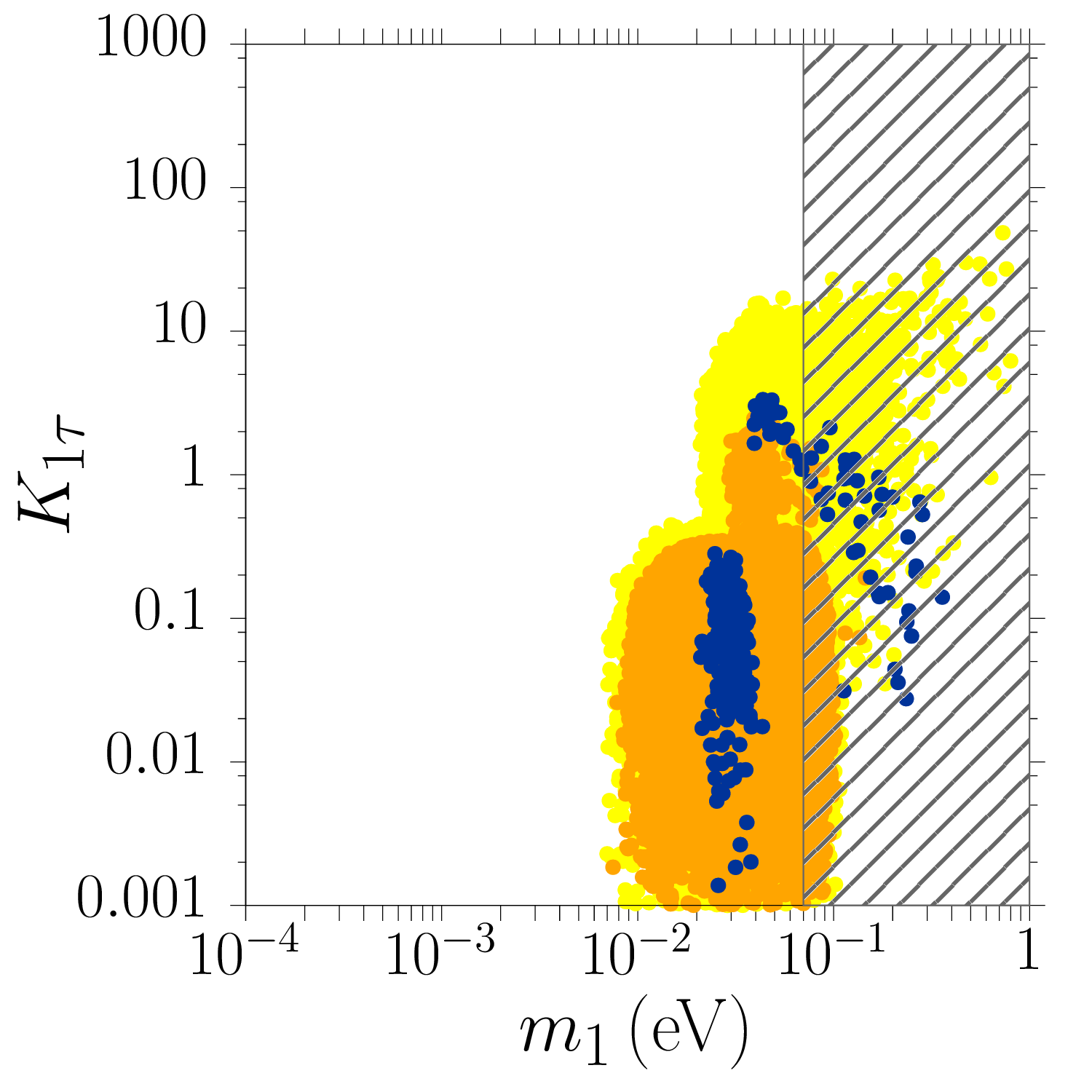,height=49mm,width=56mm}
\end{center}
\vspace{-10mm}
\caption{Scatter plots as in Fig.~1 but for IO and $\tan\b=5$.}
\label{constrIO}
\end{figure}
\begin{figure}
\begin{center}
\psfig{file=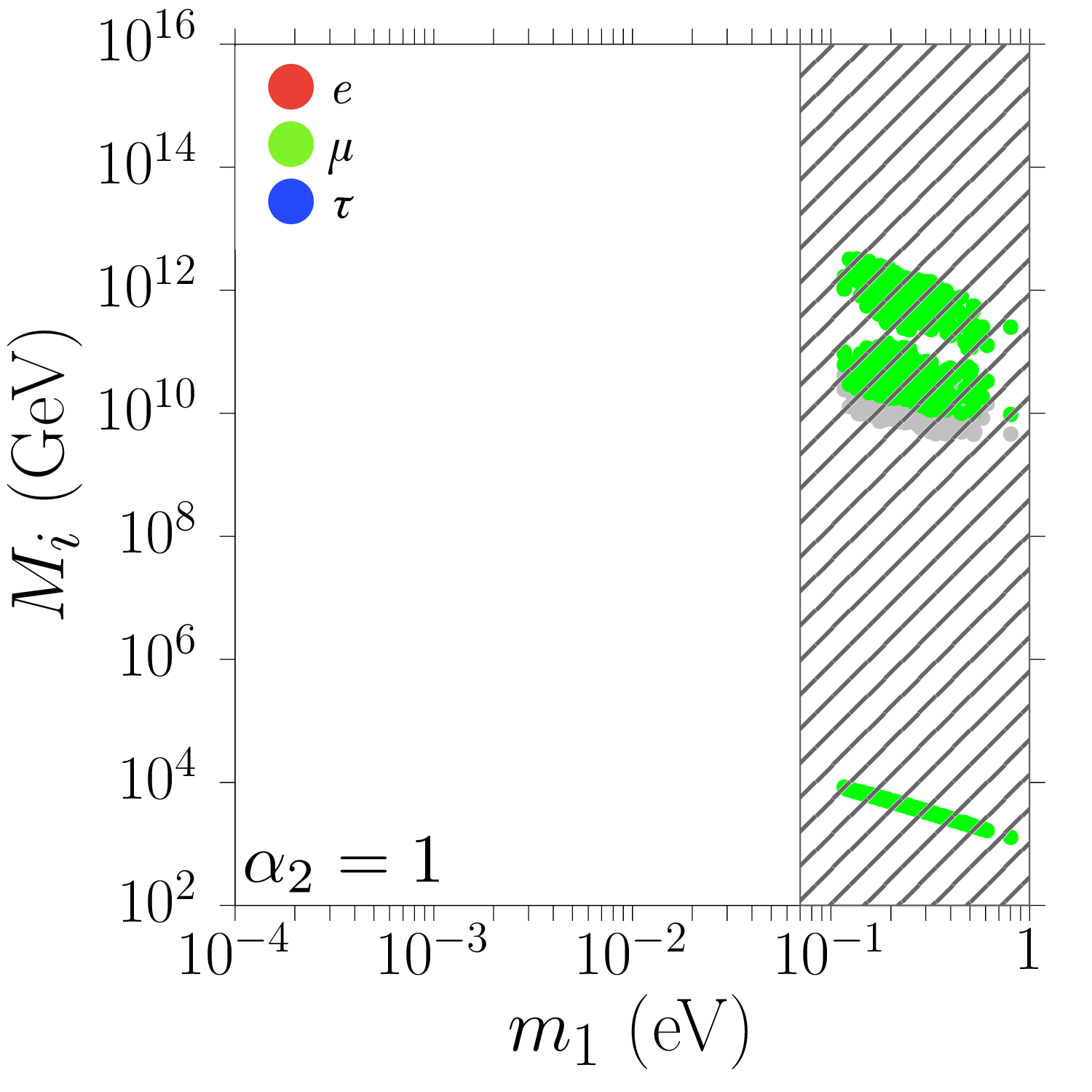,height=49mm,width=56mm}
\hspace{-7mm}
\psfig{file=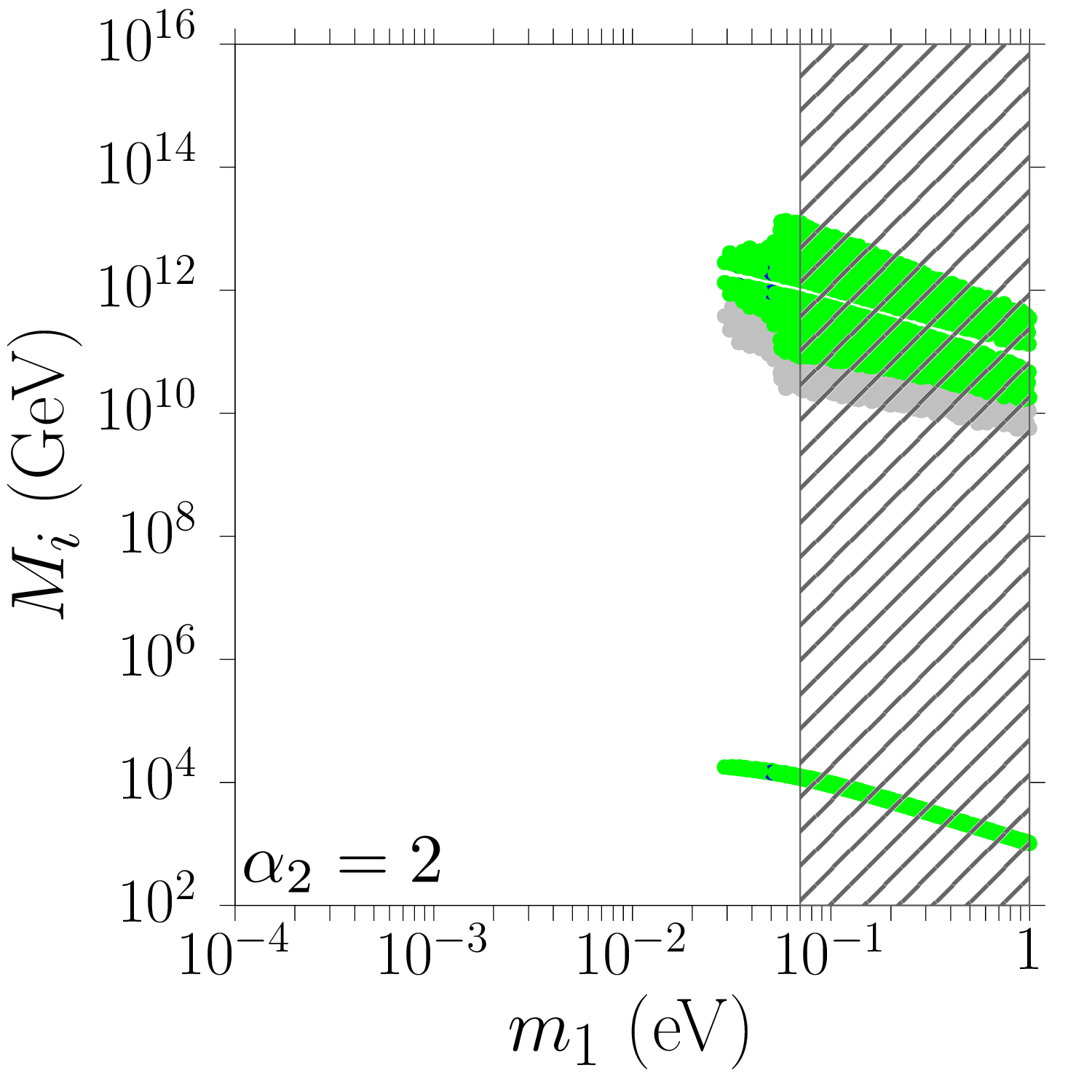,height=49mm,width=56mm}
\hspace{-7mm}
\psfig{file=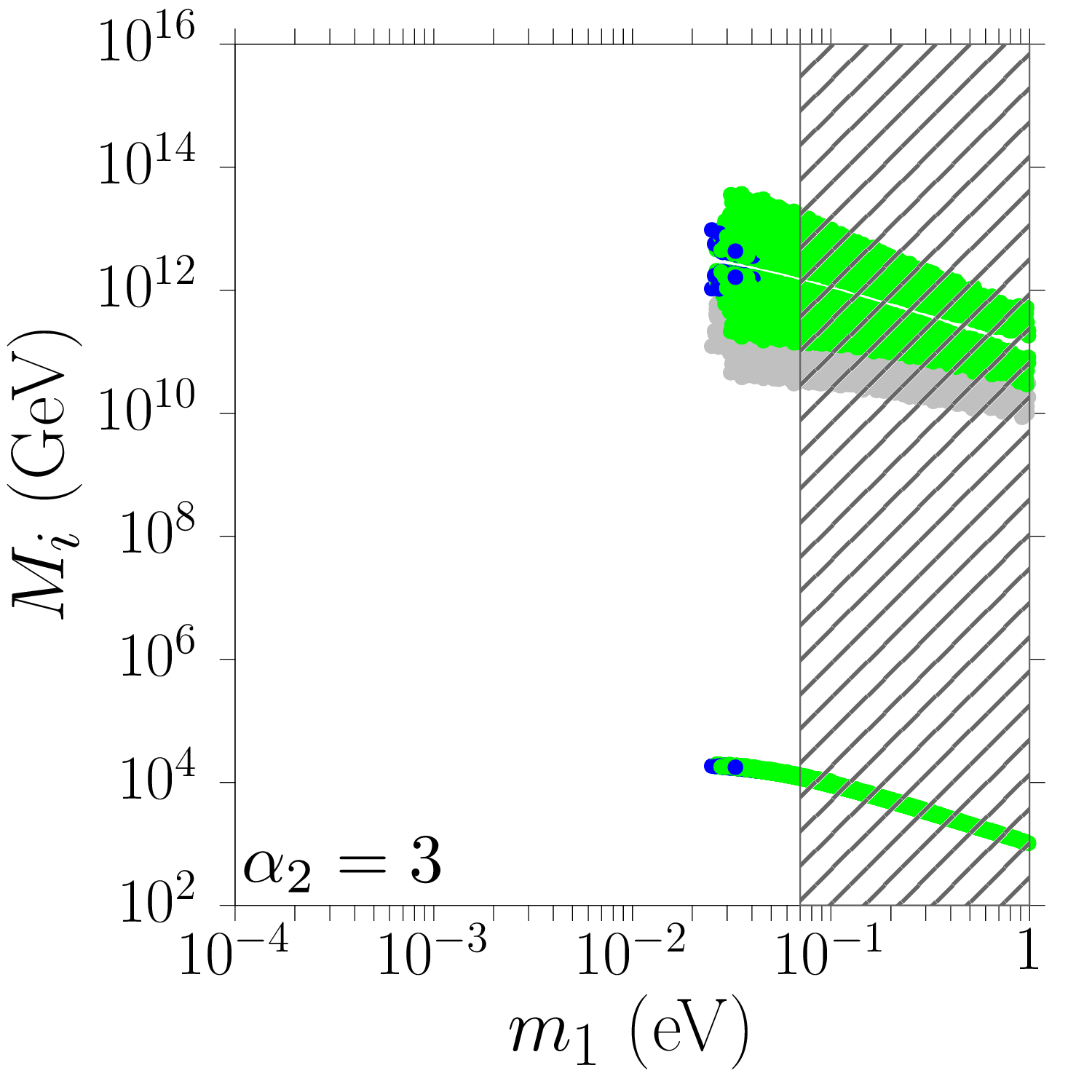,height=49mm,width=56mm}  \\
%\hspace{-7mm}
\psfig{file=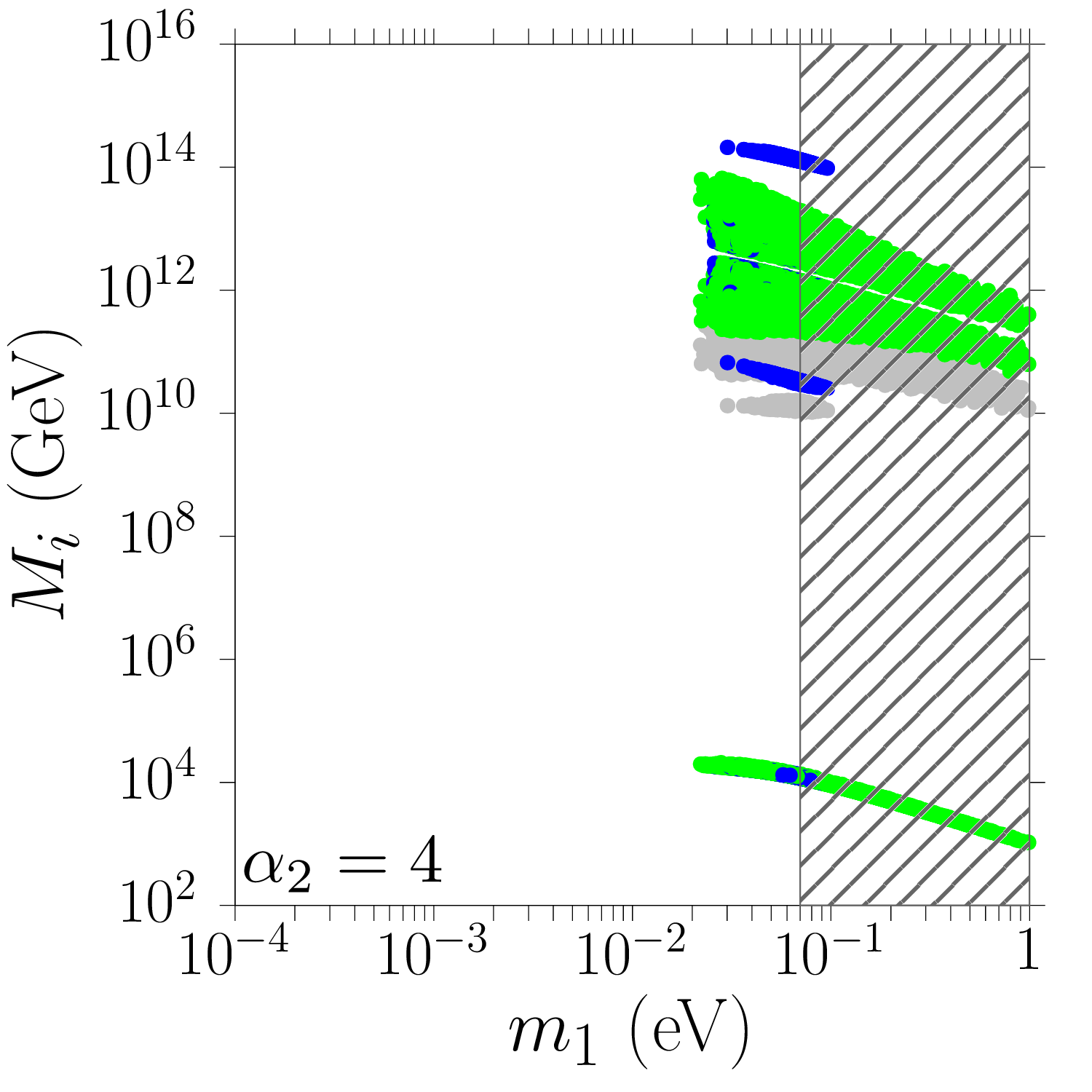,height=49mm,width=56mm}
\hspace{-7mm}
\psfig{file=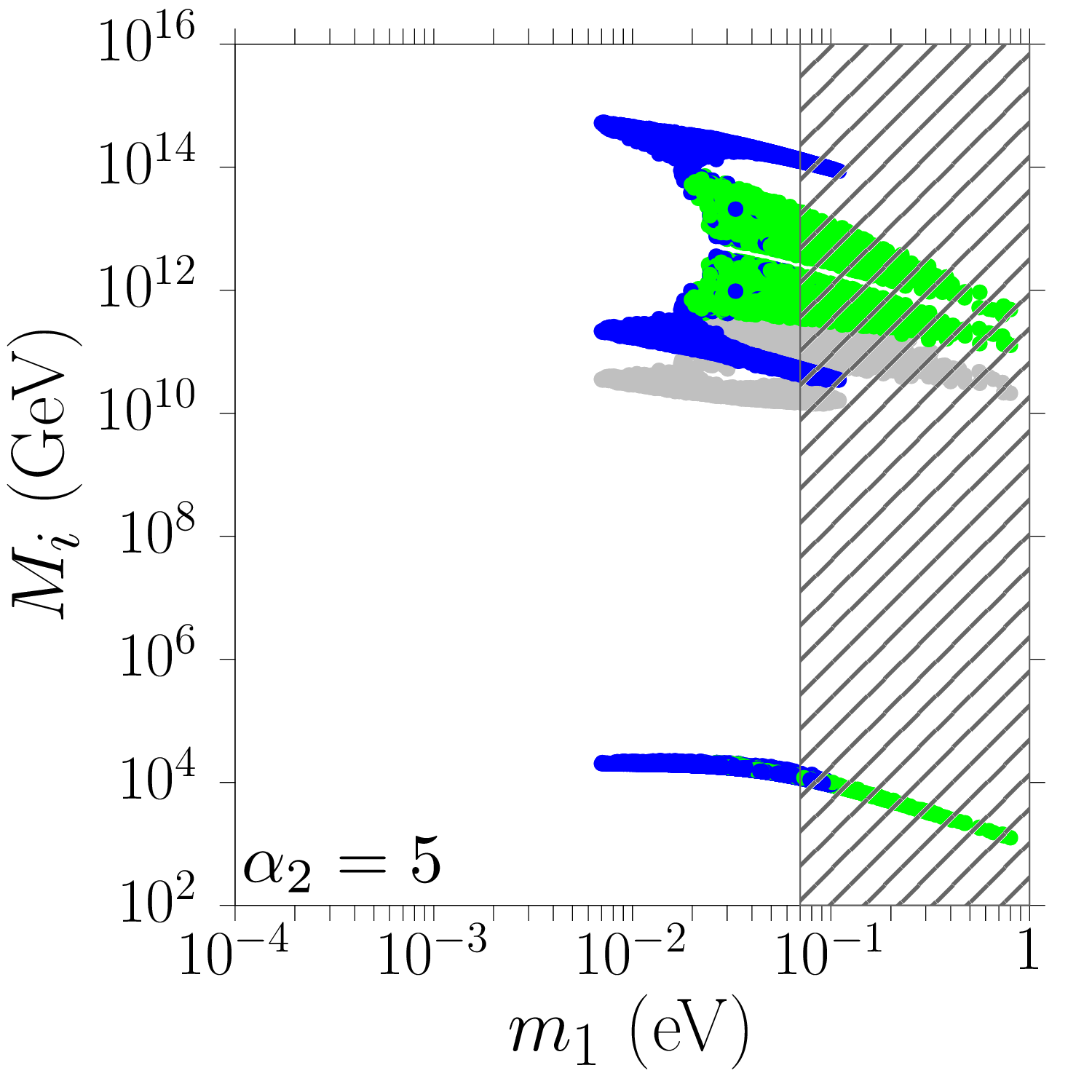,height=49mm,width=56mm}
\hspace{-7mm}
\psfig{file=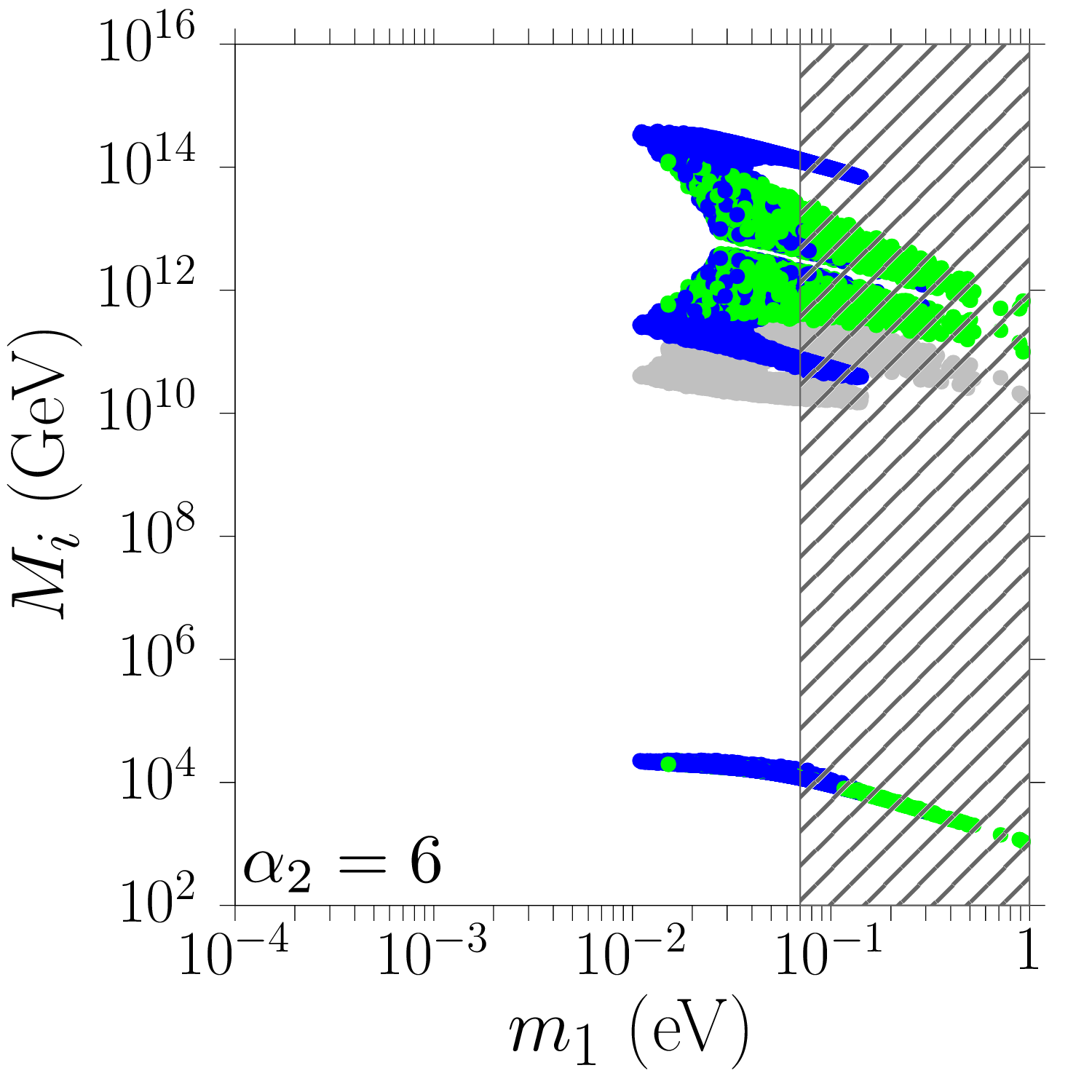,height=49mm,width=56mm}
\end{center}
\vspace{-10mm}
\caption{Scatter plots as in Fig.~2 but for  IO and $\tan\b =5$.}
\label{fldomIOhightan}
\end{figure}
In Fig.~5 we show the results again for $\tan\b=5$ and one can see in particular that:
\begin{itemize}
\item there is a lower bound $m_1 \gtrsim 10\,{\rm meV}$ corresponding to $\sum m_i \gtrsim 130\,{\rm meV}$
that will be in a close future tested by the cosmological observations;
\item this time, differently from the non-supersymmetric case, there is no lower bound on the atmospheric 
mixing angle, though values in the first octant require higher values of the absolute neutrino mass scale 
on the verge of being excluded by the cosmological observations.  
\end{itemize}
We can therefore conclude again, as in the non-supersymmetric case, that the IO case is disfavoured compared to the NO case.

In Fig.~6 we show again, with the same colour code as in Fig.~2 and 4 for NO, the solutions for various values of $\a_2$ 
indicating the flavour that dominates the final asymmetry. This time one can see that, even for
initial thermal $N_2$-abundance, there are no electron dominated solutions. 
The reason is simply that in the IO case one has 
$K_{1e}=m_{ee}/m_{\star}^{MSSM} \gtrsim 70$ \cite{SO10decription}
and, therefore, the electron asymmetry is completely washed-out by the lightest
RH neutrinos inverse processes. 
% \newpage
\subsubsection{Large $\tan \b$ values}

For large $\tan\b$ values and imposing successful leptogenesis condition, the situation is qualitatively
similar to the case of small $\tan\b$ as one can see from Fig.~7 (orange and yellow points) but simply the allowed
regions slightly further enlarge. For example now one has $m_1 \gtrsim 7\,{\rm meV}$.
 \begin{figure}
\begin{center}
\psfig{file=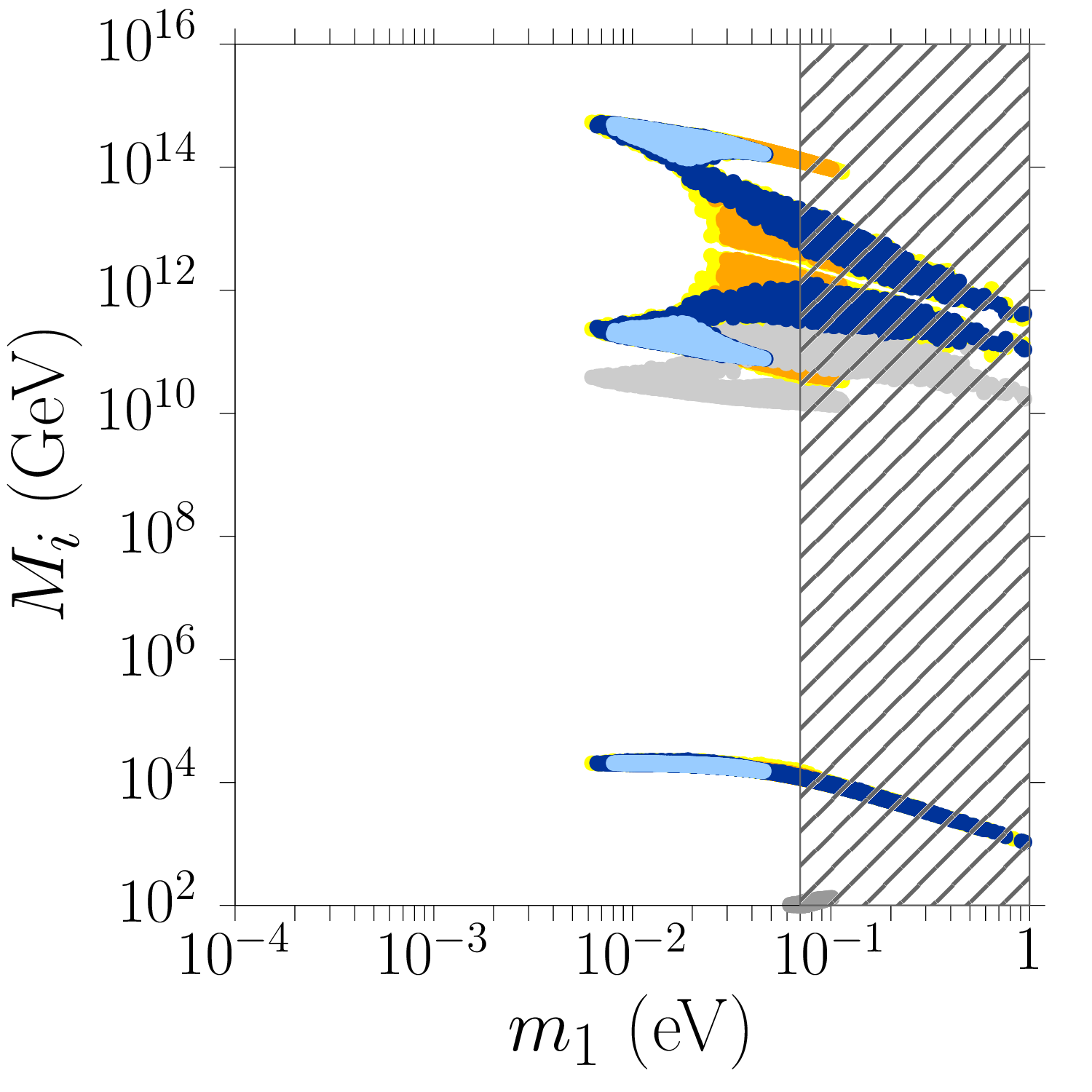,height=50mm,width=56mm}
\hspace{-7mm}
\psfig{file=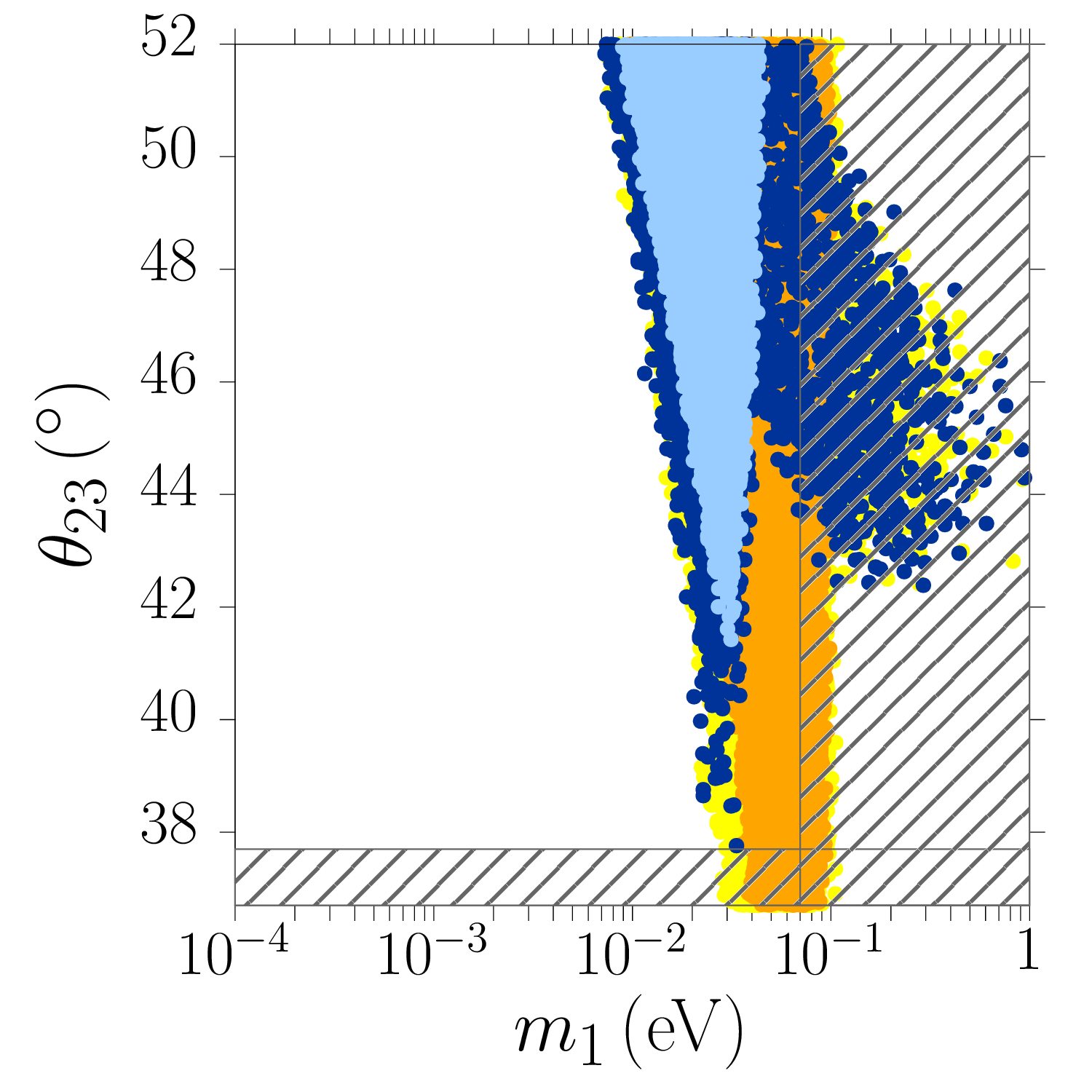,height=50mm,width=56mm}
\hspace{-7mm}
\psfig{file=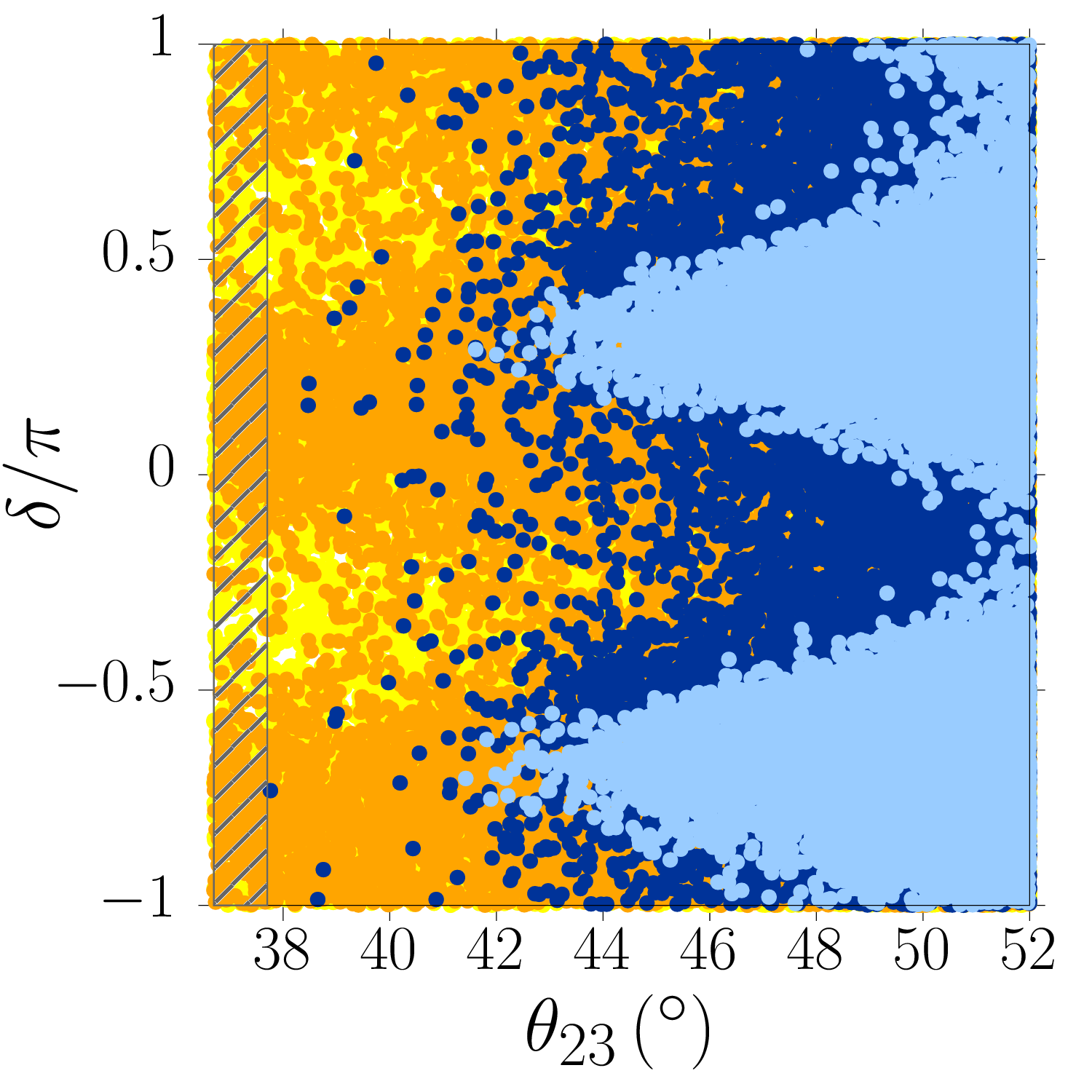,height=50mm,width=56mm} \\
\hspace{-7mm}
\psfig{file=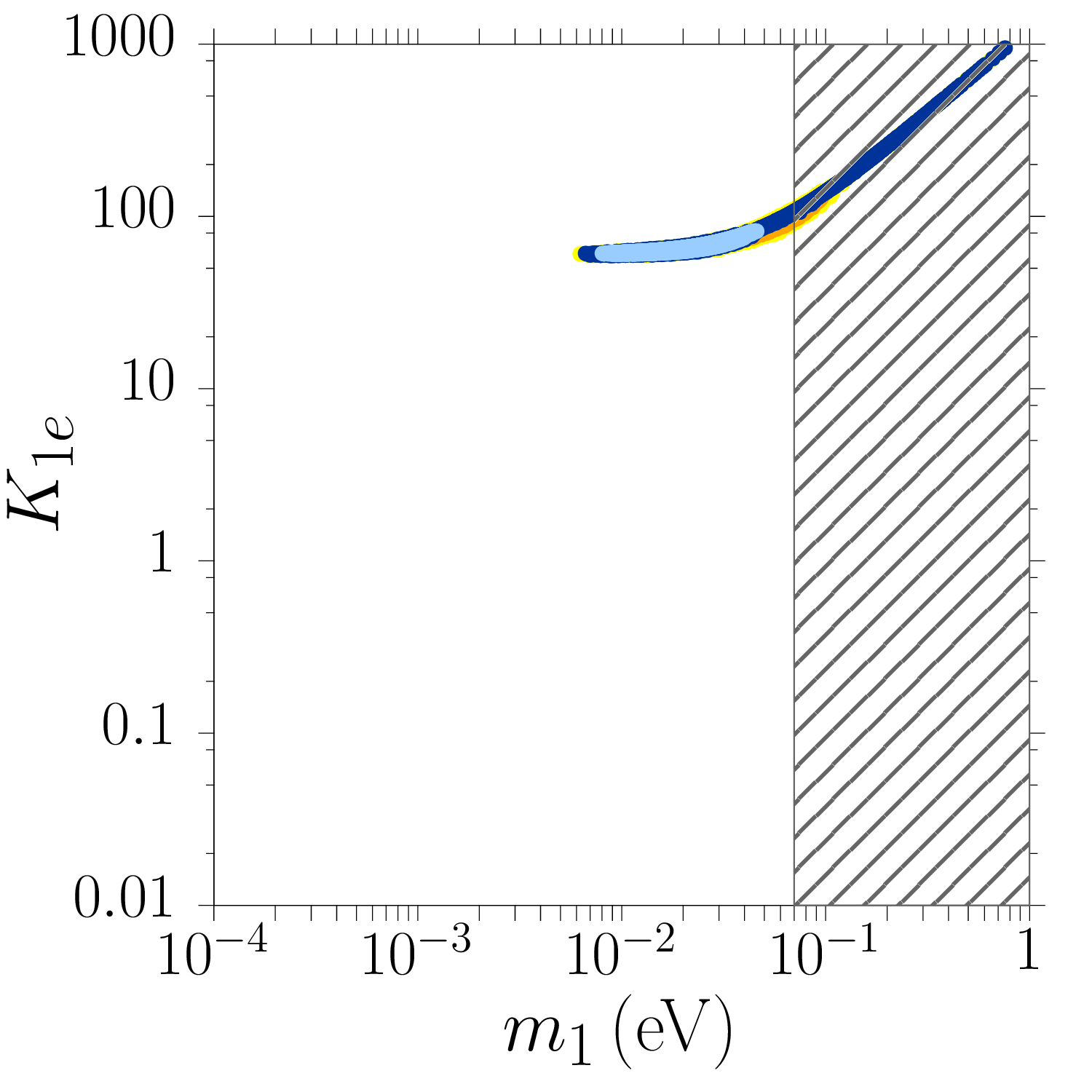,height=50mm,width=56mm}
\hspace{-7mm}
\psfig{file=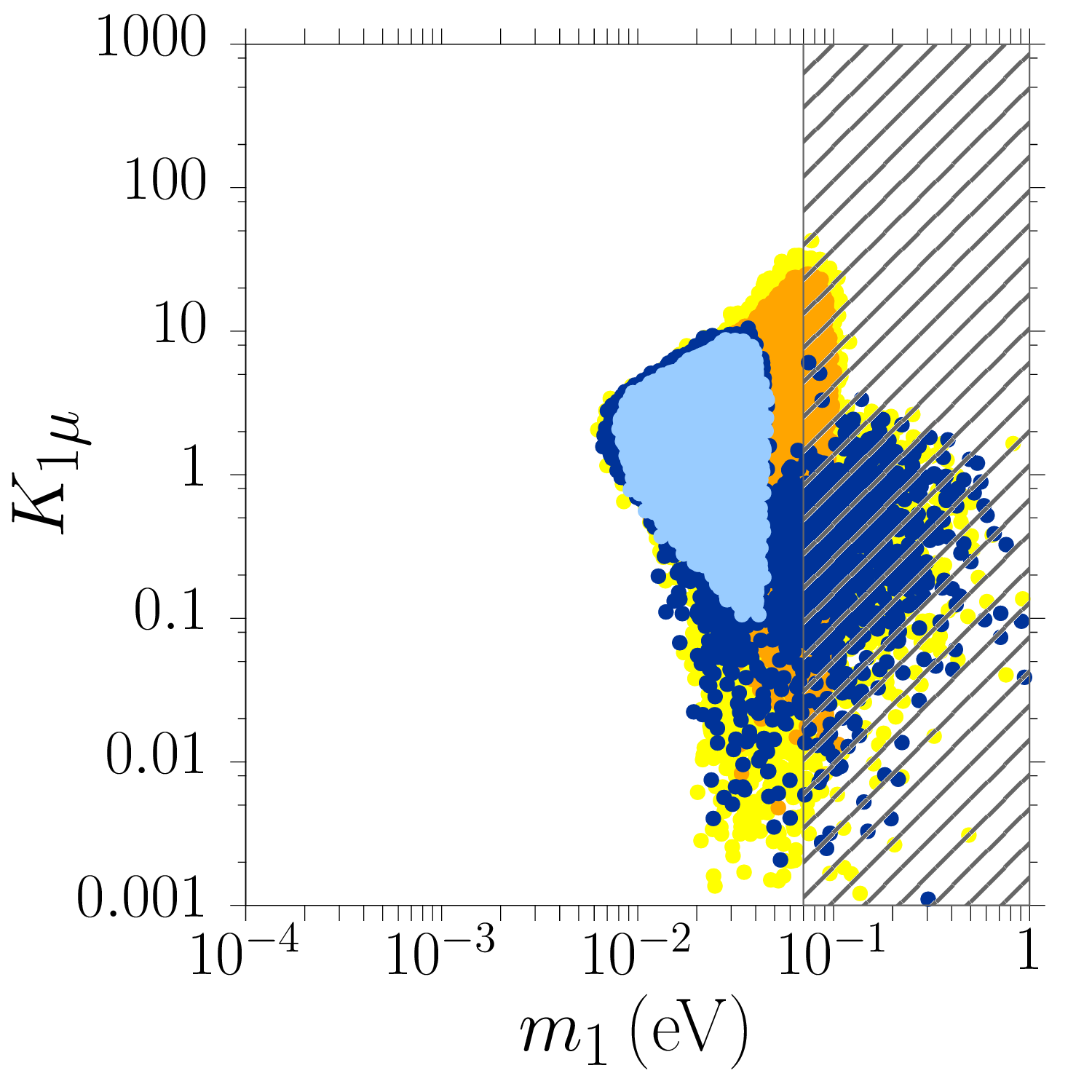,height=50mm,width=56mm}
\hspace{-7mm}
\psfig{file=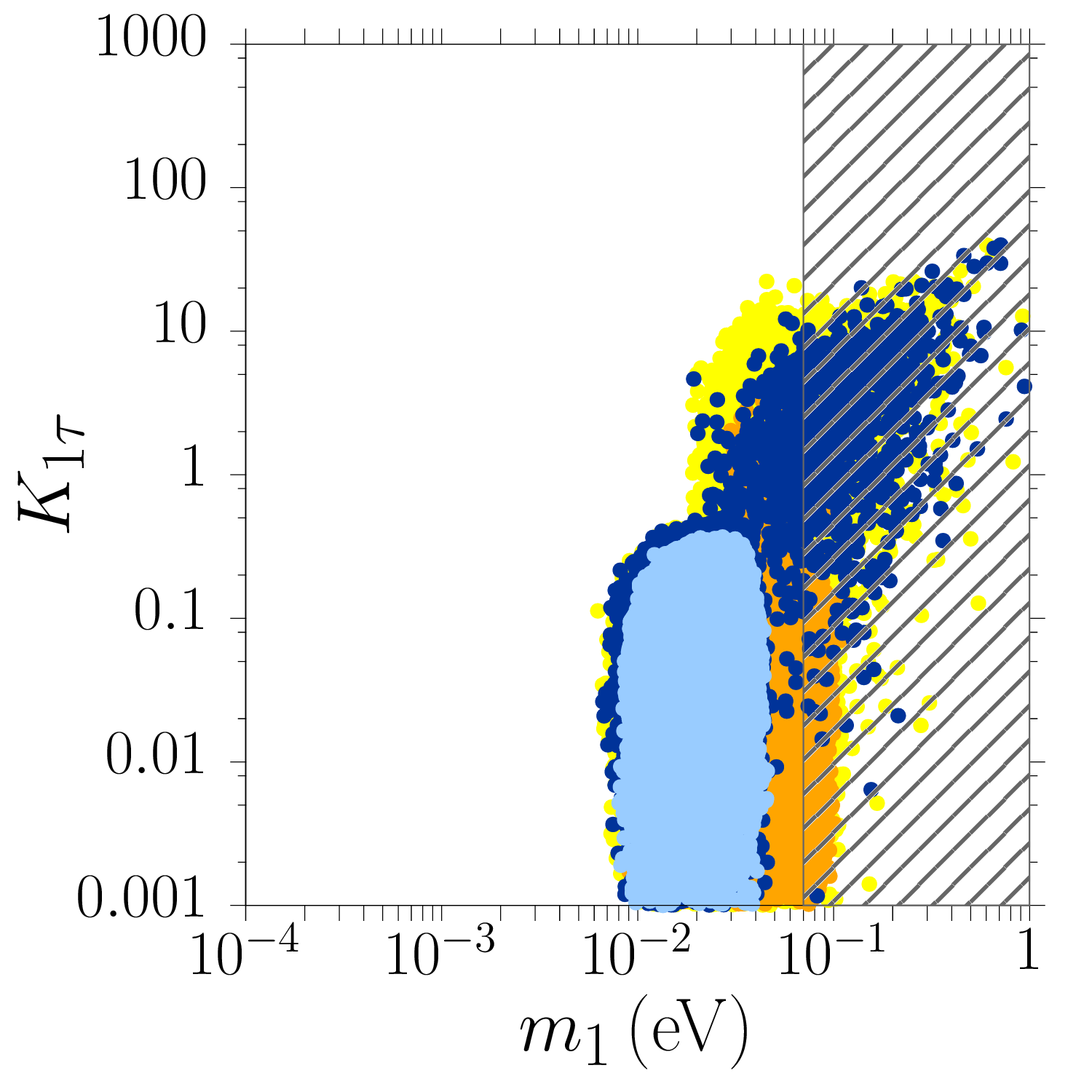,height=50mm,width=56mm}
\end{center}
\vspace{-11mm}
\caption{Scatter plots as in Fig.~1 but for IO and $\tan\b=50$.}
\label{constrIO}
\end{figure}
In the panels of Fig.~8 we show the dominant flavour and one can see that, for the same reason, there are no electron dominated solutions (no red points).
\begin{figure}
\begin{center}
\psfig{file=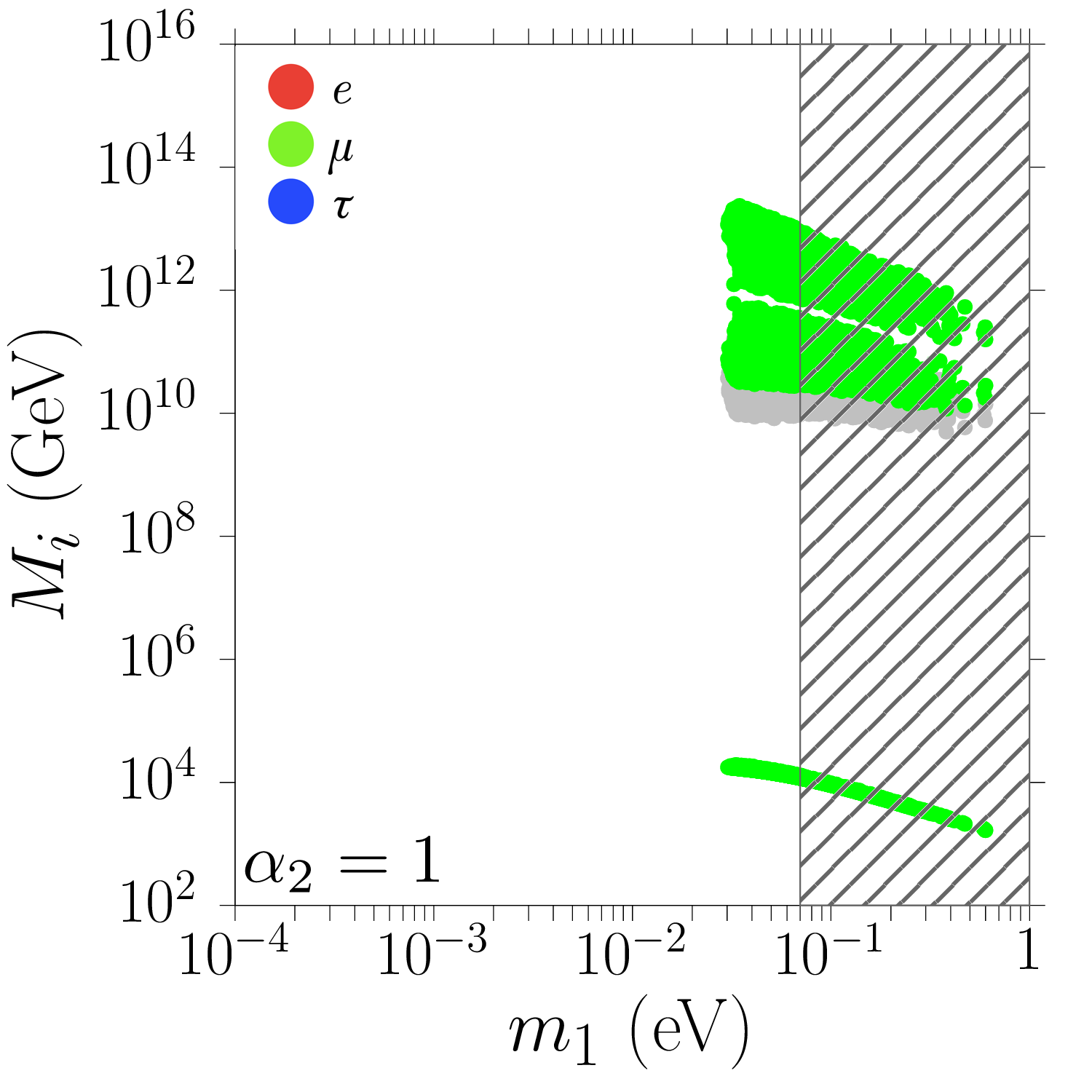,height=50mm,width=56mm}
\hspace{-7mm}
\psfig{file=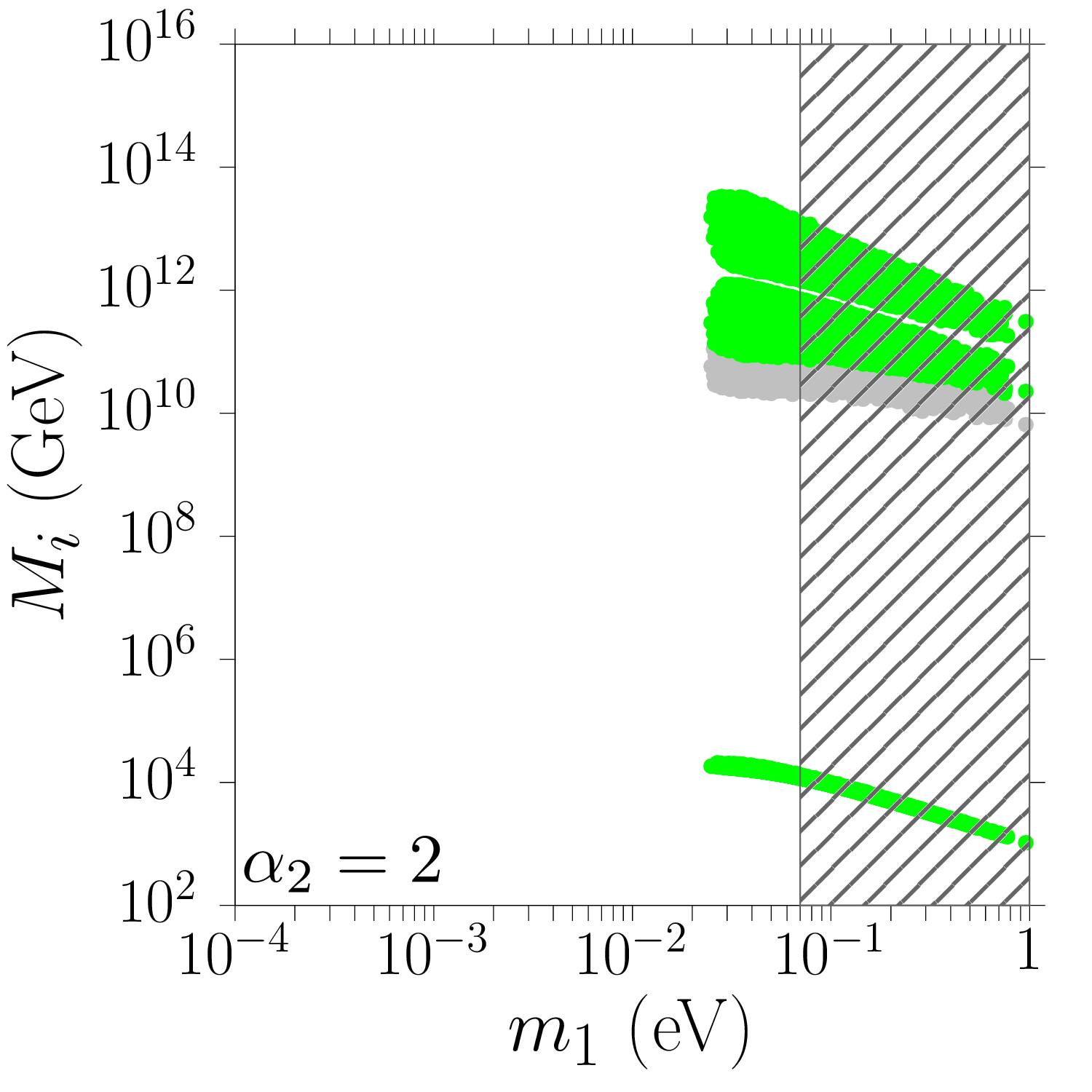,height=50mm,width=56mm}
\hspace{-7mm}
\psfig{file=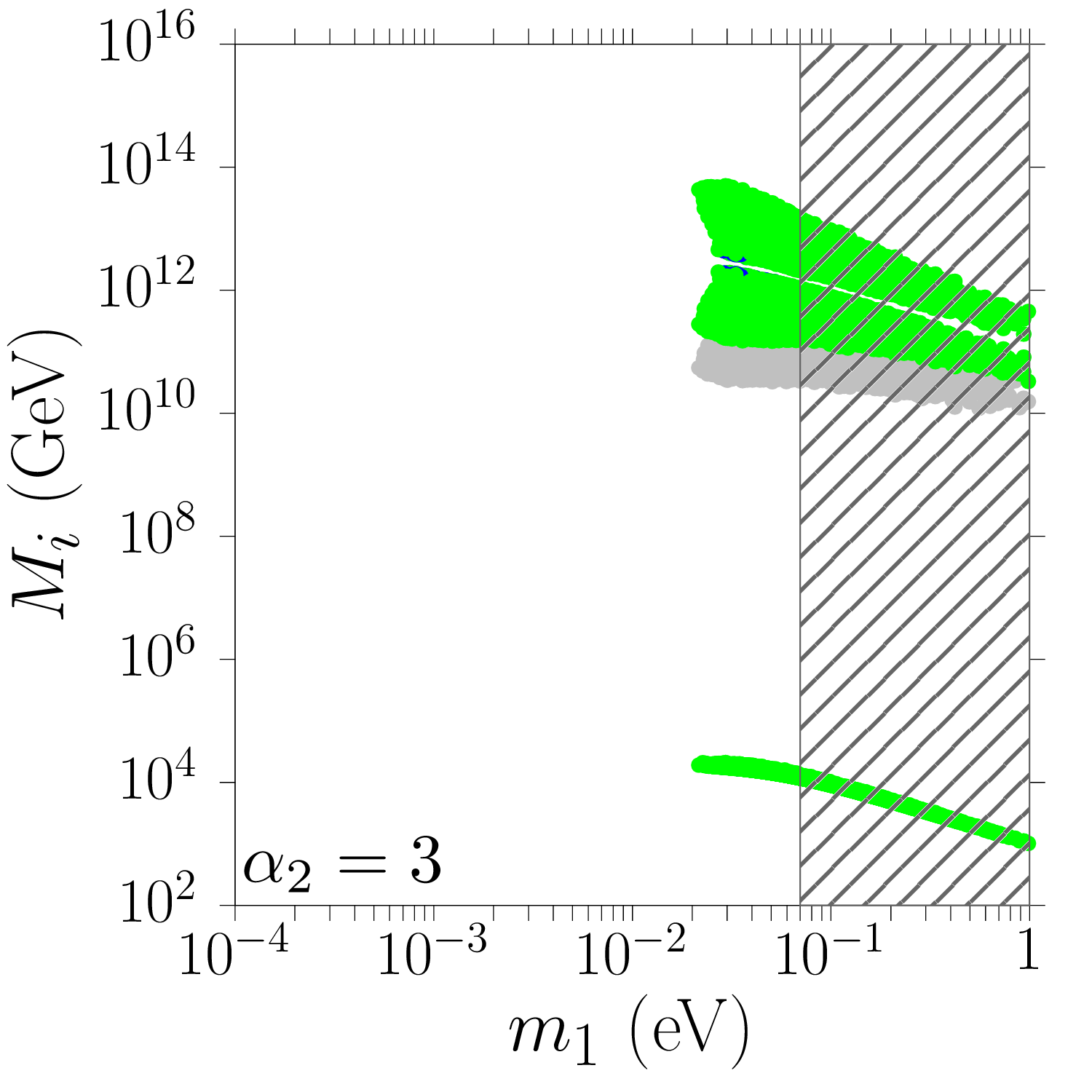,height=50mm,width=56mm}  \\
\hspace{-7mm}
\psfig{file=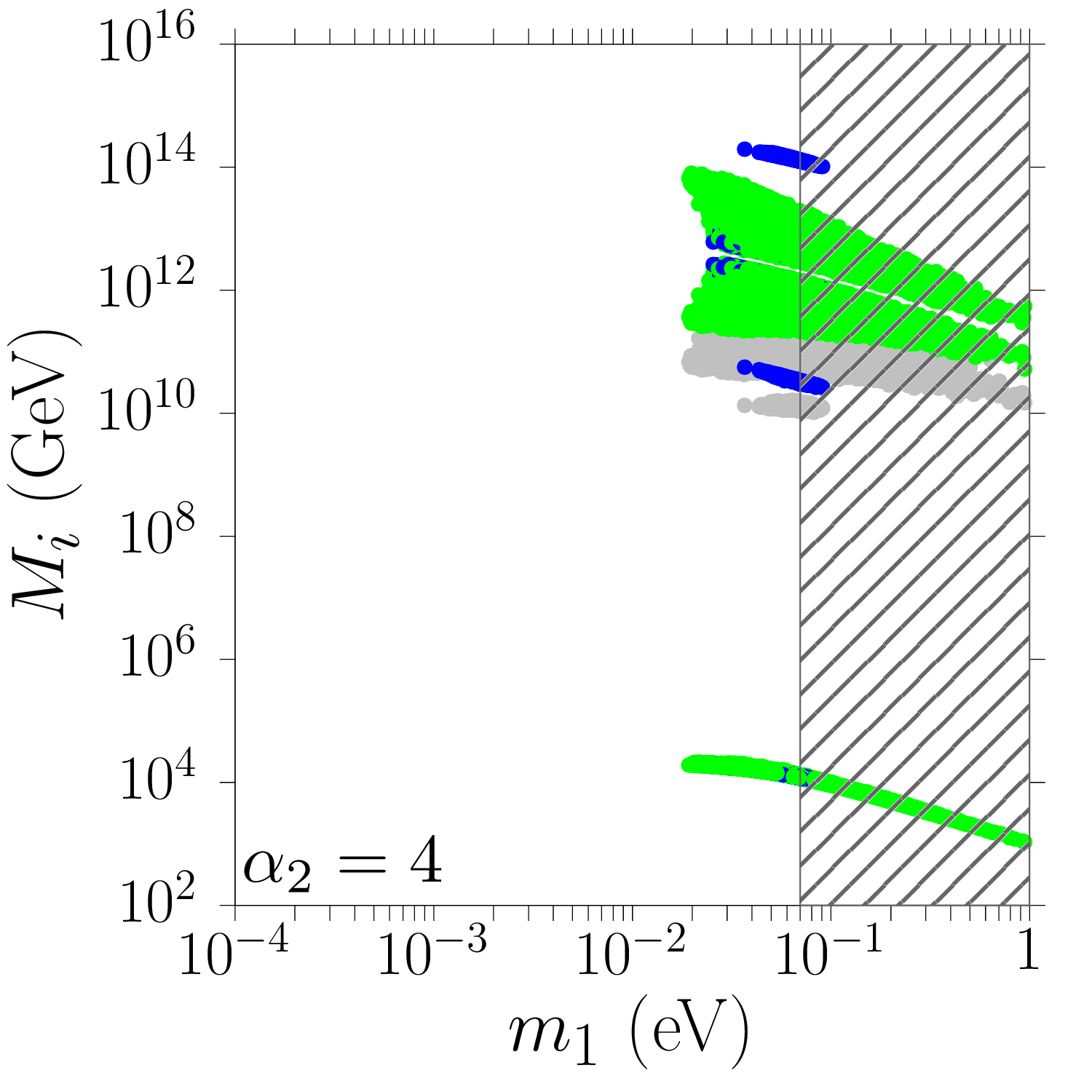,height=50mm,width=56mm}
\hspace{-7mm}
\psfig{file=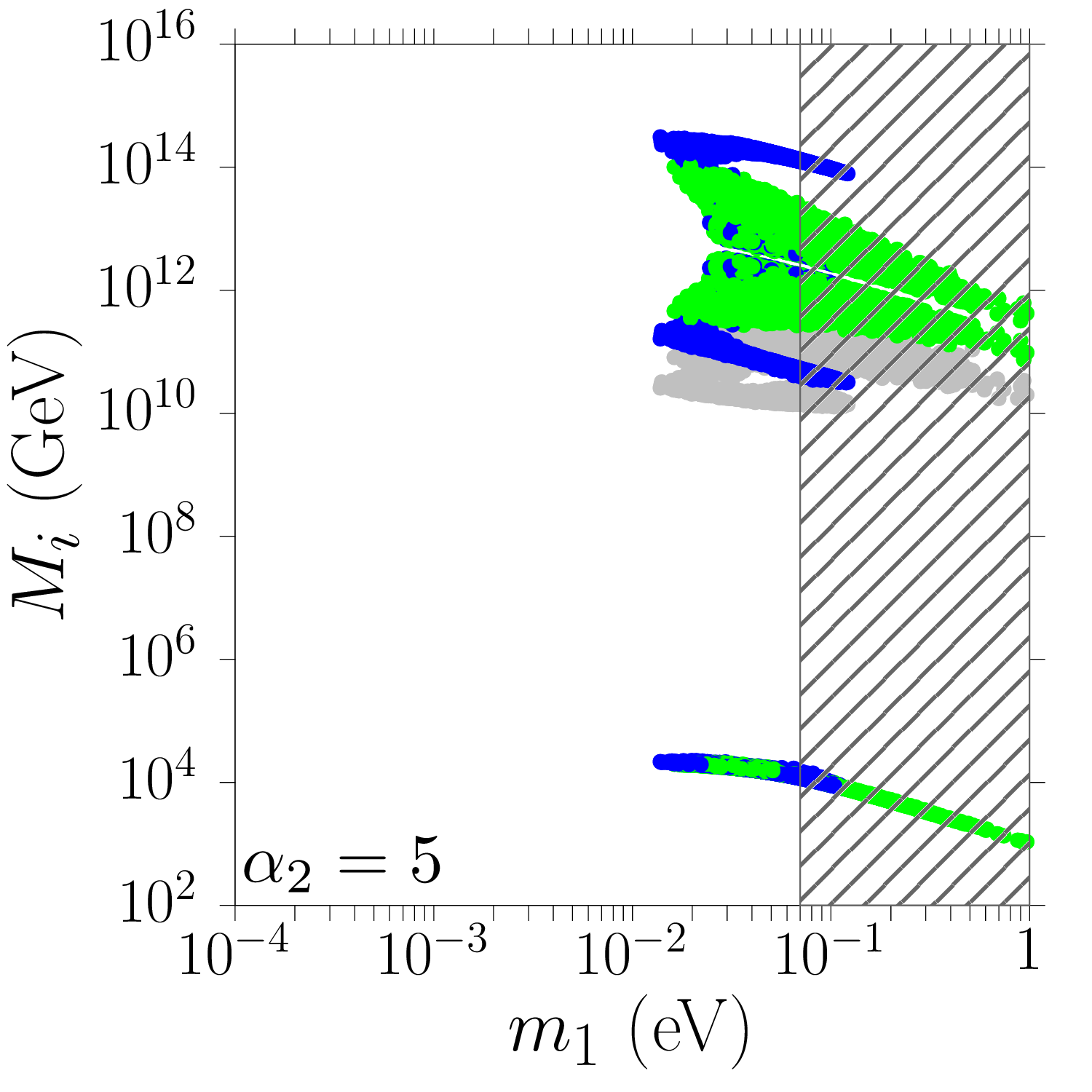,height=50mm,width=56mm}
\hspace{-7mm}
\psfig{file=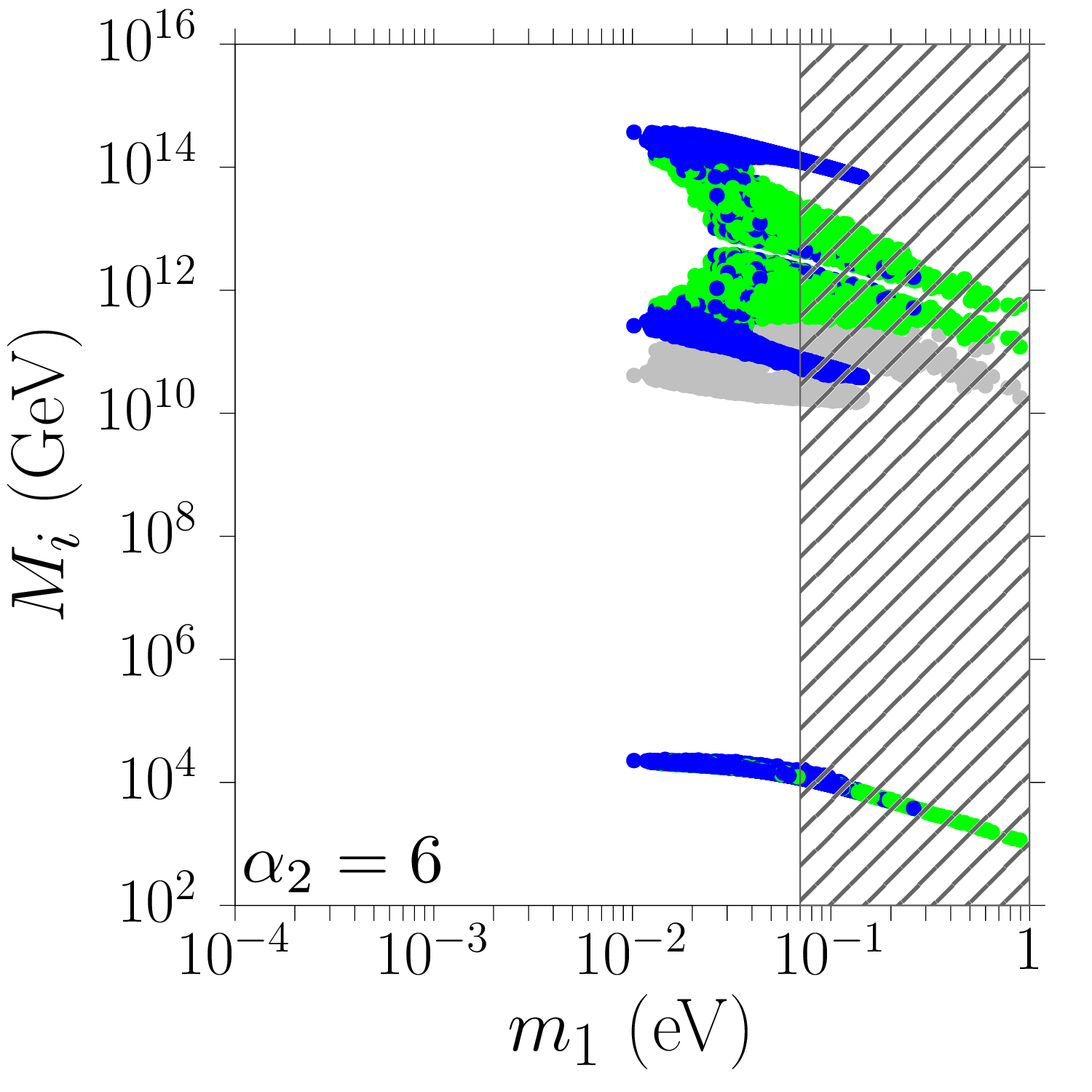,height=50mm,width=56mm}
\end{center}
\vspace{-8mm}
\caption{Scatter plots as in Fig.~2 but for  IO and $\tan\b=50$.}
\label{fldomIOhightan}
\end{figure}
The real difference is that now there is a large amount of solutions satisfying  
the strong thermal leptogenesis condition. The reason is that for large $\tan\b$
the fact that $K_{1\mu}$ tends not to be too large (see central bottom panel in Fig.~7), is not a problem, since the condition for
the wash-out of the pre-existing asymmetry now requires $K_{1\m}+K_{2\mu}\gg 1$
and it can be more easily satisfied even for low $K_{1\m}$ values. 
We can conclude that in all cases supersymmetry helps realising the strong thermal
condition. 

\section{Lower bound on $T_{\rm RH}$}

Thermal leptogenesis requires the initial temperature of the radiation dominated regime, $T_{\rm RH}$ within inflation,  to be sufficiently high for the RH neutrinos to be thermally produced before their interactions with the thermal bath, in particular their inverse decays, go out-of-equilibrium. 

For a specific solution this occurs at a temperature $T_{\rm lep}(K_{2\a}) \simeq M_2/z_B(K_{2\a})$,
where $\a$ is the flavour in equilibrium that dominates the asymmetry,
either $\tau$ or $\tau_2^{\bot}$ in the two fully flavoured regime, or
$\a=e,\m$ or $\t$ in the three fully flavoured regime.
\footnote{Of course there could be a fine tuned situation where the contributions from different
flavours are equivalent, in this case one should have $T_{\rm RH}$ above the maximum value out of the three
$T_{\rm RH}^{\rm min}(K_{2\a})$.}

At higher temperatures, in the strong wash-out regime ($K_{2\a}\gg 1$), the produced asymmetry is efficiently washed-out, while at lower temperatures, since the RH neutrino abundance is dropping exponentially, the produced asymmetry is negligible. In this way the asymmetry, in each flavour in equilibrium,  is produced within quite a well defined range of temperatures  between $M_2/[z_B(K_{2\a})-2]$ and
$M_2/[z_B(K_{2\a})+2]$ \cite{pedestrians}. 
Therefore, for a specific solution the reheat temperature has to be greater than 
$T_{\rm RH}^{\rm min}(K_{2\a})\simeq M_2/[z_B(K_{2\a})-2] $. 
In the weak wash-out regime one cannot identify such a sharp interval of temperatures and moreover
the process of production of the asymmetry depends on the initial $N_2$ abundance.
In this case one can say that $T_{\rm RH}\gtrsim M_2$ for the final asymmetry
to be equal to the asymptotic value at high temperatures. 
An expression that interpolates quite well $T^{\rm min}_{\rm RH}(K_{2\a})$ between the
strong and weak wash-out regime is given then by \cite{pedestrians} 
\be
T_{\rm RH}^{\rm min}(K_{2\a})\simeq {M_2\over z_B(K_{2\a})-2\,e^{-{3\over K_{2\a}}}}    \,   .
\ee 
This expression gives, for each solution with specified values of $K_{2\a}$ and $M_2$, the minimum 
$T_{\rm RH}$. The lower bound on $T_{\rm RH}$ can then be calculated minimising over all
the found solutions, i.e. $T_{\rm RH}^{\rm min} \equiv {\rm min}[T_{\rm RH}^{\rm min}(K_{2\a})]$.

In the non-supersymmetric case it was obtained
 $T_{\rm RH}^{\rm min}\simeq 1\times 10^{10}\,{\rm GeV}$ for $\a_2=1$  \cite{riotto2}, 
a lower bound that cannot by excluded by any experimental observation or theoretical
argument. However, this result could somehow suggest  that also
in the supersymmetric case one can expect a similar or even more stringent lower bound
because of the increased wash-out, leading in this case to a 
tension with the gravitino problem upper bound that, as discussed in the introduction, 
in a conservative way can be assumed to be $T_{\rm RH}\lesssim 10^{10}\,{\rm GeV}$
in order not to overproduce the gravitino abundance.
\footnote{The exact value depends on the neutralino mass and in particular is inversely proportional
to it, values as large as $T_{\rm RH}\simeq 2\times 10^{10}\,{\rm GeV}$ are acceptable \cite{gravitino}.}

This potential tension was confirmed by a dedicated analysis made in the supersymmetric case \cite{marfatia}. Here 
it was obtained (for $\a_2=5$) $T_{\rm RH}\gtrsim 10^{11}\,{\rm GeV}$, a result that would
suggest that $SO(10)$-inspired thermal leptogenesis is incompatible with 
the upper bound from the gravitino problem unless, as discussed in the introduction, one assumes 
very specific supersymmetric models.

We  plotted $T^{\rm min}_{\rm RH}(K_{2\a})$
for each point satisfying successful leptogenesis. The values of $T^{\rm min}_{\rm RH}(K_{2\a})$
 are shown with grey points in all plots where also the RH neutrino masses are plotted. 
These plots are shown in Figs.~ 2, 4, 6, 8 for NO low $\tan\b$, NO high $\tan \b$, IO low 
$\tan \b$ and IO high $\tan\b$ respectively for six specific integer values of  $\alpha_2$ (from 1 to 6) since one can expect a non trivial dependence on $\a_2$. This is because for decreasing $\alpha_2$ one has that $M_2$ decreases and this would go into the direction to lower $T_{\rm RH}$. 
On the other hand the final asymmetry decreases as $\propto \a_2^2$ so that there is also a lower bound on $\alpha_2$ coming from
successful leptogenesis.  
In these figures one can see indeed how the allowed range of values 
for $T_{\rm RH}^{\rm min}(K_{2\a})$ depends on $\a_2$. 

Finally, in Fig.~9 we summarised the results plotting the lower bound $T_{\rm RH}^{\rm min}$ as a function  of $\a_2$ indicating, with the same colour code
as in Figures 2, 4, 6 and 8, which flavour dominates the asymmetry for each value of $\a_2$. The results are shown both for
initial thermal $N_2$ abundance (thin lines) and for vanishing initial $N_2$ abundance (thick lines). 
The main difference is that in the second case there are no electron-dominated solutions since these all
have weak wash-out at the production ($K_{2e}\lesssim 1$) 
and the asymmetry is strongly suppressed in the case of initial $N_2$ vanishing abundance.
In the left (right) panels we show the results for low (high) values of $\tan\b$, in the
top (bottom) panels the results for NO (IO). In the case of low $\tan\b$ values (left panels)
one can see how the results do not actually differ that much from those in the  non-supersymmetric case \cite{riotto2}.
There is actually even a $\sim  \sqrt{2}$ relaxation due to the fact that the asymmetry 
increases by a factor $\sim 2$ because of the doubled $C\!P$ asymmetry and the efficiency factor
decreases of a factor $\sim \sqrt{2}$ (the efficiency factor is approximately inversely proportional to the decay parameters that increase of a factor  $\sqrt{2}$).

However, in the right panels, for large $\tan\b$ values, one can see how the red branch, corresponding to the
electron flavour dominated solutions now, for $\alpha_2 =1$--$2$, allows $T^{\rm min}_{\rm RH} \simeq (5$--$10) \times 10^9\,{\rm GeV}$, showing that it is possible to go even below $10^{10}\,{\rm GeV}$. 
\begin{figure}
\begin{center}
\psfig{file=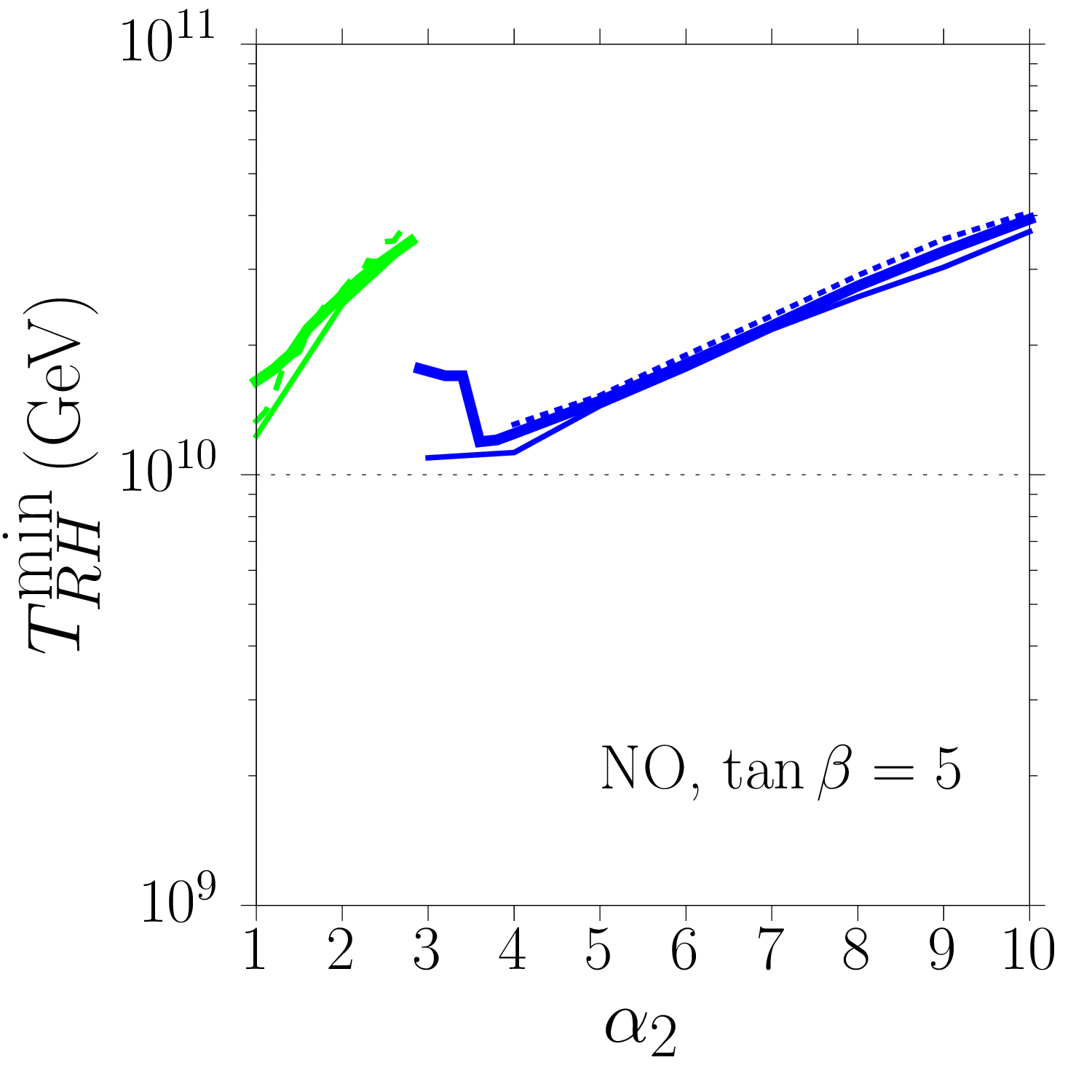,height=55mm,width=65mm}
\hspace{7mm}
\psfig{file=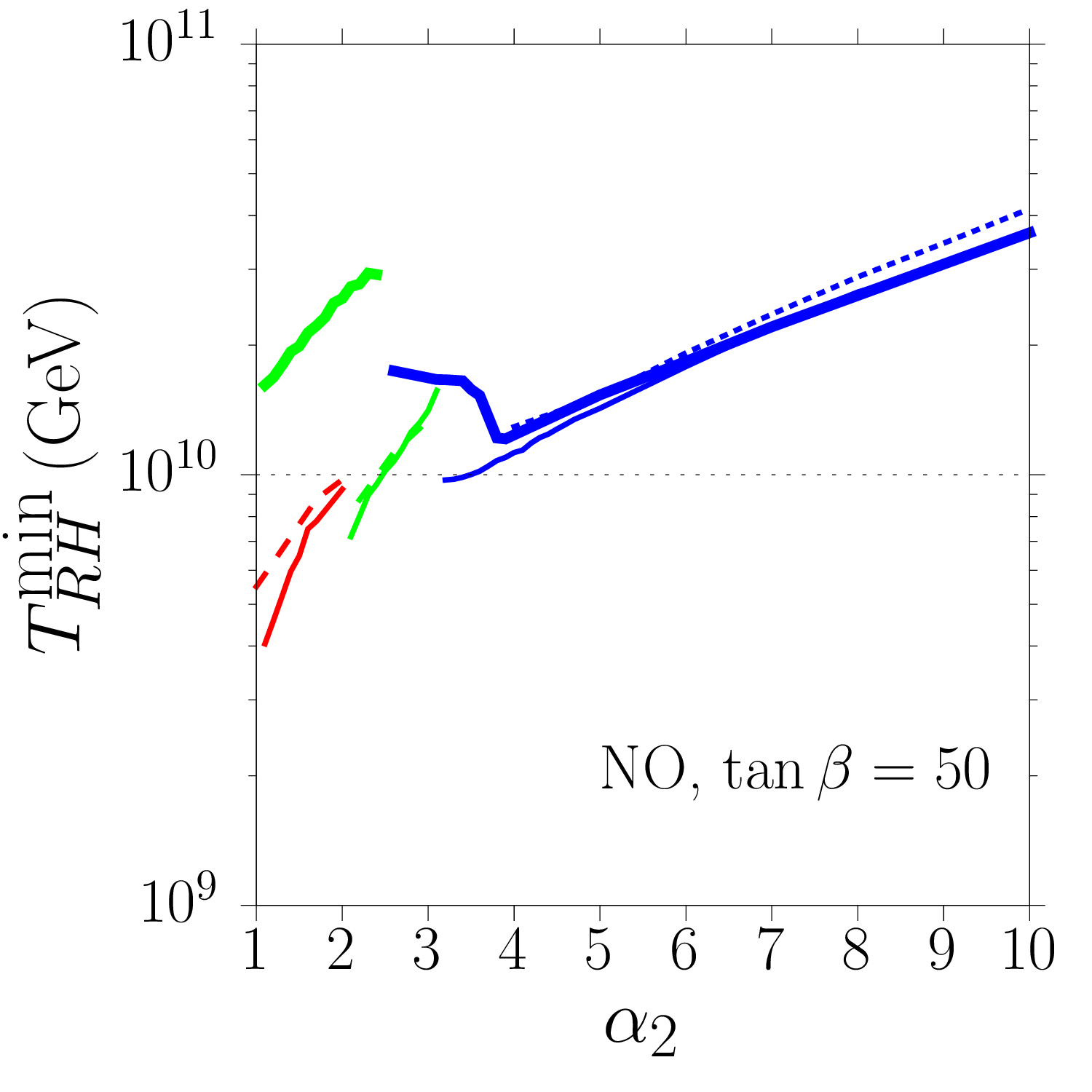,height=55mm,width=65mm}  \\
\vspace{3mm}
\psfig{file=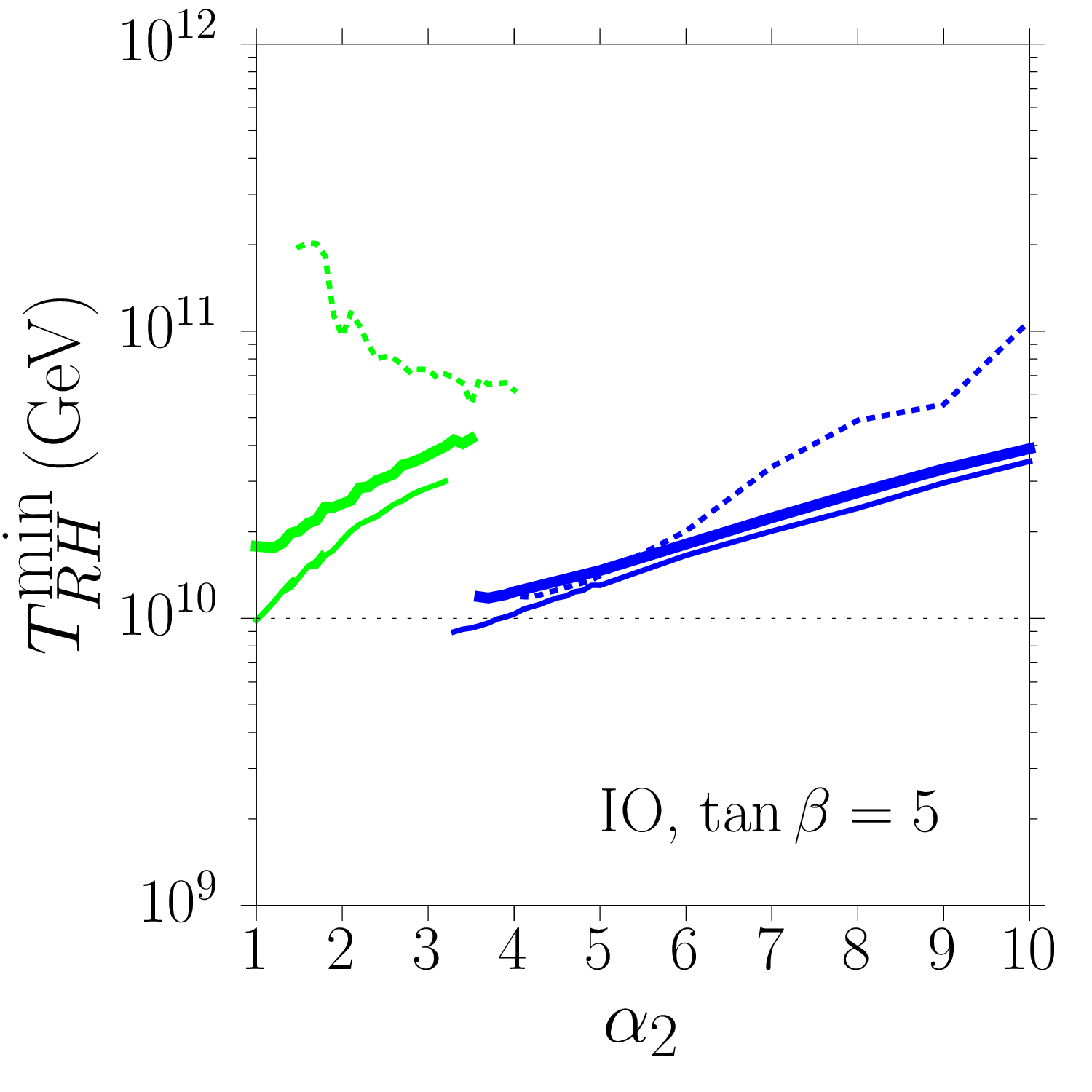,height=55mm,width=65mm}
\hspace{7mm}
\psfig{file=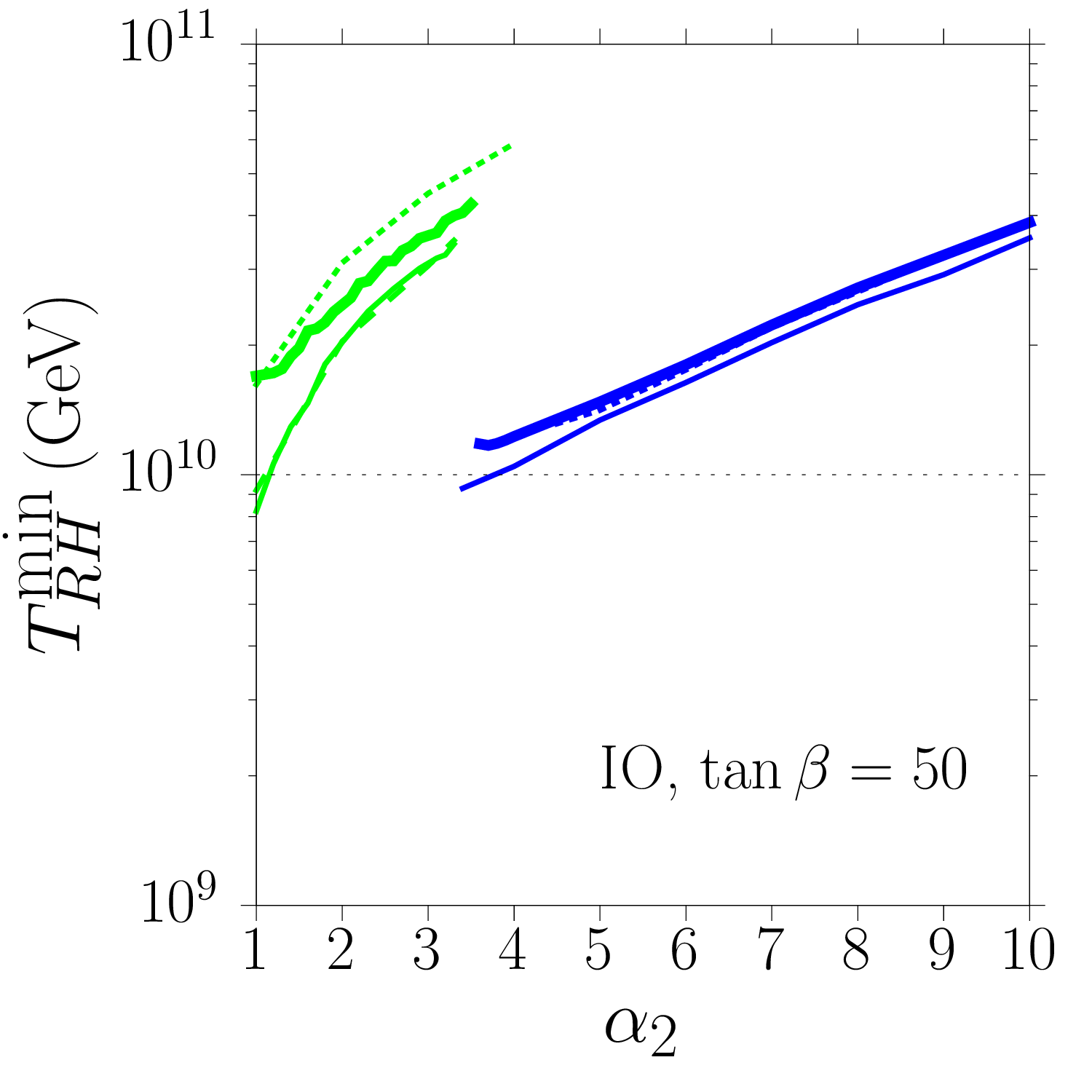,height=55mm,width=65mm}
\end{center}
\vspace{-5mm}
\caption{Lower bound on $T_{\rm RH}$ as a function of $\a_2$. The blue, green and red lines
correspond to an asymmetry tauon, muon and electron dominated respectively. The thin lines
are for initial thermal $N_2$ abundance.
The solid lines are for $I\leq V_L \leq V_{CKM}$, 
the dashed lines for $V_L =V_{CKM}$, the dotted lines for $V_L=I$.
The thick solid lines are for initial vanishing abundance and $I\leq V_L \leq V_{CKM}$.
The top (bottom) panels are for NO (IO). The left (right) panels are for $\tan\b =5\,(50)$. }
\label{TRHlb}
\end{figure}
Notice however that these electron-flavour dominated solutions have two drawbacks.
First they exist only for initial thermal $N_2$ abundance, a case that should 
justified within  models where for example the RH neutrinos are produced
by $Z'$ particles, heavier than the $N_2$'s, of a left-right symmetry after $SO(10)$ breaking \cite{plumacher2}. Moreover those solutions minimising  $T_{\rm RH}^{\rm min}$ below $10^{10}\,{\rm GeV}$ 
are characterised by large values of the squared modules of the orthogonal matrix entries (implying
strongly fine tuned  cancellations in the see-saw formula). 
This happens because the $N_2$ $C\!P$ asymmetries are not upper bounded and they are enhanced 
when $|\O_{ij}|^2 \gg 1$. In the $\a_2=2$ panel of Fig.~4 (bottom central panel) these fine tuned solutions
correspond to the dark red points. One can see how they also correspond to uplifted values of $M_1$
and reduced values of $M_2$ (they are indeed in the vicinity of crossing level solution).

For these reasons these  solutions should not be over emphasized, though they still represent a possibility
that should not be disregarded. On the other hand the muon and the tauon-flavour dominated
solutions, that give $T_{\rm RH}^{\rm min}\sim 10^{10}\,{\rm GeV}$, are not fine tuned and correspond to typical $SO(10)$-inspired solutions.  Moreover in the case of the tauon dominated solutions they are also
genuinely strong washed solutions independent of the initial $N_2$ abundance.

Notice that the lower bound on $T_{\rm RH}$ that we found  for $\a_2=5$, 
$T_{\rm RH}\gtrsim 1.5 \times 10^{10}\,{\rm GeV}$, is more than one order of magnitude
below the lower bound found in \cite{marfatia}. We cannot explain the origin of the discrepancy 
since details of the calculation of the final asymmetry (for example
how the matrix $U_R$ is calculated) are not specified in \cite{marfatia}. We can only 
report that in \cite{marfatia}  the tauon-dominated solutions that we find are absent and the result on 
the $T_{\rm RH}$ lower bound mainly relies on the electron dominated
solutions and therefore on the assumption of initial thermal $N_2$ abundance. 

Our result for the lower bound on the reheat temperature,  
$T^{\rm min}_{\rm RH}\gtrsim 1\times 10^{10}\,{\rm GeV}$, 
is approximately equal to the value that one needs in order to produce 
the Dark Matter gravitino abundance depending
on the value of the gluino masses.
 This coincidence is similar to what happens in the case of traditional $N_1$-dominated leptogenesis 
\cite{buchmuller} so that one could intriguingly relate matter-antimatter asymmetry production in thermal leptogenesis
to  gravitino Dark Matter production. On the other hand the recent LHC results on the lower bound of gluino masses \cite{diphoton}
make the upper bound on $T_{\rm RH}$ more stringent, at the level of $T_{\rm RH} \lesssim 5\times 10^9\,{\rm GeV}$ within
the pMSSM \cite{covi} and this 
seems to corner this intriguing scenario of thermal leptogenesis combined with gravitino Dark Matter . 
However, as already mentioned, for large values of the gravitino mass ($\gtrsim 30\,{\rm TeV}$)
the large $T_{\rm RH}$ required by SUSY $SO(10)$-inspired thermal leptogenesis ($T_{\rm RH}\gtrsim 10^{10}\,{\rm GeV}$) can be 
reconciled with the gravitino problem. Of course within
 specific realistic models one should verify whether the  lower bound
$T^{\rm min}_{\rm RH}\sim 10^{10}\,{\rm GeV}$ can be indeed saturated. 

There is, however, still another possibility, never considered so far, 
that can allow a relaxation of $T^{\rm min}_{\rm RH}$
even below $10^{10}\,{\rm GeV}$ for usual tauon-flavour solutions. 

\section{A new scenario of $N_2$-dominated leptogenesis}

It is usually assumed that the lightest RH neutrino mass $M_1 \gtrsim T_{\rm sph}^{\rm out} \simeq 100\,{\rm GeV}$, 
where $T_{\rm sph}^{\rm out}$ is the sphaleron freeze-out temperature \cite{kuzmin}. In this case
the lightest RH neutrino wash-out has to be taken into account.
However, if $M_1$ is below such a temperature, then the lightest RH neutrino wash-out acts only on the lepton
asymmetry but not on the frozen baryon asymmetry  produced earlier by $N_2$ out-of-equilibrium decays. 
\footnote{More precisely the $N_1$ wash-out acts in an interval of temperatures $T=[{M_1/ z_{\rm in}},{M_{1}/ z_{\rm out}}]$ with $z_{\rm in}\simeq 2/\sqrt{K_{1\a}}$ \cite{pedestrians}. Therefore, more precisely one has to impose $M_1 \lesssim z_{\rm in}\,T_{\rm sph}^{\rm out}$.}
In this case the final asymmetry is given by
the expressions eqs.~(\ref{unfl}), (\ref{twofl}), (\ref{threefl})  without
the exponentials encoding the lightest RH neutrino wash-out since this is negligible. 

We have then repeated the calculation of $T^{\rm min}_{\rm RH}$ in this scenario and
the results are shown in the four panels of Fig.~10 that correspond to the same cases of the 
panels in Fig.~9.
\begin{figure}
\begin{center}
\psfig{file=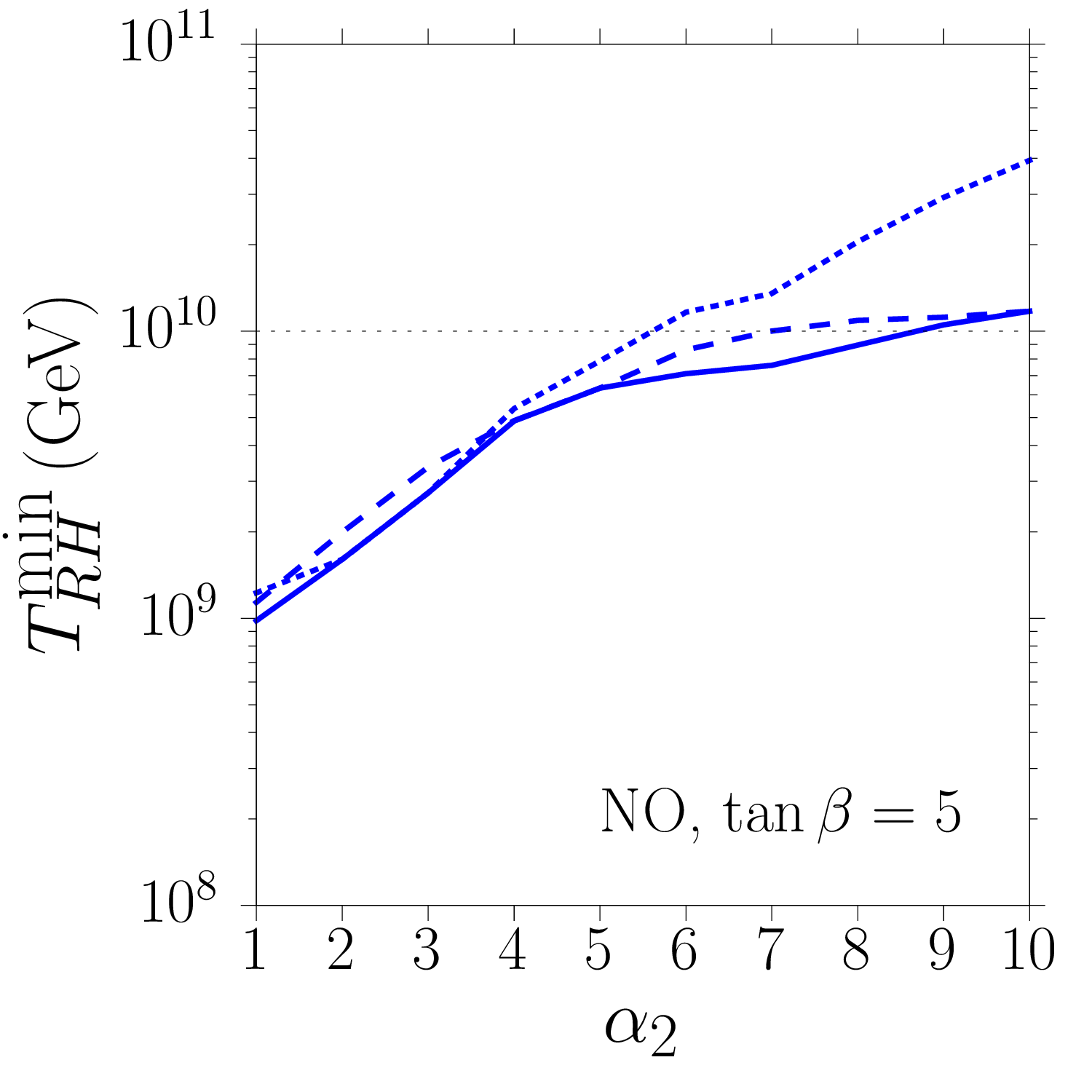,height=55mm,width=65mm}
\hspace{7mm}
\psfig{file=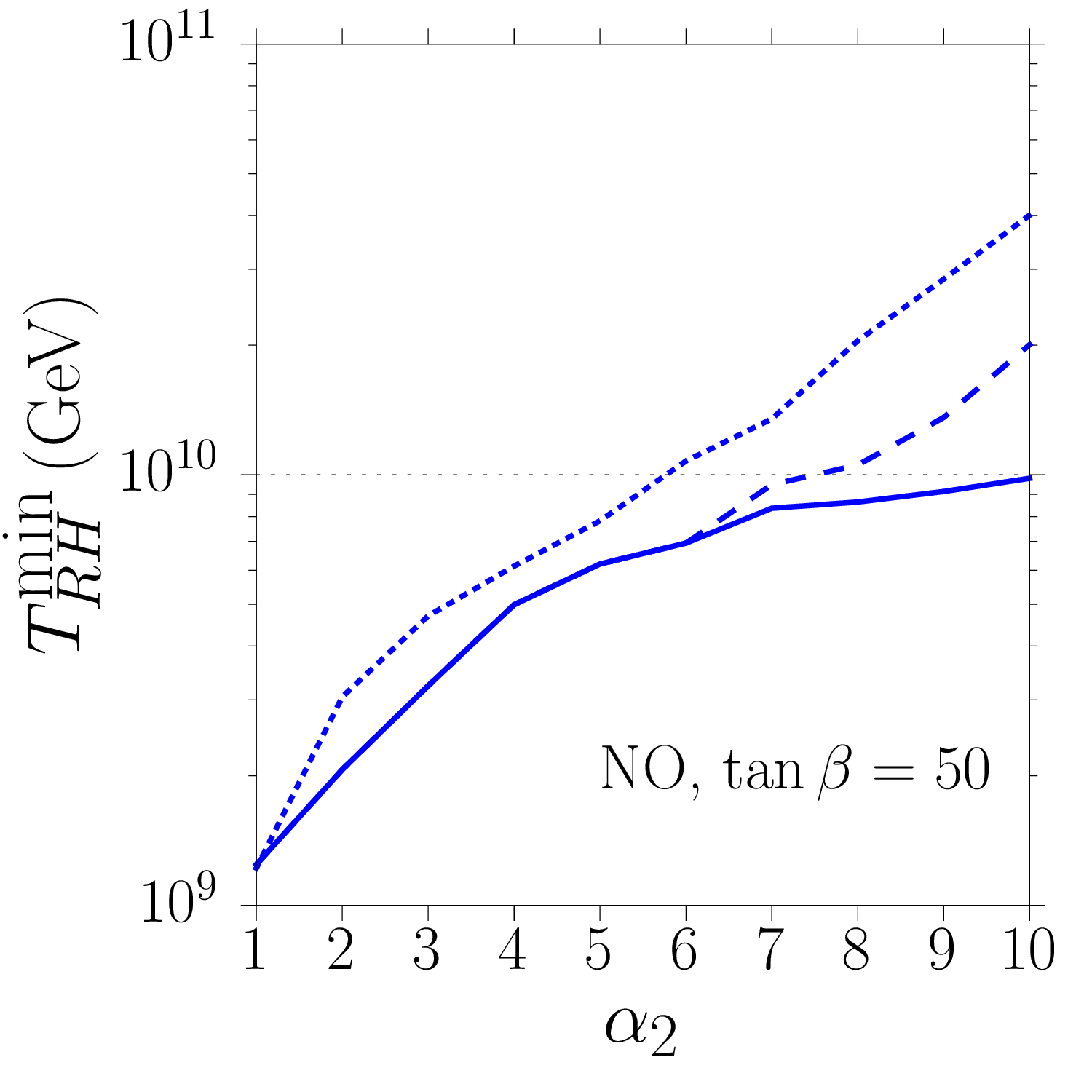,height=55mm,width=65mm} \\
\vspace{3mm}
\psfig{file=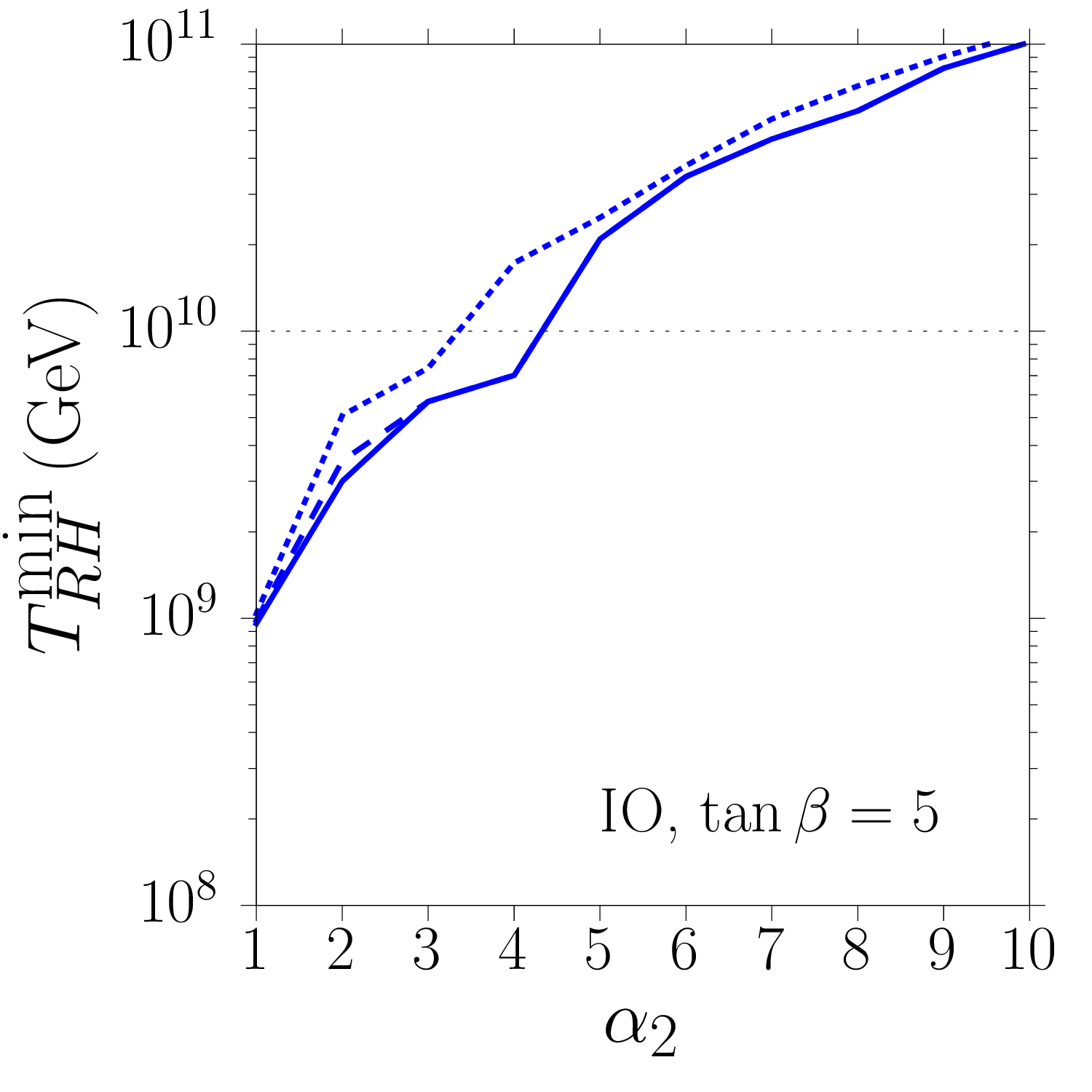,height=55mm,width=65mm}
\hspace{7mm}
\psfig{file=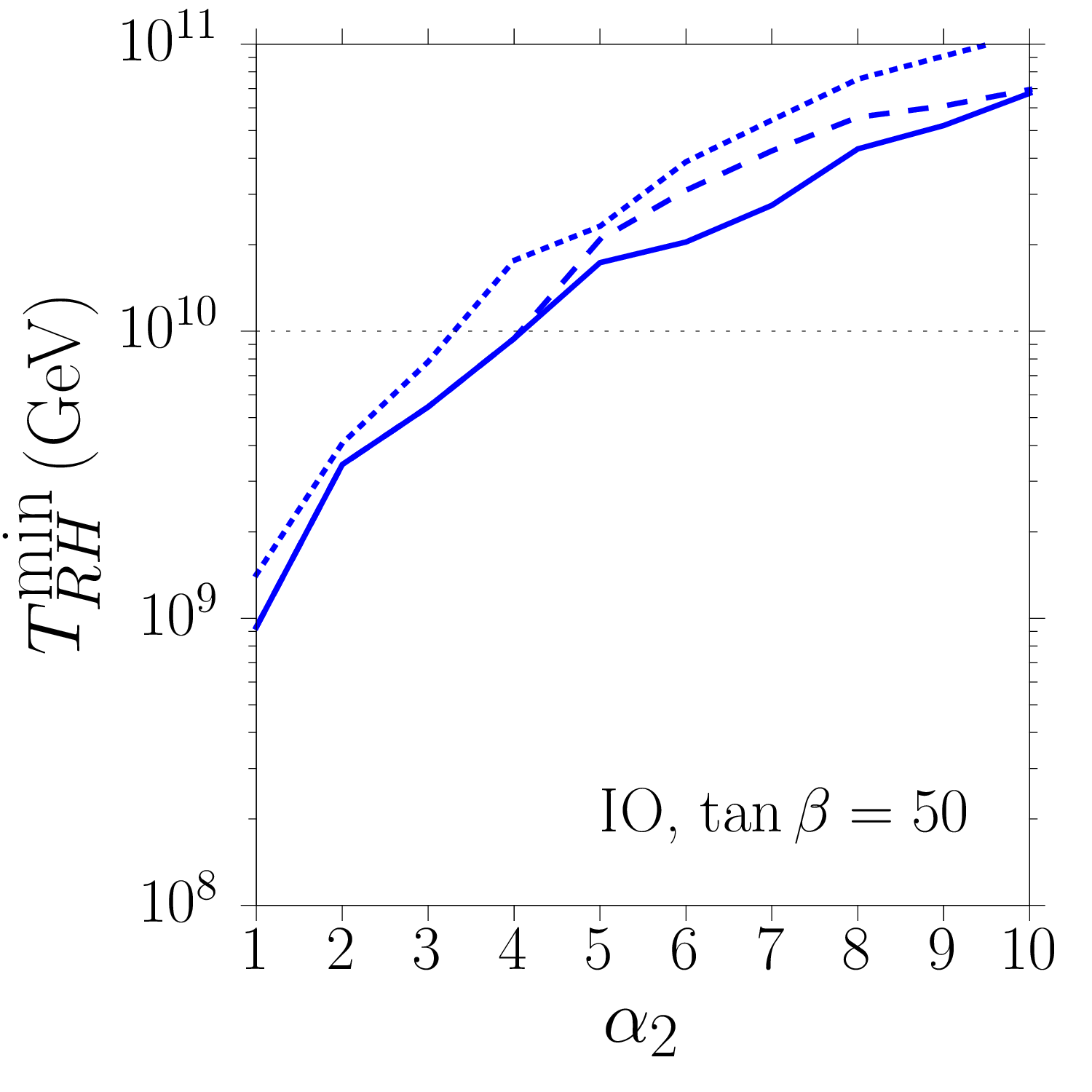,height=55mm,width=65mm}
\end{center}
\vspace{-5mm}
\caption{Lower bound on $T_{\rm RH}$ as a function of $\a_2$ in a scenario
where $M_1 \lesssim T_{\rm sph}^{\rm out} \simeq 100\,{\rm GeV}$. The left (right) panels
are for $\tan\b =5 (50)$, the top (bottom) panels are for NO (IO). Same line conventions
as in Fig.~9 with the  difference that this time there is no distinction between thin and solid lines 
since there is no dependence on the initial $N_2$ abundance.}
\label{TRHlbNS}
\end{figure}
This time the minimum is always realised by  tauon dominated solutions with strong wash-out at the production
where the final asymmetry is independent of the initial $N_2$ abundance.
It can be seen how values of $T_{\rm RH}$ as low as $10^9\,{\rm GeV}$
are possible. In this case the gravitino overabundance problem can be circumvented
for a wider range of gravitino masses compared to the traditional scenario discussed 
in the previous sections.

From the expression eq.~(\ref{Mi}) for $M_1$ one can see how this scenario requires
values $\a_1 \lesssim 0.1$ (for hierarchical neutrinos, if $m_1\gtrsim 10\,{\rm meV}$ one can have
higher values). 
This would also imply somehow that also $m_{D3} \ll T_{\rm sph}^{\rm out}\sim 100\,{\rm GeV}$ in 
order for the seesaw formula to be valid implying $\a_3 \ll 1$. 
 One can wonder whether this can be  achieved in some realistic models. Interestingly in a recent
 study of realistic $SO(10)$ models \cite{rodejohann} one of the found best fit cases, a supersymmetric
 model with $10_H, 120_H, \bar{126}_H$ Higgs representations, is realised for $M_1 \simeq 1\,{\rm TeV}$ corresponding to $\a_1 \simeq 0.3$.  
 Since this case also has a very small $\chi^2_{\rm min}\simeq 0.6$, one can wonder whether
 with some deviation from the best fit one could get $M_1\lesssim T_{\rm RH}^{\rm sph}$ with still
 an acceptable value of $\chi^2_{\rm min}$. In any case this specific example seems to suggest that 
 this scenario might be  indeed realised within some realistic model.  
 Notice that within this scenario we are not showing the low energy neutrino constraints since these simply evaporate. Indeed these constraints exist mainly because of the presence of the lightest RH neutrino wash-out,
 as stressed in previous papers \cite{riotto1,riotto2,SO10decription}.
 It should also be 
 made clear that though we are presenting this scenario in a supersymmetric framework, where it nicely allows $T_{RH}$ values below $10^{10}\,{\rm GeV}$, it might be also realised and find applications 
 within a non-supersymmetric framework. 
 
 \section{Conclusions}

We extended the study of $SO(10)$-inspired leptogenesis, previously discussed in a non-supersymmetric framework, to the supersymmetric case calculating the constraints on the low energy neutrino parameters
and the lower bound on $T_{\rm RH}$ that has a particular 
importance because of the tension with the upper bound from the gravitino problem. Our results show that,
in the usual case, where the lightest RH neutrino mass is heavier than the sphaleron freeze-out temperature
and $N_1$ wash-out is present,  values of $T_{\rm RH}$ as low as 
$T_{\rm RH}\simeq 1 \times 10^{10}\,{\rm GeV}$ are possible 
without any fine-tuning and for a final asymmetry independent of the initial $N_2$-abundance.
We have then proposed a novel scenario where $M_1$ is below the sphaleron freeze-out temperature so that the  $N_1$ wash-out is absent.
In this case without any fine-tuning reheat temperature values as low as 
$T_{\rm RH} \simeq 1 \times 10^{9}\,{\rm GeV}$ are allowed.
In our calculation  the main neglected effects  that could produce 
some significant modifications are flavour coupling effects \cite{flcoupling2} arising 
from a redistribution of the asymmetry among quarks, right handed charged leptons and above all Higgs (and of course also among the supersymmetric particles \cite{fong}). This could open new way to circumvent the $N_1$ wash-out but in any case it should be clear that an account of these effect can at most relax the reheat temperature in the scenario
where $M_1 \gtrsim T_{\rm sph}^{\rm out}$ to the minimum  value, $T_{\rm RH}^{\rm min}\simeq 1\times 10^9\,{\rm GeV}$ found in the case where  $M_1 \lesssim T_{\rm sph}^{\rm out}$.
We have also described the transition between different fully flavoured regimes 
for a changing value of $M_2$ with a step approximation, while
a full description would require solution of density matrix equations 
\cite{bcst,flavour2,densitymatrix,densitymatrix2}.
Another important effect that we neglected and that might be important in the supersymmetric case for large 
$\tan\b$ is the running of low energy neutrino parameters that might modify the constraints in this case \cite{running}. However, this effect would not 
change our main results on the lower bound on $T_{\rm RH}$. 
In conclusion, we have shown the existence of a window for the viability of thermal $SO(10)$-inspired leptogenesis in the supersymmetric case 
without gravitino overabundance problem. Independently whether supersymmetry is found at the LHC, 
these results are interesting in connection with the current debate of identifying a realistic grand-unified model able also to realise successful  leptogenesis, since supersymmetric extensions might more easily provide good fits of the parameters even if supersymmetry breaking occurs above the 
scale testable at colliders. 
With more  experimental information on the neutrino mixing parameters coming in a close future, a particular successful model (or class of models) might emerge with interesting further phenomenological predictions (e.g proton life-time) and even more specific links between leptonic and quark sector and possibly with the identification of the DM candidate and new predictions at colliders. In this exciting search, leptogenesis might play a primary role, increasing the predictive power of the model and solving the matter-antimatter asymmetry cosmological puzzle.

\vspace{-1mm}
\subsection*{Acknowledgments}

We thank Borut Bajc, Steve F. King, Luca Marzola and Rabi Mohapatra for useful comments and discussions. 
PDB acknowledges financial support  from the NExT/SEPnet Institute, 
from the STFC Consolidated Grant ST/J000396/1 and  the  
EU FP7  ITN INVISIBLES  (Marie Curie Actions, PITN- GA-2011- 289442).
MRF acknowledges financial support from the STAG Institute.


\begin{thebibliography}{99}
\bibitem{seesaw}
  P.~Minkowski,
  %``Mu $\to$ E Gamma At A Rate Of One Out Of 1-Billion Muon Decays?,''
  Phys.\ Lett.\  B {\bf 67} (1977) 421;
  %%CITATION = PHLTA,B67,421;%%
T. Yanagida, in Proceedings of the Workshop on Unified Theory and Baryon Number
of the Universe, eds. O. Sawada and A. Sugamoto (KEK, 1979) p.95;
%\cite{Ramond:1979py}
%\bibitem{Ramond:1979py}
  P.~Ramond, 
Invited talk given at Conference: C79-02-25
(Feb 1979) p.265-280, CALT-68-709,
  %``The Family Group in Grand Unified Theories,''
  hep-ph/9809459;
  %%CITATION = HEP-PH/9809459;%%
  %135 citations counted in INSPIRE as of 24 Apr 2013
 M. Gell-Mann,
P. Ramond and R. Slansky, in Supergravity, eds. P. van Niewwenhuizen and D.
Freedman (North Holland, Amsterdam, 1979) Conf.Proc. C790927 p.315, PRINT-80-0576;
%%CITATION = CONFP,C790927,315
R.~Barbieri, D.~V.~Nanopoulos, G.~Morchio and F.~Strocchi,
Phys.\ Lett.\ B {\bf 90} (1980) 91;
R.~N.~Mohapatra and G.~Senjanovic,
Phys.\ Rev.\ Lett.\  {\bf 44} (1980) 912.
 
\bibitem{diphoton}
M. Kado, ATLAS 13 TeV Results, CERN Jamboree (December 15 2015, 
https://twiki.cern.ch/twiki/bin/view/ AtlasPublic/December2015-13TeV); 
J. Olsen, CMS 13 TeV Results, CERN Jamboree (December 15 2015, 
http://cms-results.web.cern.ch/cms- results/public-results/preliminary-results/LHC- Jamboree-2015/index.html).
 
\bibitem{pietroni} 
M.~Pietroni,
  %``The Electroweak phase transition in a nonminimal supersymmetric model,''
  Nucl.\ Phys.\ B {\bf 402} (1993) 27
%  doi:10.1016/0550-3213(93)90635-3
  [hep-ph/9207227].
  %%CITATION = doi:10.1016/0550-3213(93)90635-3;%%
  %131 citations counted in INSPIRE as of 21 Dec 2015
 
\bibitem{fy}  
M.~Fukugita and T.~Yanagida,
  %``Baryogenesis Without Grand Unification,''
  Phys.\ Lett.\ B {\bf 174}, 45 (1986).

\bibitem{SO10inspired}
A.~Y.~Smirnov,
  %``Seesaw enhancement of lepton mixing,''
  Phys.\ Rev.\ D {\bf 48} (1993) 3264
  [hep-ph/9304205];
  %%CITATION = HEP-PH/9304205;%%
  %163 citations counted in INSPIRE as of 26 Aug 2015
W.~Buchmuller and M.~Plumacher,
  %``Baryon asymmetry and neutrino mixing,''
  Phys.\ Lett.\ B {\bf 389} (1996) 73 [hep-ph/9608308];
  %%CITATION = HEP-PH/9608308;%%
E.~Nezri and J.~Orloff,
  %``Neutrino oscillations versus leptogenesis in SO(10) models,''
  JHEP {\bf 0304} (2003) 020
  [hep-ph/0004227];
  %%CITATION = HEP-PH/0004227;%%
F.~Buccella, D.~Falcone and F.~Tramontano,
  %``Baryogenesis via leptogenesis in SO(10) models,''
  Phys.\ Lett.\ B {\bf 524} (2002) 241 [hep-ph/0108172];
   %%CITATION = HEP-PH/0108172;%%
G.~C.~Branco, R.~Gonzalez Felipe, F.~R.~Joaquim and M.~N.~Rebelo,
  %``Leptogenesis, CP violation and neutrino data: What can we learn?,''
  Nucl.\ Phys.\ B {\bf 640} (2002) 202 [hep-ph/0202030].
  %%CITATION = HEP-PH/0202030;%% 
 
 \bibitem{afs}
E.~K.~Akhmedov, M.~Frigerio and A.~Y.~Smirnov, JHEP {\bf 0309}, 021 (2003).

 
 \bibitem{SO10}
 H.~Georgi,
  %``The State of the Art?Gauge Theories,''
  AIP Conf.\ Proc.\  {\bf 23} (1975) 575;
  %%CITATION = APCPC,23,575;%%
  %116 citations counted in INSPIRE as of 05 août 2015
 H.~Fritzsch and P.~Minkowski,
  %``Unified Interactions of Leptons and Hadrons,''
  Annals Phys.\  {\bf 93} (1975) 193.
  %%CITATION = APNYA,93,193;%%
  %1391 citations counted in INSPIRE as of 05 août 2015
 
\bibitem{A2Z}
  S.~F.~King,
  %``A to Z of Flavour with Pati-Salam,''
  JHEP {\bf 1408} (2014) 130
  [arXiv:1406.7005 [hep-ph]].
  %%CITATION = ARXIV:1406.7005;%%
  %12 citations counted in INSPIRE as of 24 Jul 2015
  
 \bibitem{ferruccio}
  F.~Feruglio, K.~M.~Patel and D.~Vicino,
  %``A realistic pattern of fermion masses from a five-dimensional SO(10) model,''
  JHEP {\bf 1509} (2015) 040
%  doi:10.1007/JHEP09(2015)040
  [arXiv:1507.00669 [hep-ph]].
  %%CITATION = doi:10.1007/JHEP09(2015)040;%%
  %3 citations counted in INSPIRE as of 06 Jan 2016
  
 \bibitem{geometry}
P.~Di Bari,
  %``Seesaw geometry and leptogenesis,''
  Nucl.\ Phys.\ B {\bf 727} (2005) 318
  [hep-ph/0502082].
  %%CITATION = HEP-PH/0502082;%%
  %87 citations counted in INSPIRE as of 27 Jul 2015

\bibitem{bcst}
R.~Barbieri, P.~Creminelli, A.~Strumia and N.~Tetradis,
  %``Baryogenesis through leptogenesis,''
  Nucl.\ Phys.\ B {\bf 575} (2000) 61
  [hep-ph/9911315].

\bibitem{flavour1} 
 E.~Nardi, Y.~Nir, E.~Roulet and J.~Racker,
  %``The Importance of flavor in leptogenesis,''
  JHEP {\bf 0601} (2006) 164
  [hep-ph/0601084].
  %%CITATION = HEP-PH/0601084;%%
  %289 citations counted in INSPIRE as of 27 Jul 2015

\bibitem{flavour2}  
 A.~Abada, S.~Davidson, A.~Ibarra, F.-X.~Josse-Michaux, M.~Losada and A.~Riotto,
  %``Flavour Matters in Leptogenesis,''
  JHEP {\bf 0609} (2006) 010
  [hep-ph/0605281].
  
 \bibitem{vives}
O.~Vives,
  %``Flavor dependence of CP asymmetries and thermal leptogenesis with strong right-handed neutrino mass hierarchy,''
  Phys.\ Rev.\ D {\bf 73} (2006) 073006
  [hep-ph/0512160].
  %%CITATION = HEP-PH/0512160;%%
  %119 citations counted in INSPIRE as of 27 Jul 2015

\bibitem{bounds}
S.~Blanchet and P.~Di Bari,
  %``New aspects of leptogenesis bounds,''
  Nucl.\ Phys.\ B {\bf 807} (2009) 155
  [arXiv:0807.0743 [hep-ph]].
  %%CITATION = ARXIV:0807.0743;%%
  %35 citations counted in INSPIRE as of 27 Jul 2015

\bibitem{riotto1}
P.~Di Bari and A.~Riotto,
  %``Successful type I Leptogenesis with SO(10)-inspired mass relations,''
  Phys.\ Lett.\ B {\bf 671} (2009) 462
  [arXiv:0809.2285 [hep-ph]];
  %%CITATION = ARXIV:0809.2285;%%
  %30 citations counted in INSPIRE as of 27 juil. 2015
 X.~G.~He, S.~S.~C.~Law and R.~R.~Volkas,
  %``Determining the heavy seesaw neutrino mass matrix from low-energy parameters,''
  Phys.\ Rev.\ D {\bf 78} (2008) 113001
  [arXiv:0810.1104 [hep-ph]].
 
 
 \bibitem{riotto2}
 P.~Di Bari and A.~Riotto,
  %``Testing SO(10)-inspired leptogenesis with low energy neutrino experiments,''
  JCAP {\bf 1104} (2011) 037
  [arXiv:1012.2343 [hep-ph]].
  %%CITATION = ARXIV:1012.2343;%%
  %15 citations counted in INSPIRE as of 27 juil. 2015

\bibitem{SO10decription}
P.~Di Bari, L.~Marzola and M.~Re Fiorentin,
  %``Decrypting $SO(10)$-inspired leptogenesis,''
  Nucl.\ Phys.\ B {\bf 893} (2015) 122 
  [arXiv:1411.5478 [hep-ph]].

\bibitem{upperbound}
W.~Buchmuller, P.~Di Bari and M.~Plumacher,
  %``A Bound on neutrino masses from baryogenesis,''
  Phys.\ Lett.\ B {\bf 547} (2002) 128
  [hep-ph/0209301];
  %%CITATION = HEP-PH/0209301;%%
  %114 citations counted in INSPIRE as of 09 Aug 2015
 W.~Buchmuller, P.~Di Bari and M.~Plumacher,
  %``The Neutrino mass window for baryogenesis,''
  Nucl.\ Phys.\ B {\bf 665} (2003) 445
  [hep-ph/0302092];
  %%CITATION = HEP-PH/0302092;%%
  %258 citations counted in INSPIRE as of 09 Aug 2015
G.~F.~Giudice, A.~Notari, M.~Raidal, A.~Riotto and A.~Strumia,
  %``Towards a complete theory of thermal leptogenesis in the SM and MSSM,''
  Nucl.\ Phys.\ B {\bf 685} (2004) 89
  [hep-ph/0310123];
  %%CITATION = HEP-PH/0310123;%%
  %537 citations counted in INSPIRE as of 09 Aug 2015
 A.~De Simone and A.~Riotto,
  %``On the impact of flavour oscillations in leptogenesis,''
  JCAP {\bf 0702} (2007) 005
  [hep-ph/0611357].
  %%CITATION = HEP-PH/0611357;%%
  %57 citations counted in INSPIRE as of 09 Aug 2015

\bibitem{pedestrians}
W.~Buchmuller, P.~Di Bari and M.~Plumacher,
  %``Leptogenesis for pedestrians,''
  Annals Phys.\  {\bf 315} (2005) 305
  [hep-ph/0401240];
  %%CITATION = HEP-PH/0401240;%%
  %515 citations counted in INSPIRE as of 09 Aug 2015 


\bibitem{planck}
  P.~A.~R.~Ade {\it et al.} [Planck Collaboration],
  %``Planck 2015 results. XIII. Cosmological parameters,''
  arXiv:1502.01589 [astro-ph.CO].
  %%CITATION = ARXIV:1502.01589;%%
  %519 citations counted in INSPIRE as of 09 Aug 2015
  
\bibitem{strongSO10}
P.~Di Bari and L.~Marzola,
  %``SO(10)-inspired solution to the problem of the initial conditions in leptogenesis,''
  Nucl.\ Phys.\ B {\bf 877} (2013) 719
  [arXiv:1308.1107 [hep-ph]].
  %%CITATION = ARXIV:1308.1107;%%
  %8 citations counted in INSPIRE as of 09 Aug 2015

\bibitem{global}
M.~C.~Gonzalez-Garcia, M.~Maltoni and T.~Schwetz,
  %``Updated fit to three neutrino mixing: status of leptonic CP violation,''
  JHEP {\bf 1411} (2014) 052 [arXiv:1409.5439 [hep-ph]];
F.~Capozzi, G.~L.~Fogli, E.~Lisi, A.~Marrone, D.~Montanino and A.~Palazzo,
  %``Status of three-neutrino oscillation parameters, circa 2013,''
  Phys.\ Rev.\ D {\bf 89} (2014) 093018 [arXiv:1312.2878 [hep-ph]];
  %%CITATION = ARXIV:1312.2878;%%
  %187 citations counted in INSPIRE as of 25 May 2015
D.~V.~Forero, M.~Tortola and J.~W.~F.~Valle,
  %``Neutrino oscillations refitted,''
  Phys.\ Rev.\ D {\bf 90} (2014) 9,  093006
  [arXiv:1405.7540 [hep-ph]].

\bibitem{spectators} 
 W.~Buchmuller and M.~Plumacher,
  %``Spectator processes and baryogenesis,''
  Phys.\ Lett.\ B {\bf 511} (2001) 74
  [hep-ph/0104189].
  %%CITATION = HEP-PH/0104189;%%
  %77 citations counted in INSPIRE as of 27 Sep 2015

\bibitem{flcoupling1}
   %%CITATION = HEP-PH/9911315;%%
  %323 citations counted in INSPIRE as of 27 Jul 2015
  W.~Buchmuller and M.~Plumacher,
  %``Spectator processes and baryogenesis,''
  Phys.\ Lett.\ B {\bf 511} (2001) 74
  [hep-ph/0104189];
  %%CITATION = HEP-PH/0104189;%%
  %77 citations counted in INSPIRE as of 27 Jul 2015   E.~Nardi, Y.~Nir, J.~Racker and E.~Roulet,
 S.~Blanchet and P.~Di Bari,
  %``New aspects of leptogenesis bounds,''
  Nucl.\ Phys.\ B {\bf 807} (2009) 155
  [arXiv:0807.0743 [hep-ph]];
  %%CITATION = ARXIV:0807.0743;%%
  %35 citations counted in INSPIRE as of 27 juil. 2015
 F.~X.~Josse-Michaux and A.~Abada,
  %``Study of flavour dependencies in leptogenesis,''
  JCAP {\bf 0710} (2007) 009 [hep-ph/0703084].
  %%CITATION = HEP-PH/0703084;%%
  %32 citations counted in INSPIRE as of 27 juil. 2015

\bibitem{flcoupling2}
  S.~Antusch, P.~Di Bari, D.~A.~Jones and S.~F.~King,
  %``A fuller flavour treatment of N_2-dominated leptogenesis,''
  Nucl.\ Phys.\ B {\bf 856} (2012) 180
  [arXiv:1003.5132 [hep-ph]].
  %%CITATION = ARXIV:1003.5132;%%
  %26 citations counted in INSPIRE as of 19 juin 2015

\bibitem{flcoupling3}
P.~Di Bari and S.~F.~King,
  %``Successful $N_2$ leptogenesis with flavour coupling effects in realistic unified models,''
  JCAP {\bf 1510} (2015) 10,  008
%  doi:10.1088/1475-7516/2015/10/008
  [arXiv:1507.06431 [hep-ph]].
  %%CITATION = doi:10.1088/1475-7516/2015/10/008;%%
  %3 citations counted in INSPIRE as of 05 Jan 2016
\bibitem{rodejohann}
A.~Dueck and W.~Rodejohann,
  %``Fits to SO(10) Grand Unified Models,''
  JHEP {\bf 1309} (2013) 024
  [arXiv:1306.4468 [hep-ph]].
  %%CITATION = ARXIV:1306.4468;%%
  %14 citations counted in INSPIRE as of 20 juil. 2015 

\bibitem{compact}
 F.~Buccella, D.~Falcone, C.~S.~Fong, E.~Nardi and G.~Ricciardi,
  %``Squeezing out predictions with leptogenesis from SO(10),''
  Phys.\ Rev.\ D {\bf 86} (2012) 035012
  [arXiv:1203.0829 [hep-ph]];
 G.~Altarelli and D.~Meloni,
  %``A non supersymmetric SO(10) grand unified model for all the physics below $M_{GUT}$,''
  JHEP {\bf 1308} (2013) 021
  [arXiv:1305.1001 [hep-ph]]; 
 C.~S.~Fong, D.~Meloni, A.~Meroni and E.~Nardi,
  %``Leptogenesis in SO(10),''
  JHEP {\bf 1501} (2015) 111
  [arXiv:1412.4776 [hep-ph]];   
  A.~Addazi, M.~Bianchi and G.~Ricciardi,
  %``Exotic see-saw mechanism for neutrini and leptogenesis in a Pati-Salam model,''
  arXiv:1510.00243 [hep-ph].
            
\bibitem{SUSYSO10}
K.~S.~Babu and C.~Macesanu,
  %``Neutrino masses and mixings in a minimal SO(10) model,''
  Phys.\ Rev.\ D {\bf 72} (2005) 115003
  [hep-ph/0505200];
  %%CITATION = HEP-PH/0505200;%%
  %110 citations counted in INSPIRE as of 20 août 2015
A.~S.~Joshipura and K.~M.~Patel,
  %``Fermion Masses in SO(10) Models,''
  Phys.\ Rev.\ D {\bf 83} (2011) 095002
  [arXiv:1102.5148 [hep-ph]].
  %%CITATION = ARXIV:1102.5148;%%
  %42 citations counted in INSPIRE as of 20 Aug 2015

\bibitem{gravitino}
M.~Y.~Khlopov and A.~D.~Linde, Phys.\ Lett.\ B {\bf 138} (1984) 265;
J.~R.~Ellis, J.~E.~Kim and D.~V.~Nanopoulos, Phys.\ Lett.\ B {\bf 145} (1984) 181;
K.~Kohri, T.~Moroi and A.~Yotsuyanagi, Phys.\ Rev.\ D {\bf 73} (2006) 123511;
 M.~Kawasaki, K.~Kohri, T.~Moroi and A.~Yotsuyanagi,
  %``Big-Bang Nucleosynthesis and Gravitino,''
  Phys.\ Rev.\ D {\bf 78} (2008) 065011
  [arXiv:0804.3745 [hep-ph]].
  %%CITATION = ARXIV:0804.3745;%%
  %298 citations counted in INSPIRE as of 27 Aug 2015

\bibitem{yanagida}
M.~Fujii and T.~Yanagida,
  %``Natural gravitino dark matter and thermal leptogenesis in gauge mediated supersymmetry breaking models,''
  Phys.\ Lett.\ B {\bf 549} (2002) 273
  [hep-ph/0208191].
  %%CITATION = HEP-PH/0208191;%%
  %92 citations counted in INSPIRE as of 27 Aug 2015

\bibitem{baer}
  H.~Baer, S.~Kraml, A.~Lessa and S.~Sekmen,
  %``Reconciling thermal leptogenesis with the gravitino problem in SUSY models with mixed axion/axino dark matter,''
  JCAP {\bf 1011} (2010) 040
%  doi:10.1088/1475-7516/2010/11/040
  [arXiv:1009.2959 [hep-ph]].
  %%CITATION = doi:10.1088/1475-7516/2010/11/040;%%
  %16 citations counted in INSPIRE as of 06 Dec 2015

\bibitem{arkani}
N.~Arkani-Hamed, S.~Dimopoulos, G.~F.~Giudice and A.~Romanino,
  %``Aspects of split supersymmetry,''
  Nucl.\ Phys.\ B {\bf 709} (2005) 3
%  doi:10.1016/j.nuclphysb.2004.12.026
  [hep-ph/0409232].
  %%CITATION = doi:10.1016/j.nuclphysb.2004.12.026;%%
  %486 citations counted in INSPIRE as of 06 Dec 2015

\bibitem{marfatia}
 S.~Blanchet, D.~Marfatia and A.~Mustafayev,
  %``Examining leptogenesis with lepton flavor violation and the dark matter abundance,''
  JHEP {\bf 1011} (2010) 038
  [arXiv:1006.2857 [hep-ph]].
  %%CITATION = ARXIV:1006.2857;%%
  %7 citations counted in INSPIRE as of 26 Aug 2015

\bibitem{sphalerons}
V.~A.~Kuzmin, V.~A.~Rubakov and M.~E.~Shaposhnikov,
  %``On the Anomalous Electroweak Baryon Number Nonconservation in the Early Universe,''
  Phys.\ Lett.\ B {\bf 155} (1985) 36.
  %%CITATION = PHLTA,B155,36;%%
  %2089 citations counted in INSPIRE as of 26 Aug 2015

  \bibitem{crv}
%\bibitem{Covi:1996wh}
  L.~Covi, E.~Roulet and F.~Vissani,
  %``CP violating decays in leptogenesis scenarios,''
  Phys.\ Lett.\ B {\bf 384} (1996) 169 [hep-ph/9605319].
  %%CITATION = HEP-PH/9605319;%%
  %610 citations counted in INSPIRE as of 12 Apr 2015
  
\bibitem{plumacher}
  M.~Plumacher,
  %``Baryon asymmetry, neutrino mixing and supersymmetric SO(10) unification,''
  Nucl.\ Phys.\ B {\bf 530} (1998) 207 [hep-ph/9704231].
  %%CITATION = HEP-PH/9704231;%%
  %157 citations counted in INSPIRE as of 18 sept. 2015

\bibitem{densitymatrix}
R.~Barbieri, P.~Creminelli, A.~Strumia and N.~Tetradis,
  %``Baryogenesis through leptogenesis,''
  Nucl.\ Phys.\ B {\bf 575} (2000) 61
  [hep-ph/9911315];
  %%CITATION = HEP-PH/9911315;%%
  %327 citations counted in INSPIRE as of 25 Sep 2015
A.~Abada, S.~Davidson, F.~X.~Josse-Michaux, M.~Losada and A.~Riotto,
  %``Flavor issues in leptogenesis,''
  JCAP {\bf 0604} (2006) 004
  [hep-ph/0601083];
   %%CITATION = HEP-PH/0601083;%%
  %292 citations counted in INSPIRE as of 25 Sep 2015
S.~Blanchet, P.~Di Bari and G.~G.~Raffelt,
  %``Quantum Zeno effect and the impact of flavor in leptogenesis,''
  JCAP {\bf 0703} (2007) 012
  [hep-ph/0611337];
  %%CITATION = HEP-PH/0611337;%%
  %75 citations counted in INSPIRE as of 25 Sep 2015
A.~De Simone and A.~Riotto,
  %``On the impact of flavour oscillations in leptogenesis,''
  JCAP {\bf 0702} (2007) 005
  [hep-ph/0611357];
  %%CITATION = HEP-PH/0611357;%%
  %57 citations counted in INSPIRE as of 25 Sep 2015  
M.~Beneke, B.~Garbrecht, C.~Fidler, M.~Herranen and P.~Schwaller,
  %``Flavoured Leptogenesis in the CTP Formalism,''
  Nucl.\ Phys.\ B {\bf 843} (2011) 177
%  doi:10.1016/j.nuclphysb.2010.10.001
  [arXiv:1007.4783 [hep-ph]];
  %%CITATION = doi:10.1016/j.nuclphysb.2010.10.001;%%
  %77 citations counted in INSPIRE as of 31 Dec 2015
P.~S.~Bhupal Dev, P.~Millington, A.~Pilaftsis and D.~Teresi,
  %``Flavour Covariant Transport Equations: an Application to Resonant Leptogenesis,''
  Nucl.\ Phys.\ B {\bf 886} (2014) 569
%  doi:10.1016/j.nuclphysb.2014.06.020
  [arXiv:1404.1003 [hep-ph]].  

\bibitem{densitymatrix2}
S.~Blanchet, P.~Di Bari, D.~A.~Jones and L.~Marzola,
  %``Leptogenesis with heavy neutrino flavours: from density matrix to Boltzmann equations,''
  JCAP {\bf 1301} (2013) 041
  [arXiv:1112.4528 [hep-ph]];
  %%CITATION = ARXIV:1112.4528;%%
  %22 citations counted in INSPIRE as of 25 Sep 2015

\bibitem{fong}
C.~S.~Fong, M.~C.~Gonzalez-Garcia, E.~Nardi and J.~Racker,
  %``Supersymmetric Leptogenesis,''
  JCAP {\bf 1012} (2010) 013
%  doi:10.1088/1475-7516/2010/12/013
  [arXiv:1009.0003 [hep-ph]].
  %%CITATION = doi:10.1088/1475-7516/2010/12/013;%%
  %20 citations counted in INSPIRE as of 06 Jan 2016
  
\bibitem{soft}
G.~D'Ambrosio, G.~F.~Giudice and M.~Raidal,
  %``Soft leptogenesis,''
  Phys.\ Lett.\ B {\bf 575} (2003) 75
  doi:10.1016/j.physletb.2003.09.037
  [hep-ph/0308031];
  %%CITATION = doi:10.1016/j.physletb.2003.09.037;%%
  %125 citations counted in INSPIRE as of 07 Jan 2016
 L.~Boubekeur, T.~Hambye and G.~Senjanovic,
  %``Low scale leptogenesis and soft supersymmetry breaking,''
  Phys.\ Rev.\ Lett.\  {\bf 93} (2004) 111601
  doi:10.1103/PhysRevLett.93.111601
  [hep-ph/0404038];
  %%CITATION = doi:10.1103/PhysRevLett.93.111601;%%
  %60 citations counted in INSPIRE as of 07 Jan 2016
Y.~Grossman, T.~Kashti, Y.~Nir and E.~Roulet,
  %``New ways to soft leptogenesis,''
  JHEP {\bf 0411} (2004) 080
  doi:10.1088/1126-6708/2004/11/080
  [hep-ph/0407063];
  %%CITATION = doi:10.1088/1126-6708/2004/11/080;%%
  %65 citations counted in INSPIRE as of 07 Jan 2016
M.~C.~Chen and K.~T.~Mahanthappa,
  %``Lepton flavor violating decays, soft leptogenesis and SUSY SO(10),''
  Phys.\ Rev.\ D {\bf 70} (2004) 113013
  doi:10.1103/PhysRevD.70.113013
  [hep-ph/0409096].
  %%CITATION = doi:10.1103/PhysRevD.70.113013;%%
  %53 citations counted in INSPIRE as of 07 Jan 2016

\bibitem{proceedings}
  P.~Di Bari,
  %``Leptogenesis, neutrino mixing data and the absolute neutrino mass scale,''
  Proceedings of the 39th Recontres de Moriond, 04 Electroweak interactions and unified theories: 
  La Thuile, Aosta, Italy, Mar 21-28, 2004, [hep-ph/0406115].
  %%CITATION = HEP-PH/0406115;%%
  %39 citations counted in INSPIRE as of 04 sept. 2015

\bibitem{predictions}
S.~Blanchet and P.~Di Bari,
  %``Flavor effects on leptogenesis predictions,''
  JCAP {\bf 0703} (2007) 018
  [hep-ph/0607330].
  %%CITATION = HEP-PH/0607330;%%
  %156 citations counted in INSPIRE as of 27 Sep 2015
      
\bibitem{2RHN}
S.~Antusch, P.~Di Bari, D.~A.~Jones and S.~F.~King,
  %``Leptogenesis in the Two Right-Handed Neutrino Model Revisited,''
  Phys.\ Rev.\ D {\bf 86} (2012) 023516 [arXiv:1107.6002 [hep-ph]].
  %%CITATION = ARXIV:1107.6002;%%
  %16 citations counted in INSPIRE as of 18 May 2015
  
 \bibitem{ks88}
S.~Yu.~Khlebnikov, M.~E.~Shaposhnikov, \np{308}{1988}{885};\\
J.~A.~Harvey, M.~S.~Turner, \pr{42}{1990}{3344}

\bibitem{engelhard}
G.~Engelhard, Y.~Grossman, E.~Nardi and Y.~Nir,
  %``The Importance of N2 leptogenesis,''
  Phys.\ Rev.\ Lett.\  {\bf 99} (2007) 081802
%  doi:10.1103/PhysRevLett.99.081802 [hep-ph/0612187].
  %%CITATION = doi:10.1103/PhysRevLett.99.081802;%%
  %80 citations counted in INSPIRE as of 01 Jan 2016

\bibitem{problem}
E.~Bertuzzo, P.~Di Bari and L.~Marzola,
  %``The problem of the initial conditions in flavoured leptogenesis and the tauon N_2-dominated scenario,''
  Nucl.\ Phys.\ B {\bf 849} (2011) 521
%  doi:10.1016/j.nuclphysb.2011.03.027
  [arXiv:1007.1641 [hep-ph]].
  %%CITATION = doi:10.1016/j.nuclphysb.2011.03.027;%%
  %17 citations counted in INSPIRE as of 01 Jan 2016

\bibitem{fusaoka}
H.~Fusaoka and Y.~Koide,
  %``Updated estimate of running quark masses,''
  Phys.\ Rev.\ D {\bf 57} (1998) 3986
%  doi:10.1103/PhysRevD.57.3986 
[hep-ph/9712201].
  %%CITATION = doi:10.1103/PhysRevD.57.3986;%%
  %308 citations counted in INSPIRE as of 05 Jan 2016

\bibitem{foglilisi2013}    
 F.~Capozzi, G.~L.~Fogli, E.~Lisi, A.~Marrone, D.~Montanino and A.~Palazzo,
  %``Status of three-neutrino oscillation parameters, circa 2013,''
  Phys.\ Rev.\ D {\bf 89} (2014) 093018
  [arXiv:1312.2878 [hep-ph]].
  %%CITATION = ARXIV:1312.2878;%%
  %261 citations counted in INSPIRE as of 09 Oct 2015   

\bibitem{wong}
J.~Hamann, S.~Hannestad and Y.~Y.~Y.~Wong,
  %``Measuring neutrino masses with a future galaxy survey,''
  JCAP {\bf 1211} (2012) 052
%  doi:10.1088/1475-7516/2012/11/052
  [arXiv:1209.1043 [astro-ph.CO]].
  %%CITATION = doi:10.1088/1475-7516/2012/11/052;%%
  %42 citations counted in INSPIRE as of 18 Dec 2015
  
  \bibitem{lowerbound}
P.~Di Bari, S.~King and M.~Re Fiorentin,
  %``Strong thermal leptogenesis and the absolute neutrino mass scale,''
  JCAP {\bf 1403} (2014) 050
 % doi:10.1088/1475-7516/2014/03/050
  [arXiv:1401.6185 [hep-ph]].
  %%CITATION = doi:10.1088/1475-7516/2014/03/050;%%
  %7 citations counted in INSPIRE as of 18 Dec 2015

\bibitem{plumacher2}
M.~Plumacher,
  %``Baryogenesis and lepton number violation,''
  Z.\ Phys.\ C {\bf 74} (1997) 549
 % doi:10.1007/s002880050418
  [hep-ph/9604229].
  %%CITATION = doi:10.1007/s002880050418;%%
  %244 citations counted in INSPIRE as of 08 Feb 2016

\bibitem{buchmuller}
M.~Bolz, W.~Buchmuller and M.~Plumacher,
  %``Baryon asymmetry and dark matter,''
  Phys.\ Lett.\ B {\bf 443} (1998) 209
%  doi:10.1016/S0370-2693(98)01342-2
  [hep-ph/9809381].
  %%CITATION = doi:10.1016/S0370-2693(98)01342-2;%%
  %153 citations counted in INSPIRE as of 18 Dec 2015


\bibitem{covi}
A.~Arbey, M.~Battaglia, L.~Covi, J.~Hasenkamp and F.~Mahmoudi,
  %``LHC constraints on Gravitino Dark Matter,''
  Phys.\ Rev.\ D {\bf 92} (2015) 11,  115008
%  doi:10.1103/PhysRevD.92.115008
  [arXiv:1505.04595 [hep-ph]].
  %%CITATION = doi:10.1103/PhysRevD.92.115008;%%

\bibitem{kuzmin}    
V.~A.~Kuzmin, V.~A.~Rubakov and M.~E.~Shaposhnikov,
  %``On the Anomalous Electroweak Baryon Number Nonconservation in the Early Universe,''
  Phys.\ Lett.\ B {\bf 155} (1985) 36.
%  doi:10.1016/0370-2693(85)91028-7
  %%CITATION = doi:10.1016/0370-2693(85)91028-7;%%
  %2146 citations counted in INSPIRE as of 17 Dec 2015    

\bibitem{running}
K.~S.~Babu, C.~N.~Leung, J.~T.~Pantaleone,
  %``Renormalization of the neutrino mass operator,''
  Phys.\ Lett.\ B {\bf 319} (1993) 191;
  %%CITATION = HEP-PH/9309223;%%
  %293 citations counted in INSPIRE as of 09 Jul 2013
S.~Antusch, J.~Kersten, M.~Lindner, M.~Ratz, M.~A.~Schmidt,
  %``Running neutrino mass parameters in see-saw scenarios,''
  JHEP {\bf 0503} (2005) 024.
  %%CITATION = HEP-PH/0501272;%%
  %197 citations counted in INSPIRE as of 09 Jul 2013
 
  
\end{thebibliography}
\end{document}